\documentclass[applsci,article,submit,pdftex,moreauthors]{Definitions/mdpi} 
\preto{\abstractkeywords}{\nolinenumbers} 

\firstpage{1} 
\pubvolume{1}
\issuenum{1}
\articlenumber{0}
\pubyear{2023}
\copyrightyear{2023}
\datereceived{ } 
\daterevised{ } 
\dateaccepted{ } 
\datepublished{ } 
\hreflink{https://doi.org/} 
\pdfoutput=1 

\usepackage{natbib}
\usepackage{amsmath,amssymb,amsfonts,bm}
\usepackage{algorithmic}
\usepackage{graphicx}
\usepackage{subfig}
\usepackage{textcomp}
\usepackage{booktabs}
\usepackage{xcolor}
\usepackage{makecell}
\usepackage{comment}
\usepackage{epstopdf}
\usepackage{appendix}

\Title{Resource Optimization Using A Step-by-step Scheme in Wireless Sensing and Localization Networks}

\TitleCitation{Resource Optimization Using A Step-by-step Scheme in Wireless Sensing and Localization Networks}

\Author{Ruihang Zhang $^{1}$\orcidA{}, Jiayan Yang $^{1}$, Mu Jia $^{1}$ and Tingting Zhang $^{1,}$*}

\AuthorNames{Ruihang Zhang, Jiayan Yang, Mu Jia and Tingting Zhang}

\AuthorCitation{Zhang, R.; Yang, J.; Jia, M.; Zhang, T.}

\address[1]{%
$^{1}$ \quad School of Electronics and Information Engineering, Harbin Institute of Technology, Shenzhen, P. R. China; 190210523@stu.hit.edu.cn; 249023600@qq.com; 20s152109@stu.hit.edu.cn\\}

\corres{Correspondence: zhangtt@hit.edu.cn}

\conference{VTC2023-spring} 

\abstract{Due to the lack of wireless spectrum resources, people are focusing on the versatile wireless networks. Wireless localization and target sensing both rely on precise extraction of parameters such as signal amplitude, propagation delay and Doppler shift from the received signals. Due to the high multi-path resolution and strong penetration of UWB signals, both localization and sensing can be achieved through the same UWB waveform. Practical networks are often resource-constrained, in order to improve the accuracy of integrated networks, we need to optimize the allocation of resources in the networks. Considering the complexity of the multi-slot networks, this paper derives the Fisher Information Matrix (FIM) expressions for single-slot and dual-slot integrated sensing and localization (ISAL) networks respectively, and proposes two resource optimization schemes, namely step-by-step scheme and integrated scheme. The numerical results show that: (i) for the sensing-resource-deficient networks with relatively uniform node distribution, the energy allocated to each step in the step-by-step scheme satisfies the relationship: energy for clock offset $<$ energy for radar localization $<$ energy for target sensing. (ii) In the multi-slot ISAL networks, the system will allocate more energy to the time slots where the networks are relatively sensing-resource-deficient. (iii) The step-by-step scheme is more suitable for the sensing-resource-abundant networks, while the integrated scheme is more suitable for the sensing-resource-deficient networks.}


\begin{document}

\section{Introduction}
\subsection{Background and Motivation}
With the development of the Internet of Things (IoTs) and massive machine type communications (mMTCs), more and more wireless devices require wireless connections \cite{hanying2022signalmultiplexing}. The surge in the number of wireless devices has led to a shortage of wireless spectrum resources. In recent years, people have found that reusing the functions of communication node networks has many advantages, such as improving spectrum utilization and reducing hardware equipment costs \cite{haojian2009cooperative}. Traditional radar positioning networks typically include two types of nodes: anchors with known positions and radars with unknown positions. The positions of radars are determined through measurements with other active nodes, including time of arrival (TOA) \cite{shen2010toa, wang2020toa}, time difference of arrival (TDOA) \cite{laoudias2018tdoa, zafari2019tdoa}, angle of arrival (AOA) \cite{gezici2005aoa, an2018aoa}, received signal strength (RSS) \cite{pavani2006rss, yaming2014rss} and so on. For passive target sensing networks, there are usually two types of nodes: active nodes with known positions and passive targets with unknown positions, where the positions of the targets are determined through time sum of arrival (TSOA) \cite{zhilong2012tsoa, xiaoneng2013tsoa} with other active nodes. The realization of wireless localization and target sensing relies on accurately extracting relevant parameters such as signal amplitude, propagation delay and Doppler shift from the received signals, and both can be achieved with high accuracy through UWB waveforms. The similarity between wireless localization and target sensing provides a prerequisite for the integration of localization and sensing networks.\par
Due to the fact that both wireless localization and target sensing are parameter estimation issues, before designing a parameter estimation algorithm for wireless localization or target sensing, we need to firstly obtain the theoretical estimation error lower bound to judge the accuracy of the designed algorithm. The authors in \cite{win2018navigation} show that inter-node measurements and intra-node measurements can bring spatial and temporal cooperation gains to the accuracy of localization systems, respectively. In addition, due to the fact that practical networks are often resource-constrained, in order to improve the accuracy of localization and sensing as much as possible, it is necessary to optimize the allocation of resources in the networks.
\subsection{Related Works}
Localization and positioning will be interchangeably used for estimation of the state (position, orientation) of a connected device in a global frame of reference \cite{wymeersch2022radio}. For the investigations of wireless localization, the authors in \cite{larsson2004crlb} show that the Cramer-Rao lower bound (CRLB) is commonly used as the lower bound of the variance of unbiased estimation parameters in wireless localization networks. The authors in \cite{shen2010fund1} introduce some basic concepts such as Fisher information matrix (FIM), equivalent Fisher information matrix (EFIM) and so on, and also provide the derivation of the localization FIM expressions in wireless localization networks. The authors in \cite{shen2010fund2} provide analysis for the error lower bound of localization in non-cooperative and spatial cooperative wireless localization networks, respectively. For radar localization networks which are difficult to meet trilateration, a method is provided to assist anchors by mutual ranging between radars to achieve accurate localization of radars. The authors in \cite{win2018navigation} provide specific FIM expressions of non-cooperative, spatial cooperative, and spatiotemporal cooperative wireless localization networks, respectively. It also provides another two lower bounds: Ziv-Zakai Lower Bound and Weiss-Weistein Lower Bound, and compares their performance with traditional CRLB's in the wireless localization networks. \par
Compared with localization and positioning, sensing has a broader scope covering from channel parameter estimation and carrier sensing to presence detection \cite{chaccour2022thz}. For the investigations of target sensing, the authors in \cite{wymeersch2022radio} discuss the geometric information brought by the different components in the sensing channel decomposition, such as the channel gain, the AOA angle, the TSOA delay and so on. The authors in \cite{liu2016passive} state the impossibility of equipping each target with communication devices in some circumstances such as intrusion detection and wildlife monitoring, and propose an inexpensive and efficient target sensing approach called RSS distribution-based localization (RDL), compared with other target sensing techniques, such as GPS and Channel State Information (CSI) \cite{xiao2013passive}. Although the aforementioned achievements include the geometric information of the sensing links and the implementation methods of target sensing, there is little research w.r.t analysis of the fundamental limits in target sensing networks. So far, the majority of research on fundamental limits has focused on wireless localization networks, and there is still little research on the fundamental limits of integrated sensing and localization (ISAL) networks. \par
For the research on resource optimization, the authors in \cite{shen2014power} proposed power optimization schemes between nodes for wireless localization networks and target sensing networks, respectively. The authors in \cite{william2012sdp} show that in wireless localization networks, the problem of anchor power optimization can be transformed into a semi-definite program (SDP), and specific SDP solutions are provided. The authors in \cite{jiamu2022power} find the regularity of node power optimization in single-slot synchronous ISAL networks, indicating that in the power-limited networks, in order to minimize target sensing error, when the anchor power allocation in the network remains unchanged, more power will be allocated to the radars closer to the target. Additionally, in \cite{jiamu2022power}, the optimization of power in ISAL networks is achieved by simultaneously optimizing all the variables. However, when the number of variables is large, the optimization problem is often too complicated to solve. Therefore, we need to explore a new optimization solution, and there has been no relevant research on this issue so far.
\subsection{Main Contributions}
The main contributions of this paper are as follows.\par
$\bullet$\quad We provide a single-slot ISAL network model with spatial cooperation information and a dual-slot ISAL network model with spatiotemporal cooperation information respectively, and propose the means of obtaining temporal cooperation.\par
$\bullet$\quad We give the derivation the fundamental limits for both synchronous and asynchronous ISAL networks, and provide the expressions for the FIMs in single-slot and dual-slot networks.\par
$\bullet$\quad We propose two energy and power optimization schemes for resource-constrained ISAL networks: a step-by-step scheme and an integrated scheme. We also propose an improvement method from the perspective of energy allocation to solve the problem of the high time complexity of the step-by-step scheme. By comparing the optimization results of the step-by-step scheme and the integrated scheme, we summarize the suitable scenarios of each scheme.\par
\textit{Notations:} We use lowercase and uppercase bold symbols to denote vectors and matrices, respectively. The lowercase bold subscript symbols (e.g., $\mathbf{a}$ in $(\mathbf{A})_{\mathbf{a}}$) represent the operation of taking submatrix corresponding to the parameter vector (i.e., $\mathbf{a}$ in the example) from the matrix (i.e., $\mathbf{A}$ in the example). The notation $\Vert \cdot \Vert$ is the Euclidean norm of its argument; $\text{tr}(\cdot)$ is the trace of a square matrix; $\mathbb{E}(\cdot)$ is the expectation operator of its argument; $|\cdot|$ represents the cardinality of the set; $\mathbf{A} \succeq \mathbf{B}$ denotes that the matrix $\mathbf{A} - \mathbf{B}$ is positive semi-definite.

\section{System Model}
When it comes to the system model of localization and sensing, there are four types of networks to be discussed, namely wireless localization networks, target sensing networks, single-slot static integrated sensing and localization networks, and multi-slot dynamic integrated sensing and localization networks. Considering the complexity of the multi-slot networks, this part discusses the dual-slot ($N=2$) dynamic ISAL networks instead of the multi-slot ones. \par
The general system description can be seen from Figure \ref{Wireless_Localization_Network} to Figure \ref{Dual-slot_Dynamic_ISAL_Network}. Assume that there are $N_{\rm a}$ anchors, $N_{\rm r}$ radars and $N_{\rm t}$ targets in each network, where the anchor positions are known and determined. In the dual-slot dynamic ISAL networks, radar and target positions remain unchanged in a single time slot, but change in different time slots. Anchors and radars called active nodes are both considered to be able to transmit and receive signals, while targets called passive nodes can only reflect signals.

\subsection{Wireless Localization Network}

\begin{figure}[htbp]
\centerline{\includegraphics[width=0.6\textwidth]{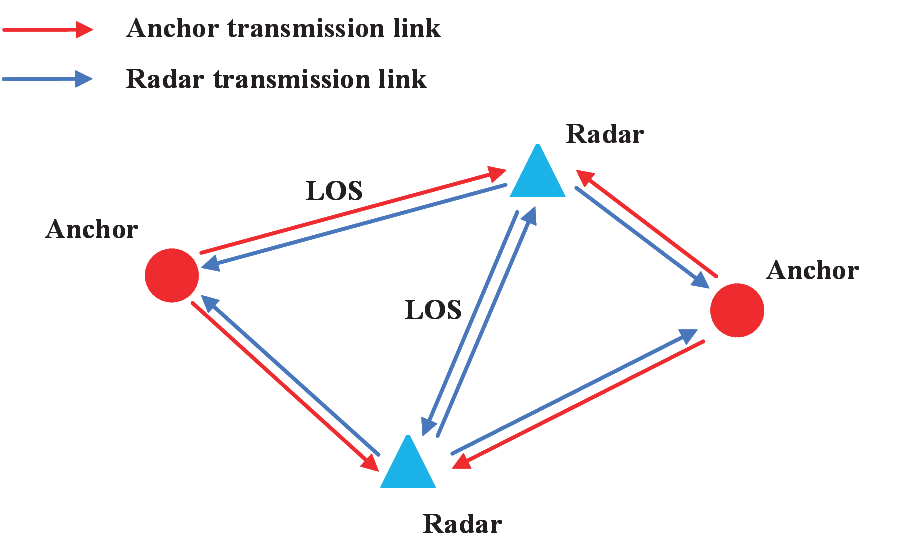}}
\caption{Wireless localization system model with spatial cooperation, where the red and blue solid line respectively represent the anchor transmission link and the radar transmission link.}
\label{Wireless_Localization_Network}
\end{figure}

As shown in Figure \ref{Wireless_Localization_Network}, there are two kinds of nodes in the network: anchors and radars. The anchors are nodes with known positions, while the positions of radars need to be estimated. Anchors and radars can obtain time of arrival (TOA) measurements by receiving line-of-sight (LOS) signals. At the same time, considering that radars can receive and transmit signals, we have introduced spatial cooperation between radars to assist in positioning radars in those scenarios where the number of anchors is limited.
\subsection{Target Sensing Network}

\begin{figure}[htbp]
\centerline{\includegraphics[width=0.6\textwidth]{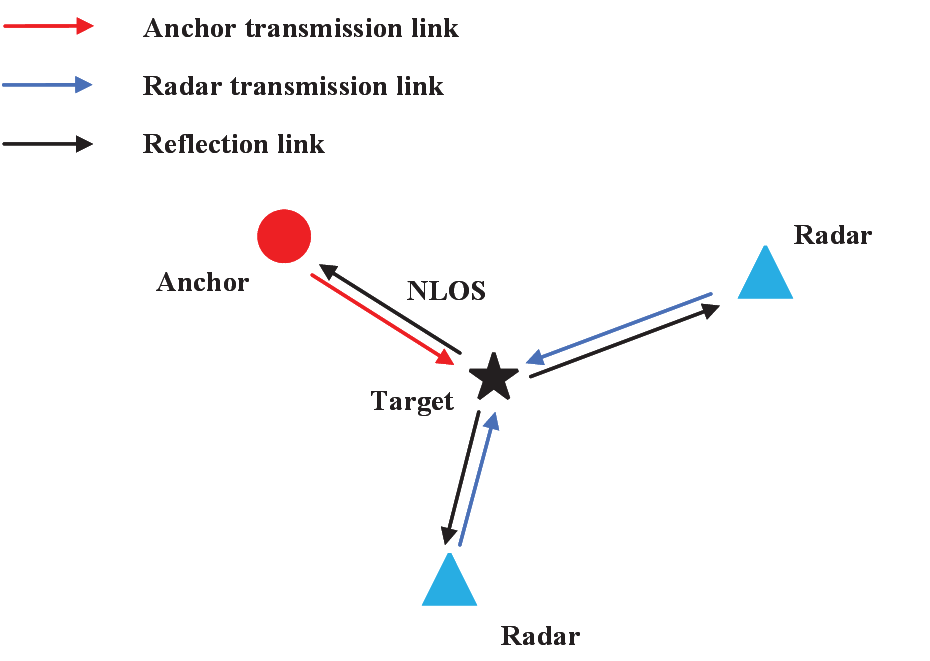}}
\caption{Target sensing system model, where the red, blue, and black solid line respectively represent the anchor transmission link, the radar transmission link and the reflection link.}
\label{Target_Sensing_Network}
\end{figure}

The target sensing network model is shown in Figure \ref{Target_Sensing_Network}. There are three kinds of nodes in the network: anchors, radars and targets, where the targets' positions need to be estimated. Anchors and radars can obtain time sum of arrival (TSOA) measurements by receiving non-line-of-sight (NLOS) signals reflected by the targets.
\subsection{Single-slot Static Integrated Sensing and Localization Network}

\begin{figure}[htbp]
\centerline{\includegraphics[width=0.65\textwidth]{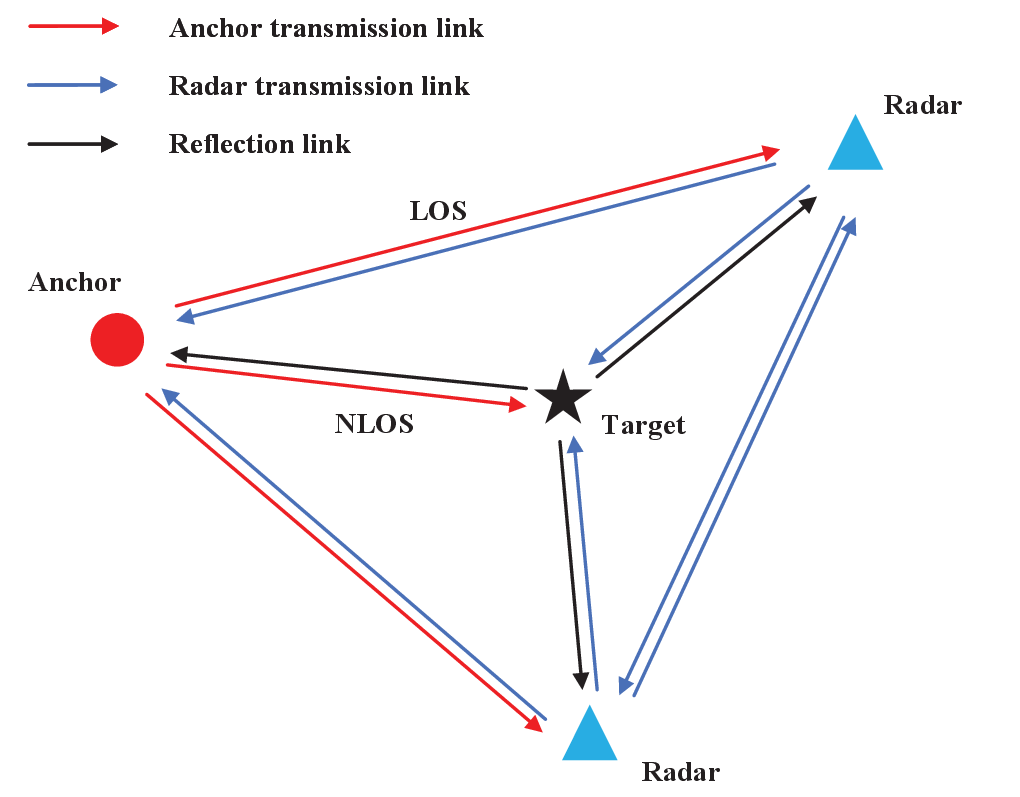}}
\caption{Integrated sensing and localization system model with spatial cooperation, where the red, blue, and black solid line respectively represent the anchor transmission link, the radar transmission link and the reflection link.}
\label{Single-slot_Static_ISAL_Network}
\end{figure}

Figure \ref{Single-slot_Static_ISAL_Network} introduces the single-slot static integrated sensing and localization network. In this network, localization and sensing are simultaneously considered, corresponding to the existence of both the localization transmission links and the sensing transmission links. Only the positions of anchors are known, while the positions of the other two kinds of nodes are both to be estimated. The signal propagation is as follows:\par
$\bullet$\quad LOS transmissions between anchors and radars: the anchors and radars send ranging signals to each other to position the radars with active localization. (Localization)\par
$\bullet$\quad LOS transmissions between radars themselves: the radars also send ranging signals to the other radars to help the active localization process with the spatial cooperative information contained. (Spatial cooperation for localization)\par
$\bullet$\quad NLOS transmissions via the passive targets: the anchors and radars called active nodes send sensing signals and reflected by the targets then received by the other active nodes. (Sensing)
\subsection{Dual-slot Dynamic Integrated Sensing and Localization Network}

\begin{figure}[htbp]
\centerline{\includegraphics[width=0.8\textwidth]{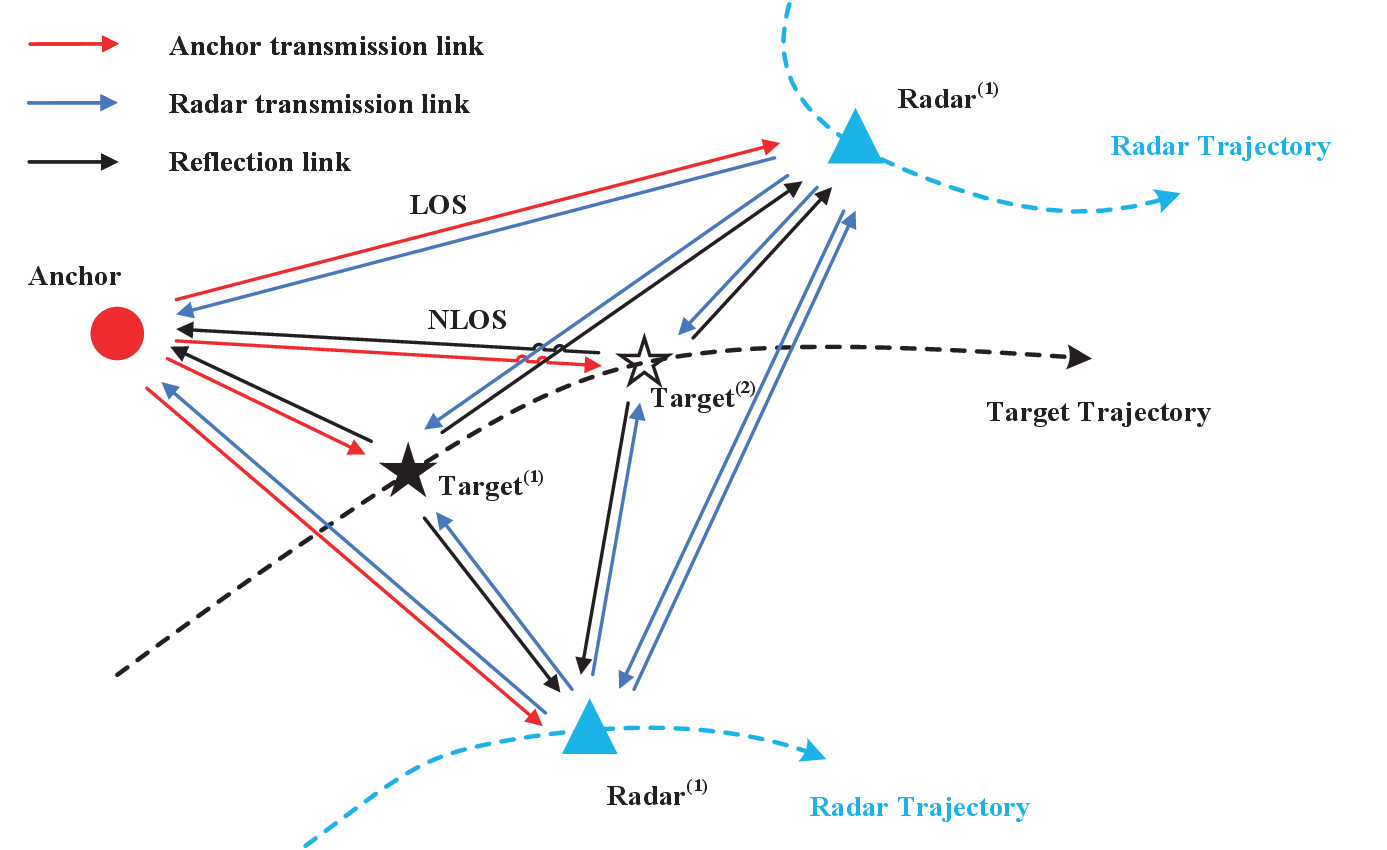}}
\caption{Integrated sensing and localization system model with spatiotemporal cooperation, where the red, blue, and black solid line respectively represent the anchor transmission link, the radar transmission link and the reflection link. The black dashed line and the blue dashed line are respectively the trajectories of the moving target and radars. The superscript $(n), n=1,2,\ldots,N$ represents the $n^{\text{th}}$ time slot.}
\label{Dual-slot_Dynamic_ISAL_Network}
\end{figure}

In Figure \ref{Dual-slot_Dynamic_ISAL_Network}, there is the system model of the dual-slot dynamic integrated sensing and localization network. In this network, radars and targets are both considered to be moveable, while only the positions of anchors remain the same in different time slots. The signal propagation in each time slot is the same as that of the single-slot static integrated sensing and localization network. The major difference is that the mobility of radars brings the possibility of introducing temporal cooperation between different time slots based on velocity information obtained from the intra-node measurements, such as Doppler measurements \cite{win2018navigation}. The temporal cooperation between different time slots can deliver the spatial localization information in time dimension.\par
The set of anchors, radars and targets are respectively represented as $\mathcal{N}_{\rm a} = \{1,2,\ldots,N_{\rm a}\}$, $\mathcal{N}_{\rm r} = \{1,2,\ldots,N_{\rm r}\}$ and $\mathcal{N}_{\rm t} = \{1,2,\ldots,N_{\rm t}\}$. The node position is expressed as $\mathbf{p}_k=[x_k,y_k]^{\text T}$,$k\in\mathcal{N}_{\rm a}\cup\mathcal{N}_{\rm r}\cup\mathcal{N}_{\rm t}$. The distance between different nodes $i$ and $j$ is
\begin{equation}
d_{ij}=\Vert\mathbf{p}_i-\mathbf{p}_j\Vert=\sqrt{(x_i-x_j)^2+(y_i-y_j)^2}
\end{equation}
angle between different nodes $i$ and $j$ is
\begin{equation}
\phi_{ij}=\arctan(\frac{y_j-y_i}{x_j-x_i})
\end{equation}\par
There are two kinds of networks to be discussed, respectively the synchronous networks and the asynchronous ones. In this article, we mainly discuss the asynchronous networks in which there are only clock offsets between the anchors and different radars, but no clock offsets between anchors. We first consider the two kinds of networks in which one way ranging (OWR) \cite{tingting2016JPBA} is used for range estimation. Then we provide a new ranging strategy called reverse-link-modified one way ranging (RLM-OWR) to introduce temporal cooperation between the two times slots in the estimation of clock offsets.\par
In synchronous networks, without considering multi-path propagation, the localization signal received by the $j^{\text{th}}$ active node from the $i^{\text{th}}$ active node can be expressed as
\begin{equation}
r_{ij}(t)=\alpha_{ij}s_{ij}(t-\Delta t_{ij})+e_{ij}(t)\label{r_ij}
\end{equation}
where $\alpha_{ij}$ and $\Delta t_{ij}$ are respectively the transmission attenuation and the delay of the LOS signals transmitted from the $i^{\text{th}}$ active node to the $j^{\text{th}}$ active node, $s_{ij}(t)$ and $e_{ij}(t)$ respectively represent the transmitted LOS signals and the measurement error on such link.\par
In synchronous networks, the sensing signal transmitted by the $i^{\text{th}}$ active node, received by the $j^{\text{th}}$ active node and reflected by the $k^{\text{th}}$ target can be expressed as
\begin{equation}
r_{ikj}(t)=\alpha_{ikj}s_{ikj}(t-\Delta t_{ikj})+e_{ikj}(t)\label{r_ijk}
\end{equation}
where $\alpha_{ikj}$ and $\Delta t_{ikj}$ are respectively the transmission attenuation and the delay of the NLOS signals transmitted from the $i^{\text{th}}$ active node via the $k^{\text{th}}$ target to the $j^{\text{th}}$ active node, $s_{ikj}(t)$ and $e_{ikj}(t)$ respectively represent the transmitted NLOS signals and the measurement error on such link.\par
In asynchronous networks, without considering multi-path propagation, the localization signal received by the $j^{\text{th}}$ active node from the $i^{\text{th}}$ active node can be expressed as
\begin{equation}
u_{ij}(t)=\alpha_{ij}s_{ij}(t-\tau_{ij}-\Delta t_{ij})+e_{ij}(t)\label{u_ij}
\end{equation}
where $\alpha_{ij}$, $\tau_{ij}$ and $\Delta t_{ij}$ are respectively the transmission attenuation, the clock offset between active node $i$ and $j$ and the delay of the LOS signals transmitted from the $i^{\text{th}}$ active node to the $j^{\text{th}}$ active node , $s_{ij}(t)$ and $e_{ij}(t)$ respectively represent the transmitted LOS signals and the measurement error on such link.\par
In asynchronous networks, the sensing signal transmitted by the $i^{\text{th}}$ active node, received by the $j^{\text{th}}$ active node and reflected by the $k^{\text{th}}$ target can be expressed as
\begin{equation}
u_{ikj}(t)=\alpha_{ikj}s_{ikj}(t-\tau_{ij}-\Delta t_{ikj})+e_{ikj}(t)\label{u_ijk}
\end{equation}
where $\alpha_{ikj}$, $\tau_{ij}$ and $\Delta t_{ikj}$ are respectively the transmission attenuation, the clock offset between active node $i$ and $j$ and the delay of the NLOS signals transmitted from the $i^{\text{th}}$ active node via the $k^{\text{th}}$ target to the $j^{\text{th}}$ active node, $s_{ikj}(t)$ and $e_{ikj}(t)$ respectively represent the transmitted NLOS signals and the measurement error on such link. \par
\begin{figure}[htbp]
\centerline{\includegraphics[width=0.65\textwidth]{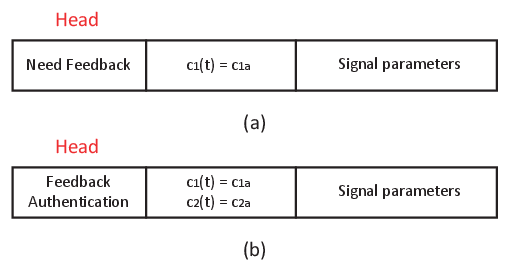}}
\caption{Reverse-link-modified one way ranging signal packet. (a) Forward link packet. (b) Reverse link packet.}
\label{RLM_OWR_packet}
\end{figure}

For the RLM-OWR, it is a ranging method that uses OWR for ranging and calibrates the signal propagation delay with the signals on the reverse link. This method can accurately obtain the relative clock drift rates between different clocks in the asynchronous ISAL networks. The schematic diagram is shown in Figure \ref{RLM_OWR} and Figure \ref{RLM_OWR_packet}:
\begin{figure}[htbp]
\centerline{\includegraphics[width=0.65\textwidth]{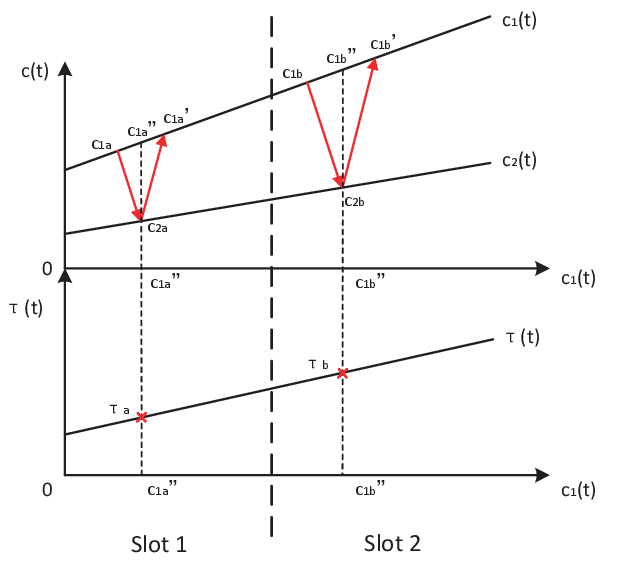}}
\caption{Reverse-link-modified one way ranging schematic diagram, where the red line represents the signal propagation between two active nodes.}
\label{RLM_OWR}
\end{figure}\par
\par
The specific realization of RLM-OWR is as follows: \par
(1) Taking the circumstance in slot 1 in Figure \ref{RLM_OWR} as an example, set the clock of node 1 $c_1(t)$ as the reference clock, record $c_{1\rm a}$ when $c_1(t) = c_{1\rm a}$, and broadcast signals with a "Need Feedback" time stamp which is shown as Figure \ref{RLM_OWR_packet}($\rm a$) to the environment by node 1; \par
(2) When the clock of node 2 $c_2(t) = c_{2\rm a}$, node 2 receives the LOS signal with a "Need Feedback" time stamp transmitted from node 1, node 2 records its time $c_{2\rm a}$ and transmits the received signal data packet to node 1 after unloading its head "Need Feedback" and adding another head "Feedback Authentication", the signal data packet transmitting to node 1 is shown as Figure \ref{RLM_OWR_packet}($\rm b$); \par
(3) When the clock of node 1 $c_1(t) = c_{1\rm a}'$, node 1 receives the LOS signal with a "Feedback Authentication" time stamp transmitted from node 2, node 1 records its time $c_{1\rm a}'$. As the time taken by the node to process the signal data packet in (2) is generally on the microsecond scale, it can be approximately assumed that the positions of the moving nodes remain unchanged during this period. Therefore, we take $c_{1\rm a}" = \frac{c_{1\rm a}+c_{1\rm a}'}{2}$ and there is $c_2(c_ {1\rm a}") = c_{2\rm a}$; \par
(4) We can get $\tau(c_{1\rm a}") = \tau_{\rm a} = c_{1\rm a}" - c_{2\rm a}$; \par
(5) Similarly in slot 2, record $c_{1\rm b}$, $c_{1\rm b}'$ and $c_{2\rm b}$, we take $c_ {1\rm b}" = \frac{c_{1\rm b}+c_{1\rm b}'}{2}$, and we can get $\tau(c_{1\rm b}") = \tau_{\rm b} = c_{1\rm b}" - c_{2\rm b}$; \par
(6) According to $\tau_{\rm a}$ and $\tau_{\rm b}$, we can obtain the slope of the line $\tau(t)-t$ in Figure \ref{RLM_OWR}, which is known as the relative clock drift rate $k_ {\tau} = \frac{\Delta \tau(t)}{\Delta t} = \frac{\tau_{\rm b} - \tau_{\rm a}}{c_{1\rm b}" - c_{1\rm a}"}$.\par

\section{Fundamental Limits}
In this section, we derive the fundamental limits for synchronous and asynchronous ISAL networks respectively. According to formula (\ref{r_ij}-\ref{u_ijk}), both localization and sensing can be considered as parameter estimation problems.
The parameters to be estimated in the $n^{\text{th}}$ time slot of the synchronous ISAL network can be written as
\begin{equation}
\bm{\theta}^{(n)}=[\mathbf{p}^{(n)\text{T}}_{\text{r}},\mathbf{p}^{(n)\text{T}}_{\text{t}}]^{\text{T}}
\label{parameters_syn}
\end{equation}
where $\mathbf{p}^{(n)}_{\text{r}}$ and $\mathbf{p}^{(n)}_{\text{t}}$ are the position vectors of radars and targets in the $n^{\text{th}}$ time slot respectively.\par
The parameters to be estimated in the $n^{\text{th}}$ time slot of the asynchronous ISAL network can be written as
\begin{equation}
\bm{\theta}^{(n)}=[\mathbf{p}^{(n)\text{T}}_{\text{r}},\bm{\tau}^{(n)\text{T}},\mathbf{p}^{(n)\text{T}}_{\text{t}}]^{\text{T}}
\label{parameters_asyn}
\end{equation}
where $\mathbf{p}^{(n)}_{\text{r}}$, $\mathbf{p}^{(n)}_{\text{t}}$ and $\bm{\tau}^{(n)}$ are the position vectors of radars and targets and the time offsets between the anchors and different radars in the $n^{\text{th}}$ time slot respectively.\par
According to \cite{larsson2004crlb}, Cramer-Rao lower bound (CRLB) is
often used as the lower bound of the variance of unbiased estimator. In the ISAL networks, we introduce the squared position error bound (SPEB) instead of CRLB as the lower bound of the mean square error (MSE), which is defined as
\begin{equation}
\mathcal{P}(\bm{\theta}_1)\triangleq \text{tr}(\mathbf{J}^{-1}_{\text{e}}(\bm{\theta}_1))\leq \mathbb{E}(\Vert\bm{\theta}_1-\hat{\bm{\theta}_1}\Vert^2)
\end{equation}
where $\bm{\theta}_1$ and $\hat{\bm{\theta}_1}$ respectively represent one of the parameters in $\bm{\theta}$ and an estimate of the parameter vector $\bm{\theta}_1$ based on an observation of the received signals, and $\mathbf{J}_{\text{e}}(\bm{\theta}_1)$ represents the equivalent Fisher Information Matrix (EFIM) of the parameter $\bm{\theta}_1$ \cite{shen2010fund1}.\par
\subsection{FIM of Synchronous Networks}
For a single objective ($N_{\rm t}=1$) synchronous ISAL network applying OWR method, another parameter vector that satisfies a mapping with the parameters $\bm{\theta} = [\mathbf{p}^{\text{T}}_{\text{r}},\mathbf{p}^{\text{T}}_{\text{t}}]^{\text{T}}$ is defined as
\begin{equation}
\bm{\gamma} = [\tau_{1,1}^{\text{sen}},\cdots,\tau_{N_{\rm r}+N_{\rm a},N_{\rm r}+N_{\rm a}}^{\text{sen}},\tau_{1,2}^{\text{ran}},\cdots,\tau_{N_{\rm r}+N_{\rm a},N_{\rm r}}^{\text{ran}}]^{\text{T}}
\label{parameter_jacob}
\end{equation}
where $\tau_{m,n}^{\text{sen}} = \frac{d_{m,tar}+d_{tar,n}}{c},m,n\in\{m,n|m,n\in\mathcal{N}_{\rm r}\cup\mathcal{N}_{\rm a}\}$,$\tau_{m,n}^{\text{ran}} = \frac{d_{m,n}}{c},m,n\in\{m,n|m\in\mathcal{N}_{\rm r},n\in\mathcal{N}_{\rm r}\cup\mathcal{N}_{\rm a},m\neq n\}\cup\{m,n|m\in\mathcal{N}_{\rm a},n\in\mathcal{N}_{\rm r}\}$.\par
Assuming that the received signals $\mathbf r = [r_{1,1}^{\text{sen}},\cdots,r_{N_{\rm r}+N_{\rm a},N_{\rm r}+N_{\rm a}}^{\text{sen}},r_{1,2}^{\text{ran}},\cdots,r_{N_{\rm r}+N_{\rm a},N_{\rm r}}^{\text{ran}}]^{\text{T}}$, due to the independence of the received signals corresponding to different links, according to the definition of FIM, the conditional probability density function
\begin{equation}
f(\mathbf r|\bm{\gamma}) = \prod_{m,n\in N_{\rm r}\cup N_{\rm a}}f(r_{m,n}^{\text{sen}}|\bm{\gamma})\prod_{m,n\in \{m,n|m\in\mathcal{N}_{\rm r},n\in\mathcal{N}_{\rm r}\cup\mathcal{N}_{\rm a},m\neq n\}\cup\{m,n|m\in\mathcal{N}_{\rm a},n\in\mathcal{N}_{\rm r}\}}f(r_{m,n}^{\text{ran}}|\bm{\gamma})
\end{equation}
The FIM for the parameter vector $\bm{\gamma}$ is
\begin{equation}
\mathbf I_{\bm{\gamma}} = \mathbb{E}_{\mathbf r|\bm{\gamma}} \{[\frac{\partial \text{ln}f(\mathbf r|\bm {\gamma})}{\partial \bm{\gamma}}][\frac{\partial \text{ln}f(\mathbf r|\bm {\gamma})}{\partial \bm{\gamma}}]^{\text{T}}\}
\end{equation}
after some derivation $\bm{\gamma}$ can be written as
\begin{equation}
\mathbf I_{\bm{\gamma}} = c^2*\left[
\begin{array}{ccc}
\lambda_{1,tar,1} & \cdots & 0\\
\vdots & \ddots & \vdots\\
0 & \cdots & \lambda_{N_{\rm r}+N_{\rm a},N_{\rm r}}
\end{array}\right]
\label{FIM_jacob}
\end{equation}
where $\lambda_{m,n}, m,n\in\{m,n|m\in\mathcal{N}_{\rm r},n\in\mathcal{N}_{\rm r}\cup\mathcal{N}_{\rm a},m\neq n\}\cup\{m,n|m\in\mathcal{N}_{\rm a},n\in\mathcal{N}_{\rm r}\}$ and $\lambda_{m,tar,n}, m,n\in\{m,n|m,n\in\mathcal{N}_{\rm r}\cup\mathcal{N}_{\rm a}\}$ respectively represent the range information intensity (RII) of the localization and sensing links, which can be expressed as
\begin{equation}
\lambda_{m,n} = \frac{8\uppi ^2B^2}{c^2} \frac{P_mG_mG_nc^2}{(4\uppi)^2d_{m,n}^2N_0BLf_c^2} = \xi_{m,n} \frac{P_m}{d_{m,n}^2}
\label{RII_loc}
\end{equation}
\begin{equation}
\lambda_{m,tar,n} = \frac{8\uppi ^2B^2}{c^2} \frac{P_mG_mG_nc^2}{(4\uppi)^2d_{m,tar}^2d_{tar,n}^2N_0BLf_c^2}\frac{\sigma}{4\uppi} = \xi_{m,tar,n} \frac{P_m}{d_{m,tar}^2d_{tar,n}^2}\frac{\sigma}{4\uppi}
\label{RII_sen}
\end{equation}
where $m,tar$ and $n$ respectively represent the $m^{\text{th}}$ transmitting node, the target node and the $n^{\text{th}}$ receiving node, $B$ is the signal bandwidth, $c$ is the speed of light, $P_m$ is the signal transmission power, $G_m, G_n$ respectively represent the antenna gain of the transmitter and receiver, $N_0$ is the power spectral density of environmental thermal noise, $L$ is the communication system loss, $f_c$ is the carrier frequency of the signal, $\sigma$ is the radar cross section, $\xi_{mn}$, $\xi_{mkn}$ are respectively the common constant coefficient in the RII of the localization link and the sensing link. \par 
By means of a bijective transformation, the FIM for the parameter vector $\bm{\theta}$ can be expressed as
\begin{equation}
\mathbf I_{\bm{\theta}} = \mathbf{J}^{\text{T}}\mathbf I_{\bm{\gamma}}\mathbf{J}
\label{jacobian_trans}
\end{equation}
where $\mathbf{J}$ is the Jacobian matrix for the transmission from $\bm{\theta}$ to $\bm{\gamma}$, which can be written as:
\begin{equation}
\mathbf{J} = \frac{\partial {\bm{\gamma}}}{\partial {\bm{\theta}}} = [\frac{\partial {\bm{\gamma}}}{\partial {\mathbf{p}_{\text{r}}}},\frac{\partial {\bm{\gamma}}}{\partial {\mathbf{p}_{\text{t}}}}]
\end{equation}
where the specific expressions of $\frac{\partial {\bm{\gamma}}}{\partial {\mathbf{p}_{\text{r}}}}$ and $\frac{\partial {\bm{\gamma}}}{\partial {\mathbf{p}_{\text{t}}}}$ are given in the Appendix A.1. \par
Based on the results above, the FIM expression for the single-slot synchronous ISAL network can be derived, which can be expressed in the following form:
\begin{equation}
\mathbf{I}_{\bm {\theta}}^{\text{syn}} = \left[
\begin{array}{cc}
\mathbf{J}^{\text{R}}_{\text{e}} & \mathbf{S}_{\text{R},\text{T}}\\
\mathbf{S}_{\text{R},\text{T}}^{\text{T}} & \mathbf{J}^{\text{T}}_{\text{e}}\\
\end{array}\right]_{2(N_{\rm r}+N_{\rm t})\times2(N_{\rm r}+N_{\rm t})}
\label{syn_singleslot}
\end{equation}
where $\mathbf{J}^{\text{R}}_{\text{e}}$, $\mathbf{J}^{\text{T}}_{\text{e}}$ and $\mathbf{S} _{\text{R},\text{T}}$ are the submatrices in $\mathbf{I}_{\bm {\theta}}^{\text{syn}}$ respectively with respect to the radar localization information, the target sensing information and the spatial cooperation information between radar localization and target sensing.\par
As for the FIM of multi-slot synchronous ISAL networks, since we can obtain the velocity information of the moving active nodes in the networks through Doppler measurement between time slots, we can predict the positions of the moving active nodes in the next time slot, corresponding to the time cooperation information of radar positioning between time slots in the FIMs. Considering the complexity of the multi-slot ISAL networks, we give the expression of the FIM of the dual-slot synchronous ISAL networks:
\begin{equation}
\mathbf{I}_{\bm {\theta}}^{\text{syn}(2)} = \left[
\begin{array}{cccc}
\mathbf{J}^{\text{R}(1)}_{\text{e}} + \mathbf{T}_{\text{Loc}}^{(1)} & \mathbf{S}_{\text{R},\text{T}}^{(1)} & \mathbf{T}_{\text{Loc}}^{(1)} & \mathbf{0}\\
\mathbf{S}_{\text{R},\text{T}}^{(1)\text{T}} & \mathbf{J}^{\text{T}(1)}_{\text{e}} & \mathbf{0} & \mathbf{0}\\
\mathbf{T}_{\text{Loc}}^{(1)} & \mathbf{0} & \mathbf{J}^{\text{R}(2)}_{\text{e}} + \mathbf{T}_{\text{Loc}}^{(1)} & \mathbf{S}_{\text{R},\text{T}}^{(2)}\\
\mathbf{0} & \mathbf{0} & \mathbf{S}_{\text{R},\text{T}}^{(2)\text{T}} & \mathbf{J}^{\text{T}(2)}_{\text{e}}\\
\end{array}\right]
\label{syn_dualslot}
\end{equation}
where superscript $(n),n = 1,2$ represents the $n^{th}$ time slot, and submatrix
$\mathbf{T}_{\text{Loc}}$ contains the temporal cooperation information on radar positioning, which is expressed as
\begin{equation}
\mathbf{T}_{\text{Loc}}=\left[
\begin{array}{cccc}
\frac{1}{\eta^2} & 0 & \cdots & 0\\
0 & \frac{1}{\eta^2} & \cdots & 0\\
\vdots & \vdots & \ddots & \vdots\\
0 & 0 & \cdots & \frac{1}{\eta^2}\\
\end{array}\right]_{2N_{\rm r}\times2N_{\rm r}}
\label{tloc}
\end{equation}
where $\eta^2$ is the variance of node velocity.

\subsection{FIM of Asynchronous Networks}
For a single objective $(N_{\rm t} = 1)$ asynchronous ISAL network applying OWR method, another parameter vector that satisfies a mapping with the parameters $\bm {\theta} = [\mathbf{p}^{\text{T}}_{\text{r}},\bm{\tau}^{\text{T}},\mathbf{p}^{\text{T}}_{\text{t}}]^{\text{T}}$ is defined as
\begin{equation}
\bm{\zeta} = [\upsilon_{1,1}^{\text{sen}},\cdots,\upsilon_{N_{\rm r}+N_{\rm a},N_{\rm r}+N_{\rm a}}^{\text{sen}},\upsilon_{1,2}^{\text{ran}},\cdots,\upsilon_{N_{\rm r}+N_{\rm a},N_{\rm r}}^{\text{ran}}]^{\text{T}}
\end{equation}
where $\upsilon_{m,n}^{\text{sen}} = \frac{d_{m,tar}+d_{tar,n}}{c} + \tau_{m,n},m,n\in\{m,n|m,n\in\mathcal{N}_{\rm r}\cup\mathcal{N}_{\rm a}\}$,$\upsilon_{m,n}^{\text{ran}} = \frac{d_{m,n}}{c} + \tau_{m,n},m,n\in\{m,n|m\in\mathcal{N}_{\rm r},n\in\mathcal{N}_{\rm r}\cup\mathcal{N}_{\rm a},m\neq n\}\cup\{m,n|m\in\mathcal{N}_{\rm a},n\in\mathcal{N}_{\rm r}\}$.\par
Similar to the synchronous ISAL networks, we can obtain the FIM of $\bm{\zeta}$ expressed as (\ref{FIM_jacob}), and the FIM of $\bm{\theta}$ can also be obtained from (\ref{jacobian_trans}).\par
The Jacobian matrix for the transformation from $\bm {\theta}$ to $\bm {\zeta}$ in the asynchronous ISAL network can be expressed as
\begin{equation}
\mathbf{J} = \frac{\partial {\bm{\zeta}}}{\partial {\bm{\theta}}} = [\frac{\partial {\bm{\zeta}}}{\partial {\mathbf{p}_{\text{r}}}},\frac{\partial {\bm{\zeta}}}{\partial {\bm{\tau}}},\frac{\partial {\bm{\zeta}}}{\partial {\mathbf{p}_{\text{t}}}}]
\end{equation}
where the specific expressions of $\frac{\partial {\bm{\zeta}}}{\partial {\mathbf{p}_{\text{r}}}}, \frac{\partial {\bm{\zeta}}}{\partial {\bm{\tau}}}$ and $\frac{\partial {\bm{\zeta}}}{\partial {\mathbf{p}_{\text{t}}}}$ are given in the Appendix A.2. \par
Based on the results above, the FIM expression of the single-slot asynchronous ISAL networks can be derived, which can be expressed in the following form:
\begin{equation}
\mathbf{I}_{\bm {\theta}}^{\text{asyn}} = \left[
\begin{array}{ccc}
\mathbf{J}^{\text{R}}_{\text{e}} & \mathbf{S}_{\text{R},\tau} & \mathbf{S}_{\text{R},\text{T}}\\
\mathbf{S}_{\text{R},\tau}^{\text{T}} & \mathbf{J}^{\tau}_{\text{e}} & \mathbf{S}_{\text{T},\tau}\\
\mathbf{S}_{\text{R},\text{T}}^{\text{T}} & \mathbf{S}_{\text{T},\tau}^{\text{T}} & \mathbf{J}^{\text{T}}_{\text{e}}\\
\end{array}\right]_{[2(N_{\rm r}+N_{\rm t})+|\mathcal{N}_{\rm r}|]\times[2(N_{\rm r}+N_{\rm t})+|\mathcal{N}_{\rm r}|]}
\label{asyn_singleslot}
\end{equation}
where $\mathbf{J}^{\tau}_{\text{e}}$, $\mathbf{S}_{\text{R},\tau}$ and $\mathbf{S}_{\text{T},\tau}$ are the submatrices in $\mathbf{I}_{\bm {\theta}}^{\text{asyn}}$ respectively with respect to the clock offset information, the cooperation information between radar localization and clock offset and the cooperation information between target sensing and clock offset.\par
According to the principle of RLM-OWR, for the FIM of multi-slot asynchronous ISAL networks, in addition to introducing radar positioning temporal cooperation information between time slots through Doppler measurement, we can also obtain relative clock drift rate through RLM-OWR, thereby further introducing temporal cooperation information for estimating node clock difference between time slots. Considering the complexity of the multi-slot ISAL networks, here we give the expression of the FIM of the dual-slot asynchronous ISAL networks applying RLM-OWR, which is shown as (\ref{asyn_dualslot}):
\begin{equation}
\mathbf{I}_{\bm {\theta}}^{\text{asyn}(2)} = \left[
\begin{array}{cccccc}
\mathbf{J}^{\text{R}(1)}_{\text{e}} + \mathbf{T}_{\text{Loc}}^{(1)} & \mathbf{S}_{\text{R},\tau}^{(1)} & \mathbf{S}_{\text{R},\text{T}}^{(1)} & \mathbf{T}_{\text{Loc}}^{(1)} & \mathbf{0} & \mathbf{0}\\
\mathbf{S}_{\text{R},\tau}^{(1)\text{T}} & \mathbf{J}^{\tau(1)}_{\text{e}} + \mathbf{T}_{\tau}^{(1)} & \mathbf{S}_{\text{T},\tau}^{(1)} & \mathbf{0} & \mathbf{T}_{\tau}^{(1)} & \mathbf{0}\\
\mathbf{S}_{\text{R},\text{T}}^{(1)\text{T}} & \mathbf{S}_{\text{T},\tau}^{(1)\text{T}} & \mathbf{J}^{\text{T}(1)}_{\text{e}} & \mathbf{0} & \mathbf{0} & \mathbf{0}\\
\mathbf{T}_{\text{Loc}}^{(1)} & \mathbf{0} & \mathbf{0} & \mathbf{J}^{\text{R}(2)}_{\text{e}} + \mathbf{T}_{\text{Loc}}^{(1)} & \mathbf{S}_{\text{R},\tau}^{(2)} & \mathbf{S}_{\text{R},\text{T}}^{(2)}\\
\mathbf{0} & \mathbf{T}_{\tau}^{(1)} & \mathbf{0} & \mathbf{S}_{\text{R},\tau}^{(2)\text{T}} &  \mathbf{J}^{\tau(2)}_{\text{e}} + \mathbf{T}_{\tau}^{(1)} & \mathbf{S}_{\text{T},\tau}^{(2)}\\
\mathbf{0} & \mathbf{0} & \mathbf{0} & \mathbf{S}_{\text{R},\text{T}}^{(2)\text{T}} & \mathbf{S}_{\text{T},\tau}^{(2)\text{T}} & \mathbf{J}^{\text{T}(2)}_{\text{e}}\\
\end{array}\right]
\label{asyn_dualslot}
\end{equation}
where submatrix $\mathbf{T}_{\tau}$ contains the temporal cooperation information on the clock offset, which is expressed as
\begin{equation}
\mathbf{T}_{\tau}=\left[
\begin{array}{cccc}
\frac{1}{\rho^2} & 0 & \cdots & 0\\
0 & \frac{1}{\rho^2} & \cdots & 0\\
\vdots & \vdots & \ddots & \vdots\\
0 & 0 & \cdots & \frac{1}{\rho^2}\\
\end{array}\right]_{|\mathcal{N}_{\rm r}|\times|\mathcal{N}_{\rm r}|}
\label{ttau}
\end{equation}
where $\rho^2$ is the variance of relative clock drift rate.

\section{Energy and Power Optimization Allocation}
In practical ISAL networks, network resources such as energy and bandwidth are often constrained. Therefore, in order to improve the accuracy of localization and sensing as much as possible, it is necessary to optimize the allocation of network resources. In this section, we mainly discussed the optimal allocation of power in the single-slot ISAL networks, as well as the optimal allocation of energy between time slots in the  multi-slot ISAL networks. We propose two resource optimization schemes: a step-by-step scheme and an integrated scheme, and summarize the suitable scenarios for each of them. \par
Firstly, we introduce the meanings of the step-by-step scheme and the integrated scheme respectively. For an ISAL network, assuming that the parameters are $\bm {\theta} = [\bm {\theta}_1^{\text{T}},\cdots,\bm {\theta}_N^{\text{T}}]$, in the integrated optimization scheme, we simultaneously optimize and allocate resources to all of the N variables in $\bm {\theta}$; in the step-by-step optimization scheme, we only optimize and allocate resources to one single variable in $\bm {\theta}$, and apply the resource optimization allocation result of the previous variable as a prior knowledge for the optimization of the next variable. This converts the multi-variable optimization problem into multiple single-variable optimization problems and solves them step by step, which can avoid the optimization difficulties caused by dealing with multiple variables at the same time, but it also comes at a cost of high time complexity. \par
Based on the statement above, we need to minimize the time complexity of the step-by-step scheme when dealing with multi-variable optimization problems such as those in ISAL networks. In addition to designing optimization algorithms with low time complexity, this article explores the resource allocation of ISAL networks with different topologies and finds the regularity of energy allocation for step-by-step optimization schemes in the networks with different topologies. This provides certain reference value for practical step-by-step optimization schemes in ISAL networks. In addition, this article also analyzes the energy allocation between time slots in the dual-slot ISAL networks.
\subsection{Integrated Optimization Scheme}
\subsubsection{Optimization for Synchronous Networks}
For the single-slot synchronous ISAL networks, the FIM of $\bm {\theta} = [\mathbf{p}_{\rm r}^{\text{T}},\mathbf{p}_{\rm t}^{\text{T}}]^{\text{T}}$ is as (\ref{syn_singleslot}), the power optimization allocation strategy is as follows:
\begin{alignat}{2}
\mathfrak{B}_{1}:\quad & \min\quad \text{tr}(\mathbf{J}^{\text{T}-1}_{\text{e}})\\
\text{s.t.}\quad & \sum\limits_{j\in\mathcal{N}_{\rm a}\cup\mathcal{N}_{\rm r}} P_j\leq E_{\text{total}}/1 \label{total_limit}\\
& 0\leq P_j\leq P_{\text{r},\text{max}}, \forall j\in\mathcal{N}_{\rm r} \label{radar_limit}\\
& 0\leq P_j\leq P_{\text{a},\text{max}}, \forall j\in\mathcal{N}_{\rm a} \label{anchor_limit}\\\notag
\end{alignat}
where $E_{\text{total}}$ is the total energy of the integrated sensing and localization network, $P_{\text{a},\text{max}}$ and $P_{\text{r},\text{max}}$ are the upper limit of power allocated to each anchor and radar respectively within a single slot.\par
According to \cite{william2012sdp}, problem $\mathfrak{B}_{1}$ can be reformulated into a standard semi-definite programming (SDP) problem as follows:
\begin{alignat}{2}
\mathfrak{A}_{1}^{\text{SDP}}:\quad & \min\quad \text{tr}(\mathbf{H})_{\mathbf{p}_{\rm t}} \\
\text{s.t.}\quad & \mathbf{H} \succeq 0\\
& \left[\begin{array}{cc}
\mathbf{H} & \mathbf{I}\\
\mathbf{I} & \mathbf{I}_{\bm{\theta}}^{\text{syn}}\\
\end{array}\right] \succeq 0 \\
& (\ref{total_limit}-\ref{anchor_limit}) \notag\\\notag
\end{alignat}
where $\mathbf{H}$ is called the auxiliary matrix of the FIM, and $\mathbf{I}$ represents the identity matrix whose dimension is the same as the FIM.\par
For the dual-slot synchronous ISAL networks, the FIM expression is as (\ref{syn_dualslot}). We search for the optimal energy allocation between two time slots through traversal, where the energy allocated to slot 1 and slot 2 is respectively $E_{\text{slot1}}$ and $E_{\text{slot2}}$, satisfying $E_{\text{total}} = E_{\text{slot1}} + E_{\text{slot2}}$. The power optimization allocation strategy within slot 1 is the same as that of the single-slot synchronous ISAL networks mentioned above. The power optimization allocation within slot 2 applies the result in slot 1 as a prior knowledge, which is shown as follows:
\begin{alignat}{2}
\mathfrak{A}_{2}^{\text{SDP}}:\quad & \min\quad \text{tr}(\mathbf{H})_{\mathbf{p}_{\rm t}^{(2)}} \\
\text{s.t.}\quad & \mathbf{H} \succeq 0\\
& \left[\begin{array}{cc}
\mathbf{H} & \mathbf{I}\\
\mathbf{I} & \mathbf{I}_{\bm{\theta}}^{\text{syn}(2)}\\
\end{array}\right] \succeq 0 \\
& \sum\limits_{j\in\mathcal{N}_{\rm a}\cup\mathcal{N}_{\rm r}} P_j\leq E_{\text{slot2}}/1 \\
& (\ref{radar_limit}-\ref{anchor_limit}) \notag\\\notag
\end{alignat}
\subsubsection{Optimization for Asynchronous Networks}
For the single-slot asynchronous ISAL networks, the FIM of $\bm {\theta} = [\mathbf{p}_{\rm r}^{\text{T}},\bm {\tau}^{\text{T}},\mathbf{p}_{\rm t}^{\text{T}}]^{\text{T}}$ is as (\ref{asyn_singleslot}), the power optimization allocation strategy is as follows:
\begin{alignat}{2}
\mathfrak{A}_{3}^{\text{SDP}}:\quad & \min\quad \text{tr}(\mathbf{H})_{\mathbf{p}_{\rm t}} \\
\text{s.t.}\quad & \mathbf{H} \succeq 0\\
& \left[\begin{array}{cc}
\mathbf{H} & \mathbf{I}\\
\mathbf{I} & \mathbf{I}_{\bm{\theta}}^{\text{asyn}}\\
\end{array}\right] \succeq 0 \\
& (\ref{total_limit}-\ref{anchor_limit}) \notag\\\notag
\end{alignat}\par
For the dual-slot asynchronous ISAL networks, the FIM is expressed as (\ref{asyn_dualslot}). Similarly to the synchronous networks, we search for the optimal energy allocation between two time slots through traversal. The power optimization allocation strategy within slot 1 is the same as that of the single-slot asynchronous networks. The power optimization allocation within slot 2 applies the result in slot 1 as a prior knowledge, which is shown as follows:
\begin{alignat}{2}
\mathfrak{A}_{4}^{\text{SDP}}:\quad & \min\quad \text{tr}(\mathbf{H})_{\mathbf{p}_{\rm t}^{(2)}} \\
\text{s.t.}\quad & \mathbf{H} \succeq 0\\
& \left[\begin{array}{cc}
\mathbf{H} & \mathbf{I}\\
\mathbf{I} & \mathbf{I}_{\bm{\theta}}^{\text{asyn}(2)}\\
\end{array}\right] \succeq 0 \\
& \sum\limits_{j\in\mathcal{N}_{\rm a}\cup\mathcal{N}_{\rm r}} P_j\leq E_{\text{slot2}}/1 \\
& (\ref{radar_limit}-\ref{anchor_limit}) \notag\\\notag
\end{alignat}
\subsection{Step-by-step Optimization Scheme}
In the step-by-step optimization scheme, each optimization step only focuses on a single parameter vector, and the optimization results of the previous step can serve as a prior knowledge for subsequent steps. Here we only give the power optimization allocation strategy for the single-slot synchronous and asynchronous ISAL networks. According to the optimization allocation method in the single-slot networks, the power optimization allocation in the dual-slot networks is easy to obtain.
\subsubsection{Optimization for Synchronous Networks}
For the single-slot synchronous ISAL networks, there are two steps in the step-by-step scheme, the power optimization allocation strategy is as follows: \par
{\bf Step 1}: Radar Localization
\begin{alignat}{2}
\mathfrak{A}_{5}^{\text{SDP}}:\quad & \min\quad \text{tr}(\mathbf{H}_{\text r})_{\mathbf{p}_{\rm r}} \\
\text{s.t.}\quad & \mathbf{H}_{\text r} \succeq 0\\
& \left[\begin{array}{cc}
\mathbf{H}_{\text r} & \mathbf{I}\\
\mathbf{I} & \mathbf{J}_{\text{e}}^{\text{R}}\\
\end{array}\right] \succeq 0 \\
& \sum\limits_{j\in\mathcal{N}_{\rm a}\cup\mathcal{N}_{\rm r}} P_j\leq E_{\text{step1}}/1 \\
& (\ref{radar_limit}-\ref{anchor_limit}) \notag\\\notag
\end{alignat}\par
{\bf Step 2}: Target Sensing
\begin{alignat}{2}
\mathfrak{A}_{6}^{\text{SDP}}:\quad & \min\quad \text{tr}(\mathbf{H}_{{\text r},{\text t}})_{\mathbf{p}_{\rm t}} \\
\text{s.t.}\quad & \mathbf{H}_{{\text r},{\text t}} \succeq 0\\
& \left[\begin{array}{cc}
\mathbf{H}_{{\text r},{\text t}} & \mathbf{I}\\
\mathbf{I} & \mathbf{I}_{\bm {\theta}}^{\text{syn}}\\
\end{array}\right] \succeq 0 \label{syn_singleslot_2step}\\
& \sum\limits_{j\in\mathcal{N}_{\rm a}\cup\mathcal{N}_{\rm r}} P_j\leq E_{\text{step2}}/1 \\
& (\ref{radar_limit}-\ref{anchor_limit}) \notag\\\notag
\end{alignat}
where $\mathbf{H}_{\text r}$ and $\mathbf{H}_{{\text r},{\text t}}$ are the auxiliary matrices, the total energy $E_{\text{total}} = E_{\text{step1}} + E_{\text{step2}}$, and there is a prior knowledge $\mathbf{J}_{\text{e}}^{\text{R}}$ obtained from step 1 in $\mathbf{I}_{\bm {\theta}}^{\text{syn}}$ in (\ref{syn_singleslot_2step}).
\subsubsection{Optimization for Asynchronous Networks}
For the single-slot asynchronous ISAL networks, there are three steps in the step-by-step scheme, the power optimization allocation strategy is as follows: \par
{\bf Step 1}: Radar Localization
\begin{alignat}{2}
\mathfrak{A}_{7}^{\text{SDP}}:\quad & \min\quad \text{tr}(\mathbf{H}_{\text r})_{\mathbf{p}_{\rm r}} \\
\text{s.t.}\quad & \mathbf{H}_{\text r} \succeq 0\\
& \left[\begin{array}{cc}
\mathbf{H}_{\text r} & \mathbf{I}\\
\mathbf{I} & \mathbf{J}_{\text{e}}^{\text{R}}\\
\end{array}\right] \succeq 0 \\
& \sum\limits_{j\in\mathcal{N}_{\rm a}\cup\mathcal{N}_{\rm r}} P_j\leq E_{\text{step1}}/1 \\
& (\ref{radar_limit}-\ref{anchor_limit}) \notag\\\notag
\end{alignat}\par
{\bf Step 2}: Clock Offset Estimation
\begin{alignat}{2}
\mathfrak{A}_{8}^{\text{SDP}}:\quad & \min\quad \text{tr}(\mathbf{H}_{{\text r},\tau})_{\bm{\tau}} \\
\text{s.t.}\quad & \mathbf{H}_{{\text r},\tau} \succeq 0\\
& \left[\begin{array}{cc}
\mathbf{H}_{{\text r},\tau} & \mathbf{I}\\
\mathbf{I} & \mathbf{J}_{\text{sub}}\\
\end{array}\right] \succeq 0 \label{asyn_singleslot_2step}\\
& \sum\limits_{j\in\mathcal{N}_{\rm a}\cup\mathcal{N}_{\rm r}} P_j\leq E_{\text{step2}}/1 \\
& (\ref{radar_limit}-\ref{anchor_limit}) \notag\\\notag
\end{alignat}\par
{\bf Step 3}: Target Sensing
\begin{alignat}{2}
\mathfrak{A}_{9}^{\text{SDP}}:\quad & \min\quad \text{tr}(\mathbf{H}_{{\text r},\tau,{\text t}})_{\mathbf{p}_{\rm t}} \\
\text{s.t.}\quad & \mathbf{H}_{{\text r},\tau,{\text t}} \succeq 0\\
& \left[\begin{array}{cc}
\mathbf{H}_{{\text r},\tau,{\text t}} & \mathbf{I}\\
\mathbf{I} & \mathbf{I}_{\bm {\theta}}^{\text{asyn}}\\
\end{array}\right] \succeq 0 \label{asyn_singleslot_3step}\\
& \sum\limits_{j\in\mathcal{N}_{\rm a}\cup\mathcal{N}_{\rm r}} P_j\leq E_{\text{step3}}/1 \\
& (\ref{radar_limit}-\ref{anchor_limit}) \notag\\\notag
\end{alignat}
where $\mathbf{H}_{\text r}, \mathbf{H}_{{\text r},\tau}$ and $\mathbf{H}_{{\text r},\tau,{\text t}}$ are the auxiliary matrices, the total energy $E_{\text{total}} = E_{\text{step1}} + E_{\text{step2}} + E_{\text{step3}}$, $\mathbf{J}_{\text{sub}} = (\mathbf{I}_{\bm {\theta}}^{\text{asyn}})_{1:2N_{\rm r}+|\mathcal{N}_{\rm r}|,1:2N_{\rm r}+|\mathcal{N}_{\rm r}|}$, and there is a prior knowledge $\mathbf{J}_{\text{e}}^{\text{R}}$ obtained from step 1 in $\mathbf{J}_{\text{sub}}$ in (\ref{asyn_singleslot_2step}), as well as a prior knowledge $\mathbf{J}_{\text{sub}}$ obtained from step 2 in $\mathbf{I}_{\bm {\theta}}^{\text{asyn}}$ in (\ref{asyn_singleslot_3step}).
\section{Numerical Results and Discussions}
From the previous section, it can be seen that the reason why the step-by-step scheme has high time complexity is that it searches for the optimal energy allocation scheme through traversal. To compensate for the high time complexity of the step-by-step scheme, this section started by exploring the optimal energy allocation for each step in a single time slot and for each time slot between multiple time slots. The regularity conclusion of energy optimization allocation for the step-by-step scheme was given. By comparing the sensing accuracy of the step-by-step scheme and the integrated scheme in the networks with different topologies, the suitable network topology for each of the two schemes was given.
\subsection{Network Settings}
\subsubsection{Single-slot Network Settings}
\begin{figure}[htbp]
\begin{adjustwidth}{-\extralength}{0cm}
\centering
\vspace{0in}
\begin{minipage}{1\linewidth}	
    \captionsetup[subfloat]{justification=centering}
	\subfloat[]{
		\label{singleslot_network1}
		\includegraphics[width=0.49\linewidth]{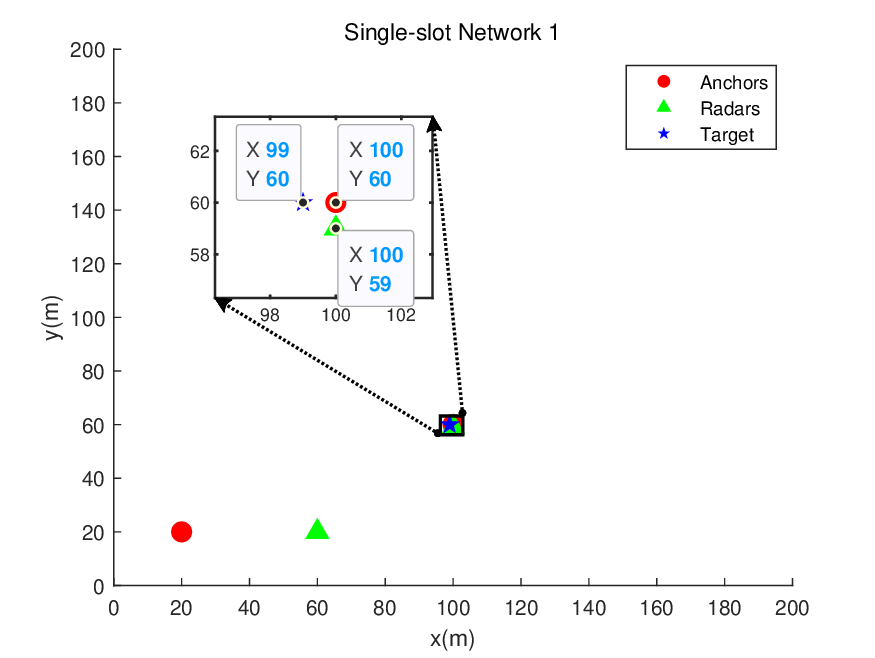}	
	}\noindent
	\subfloat[]{
		\label{singleslot_network2}
		\includegraphics[width=0.49\linewidth]{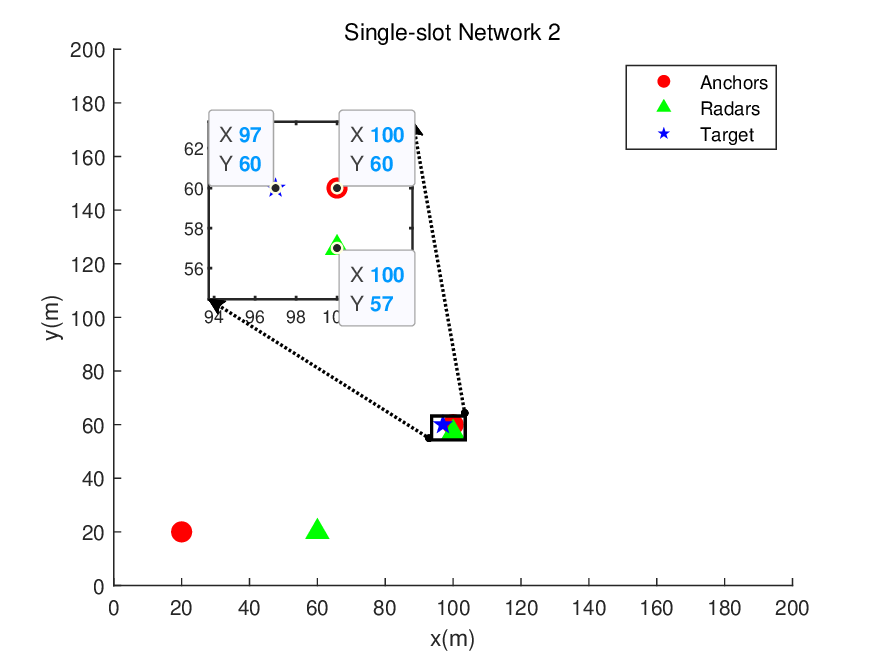}
	}
\end{minipage}
\vskip -0.3cm 
\begin{minipage}{1\linewidth }
    \captionsetup[subfloat]{justification=centering}
	\subfloat[]{
		\label{singleslot_network3}
		\includegraphics[width=0.49\linewidth]{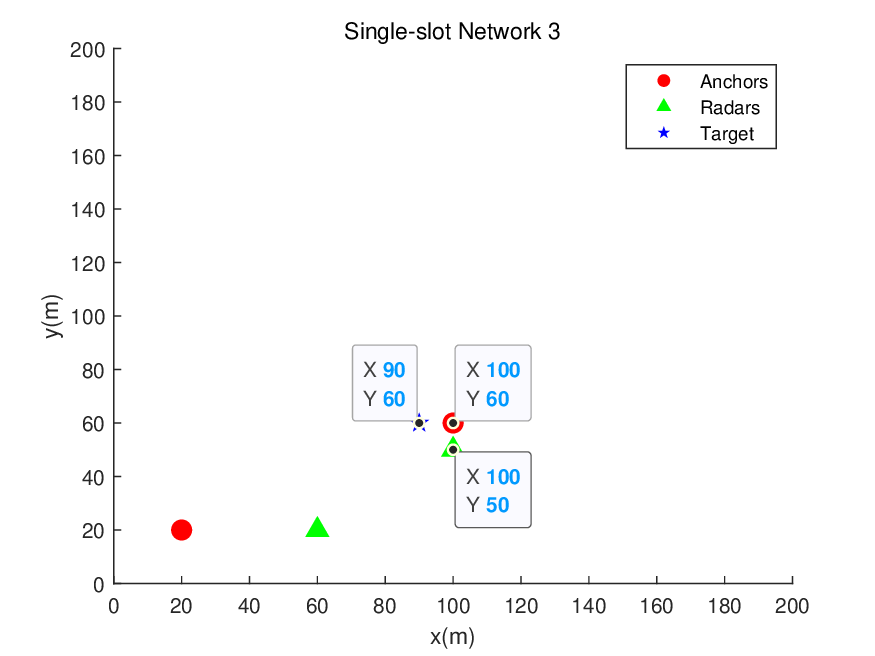
		}
	}\noindent
	\subfloat[]{
		\label{singleslot_network4}
		\includegraphics[width=0.49\linewidth]{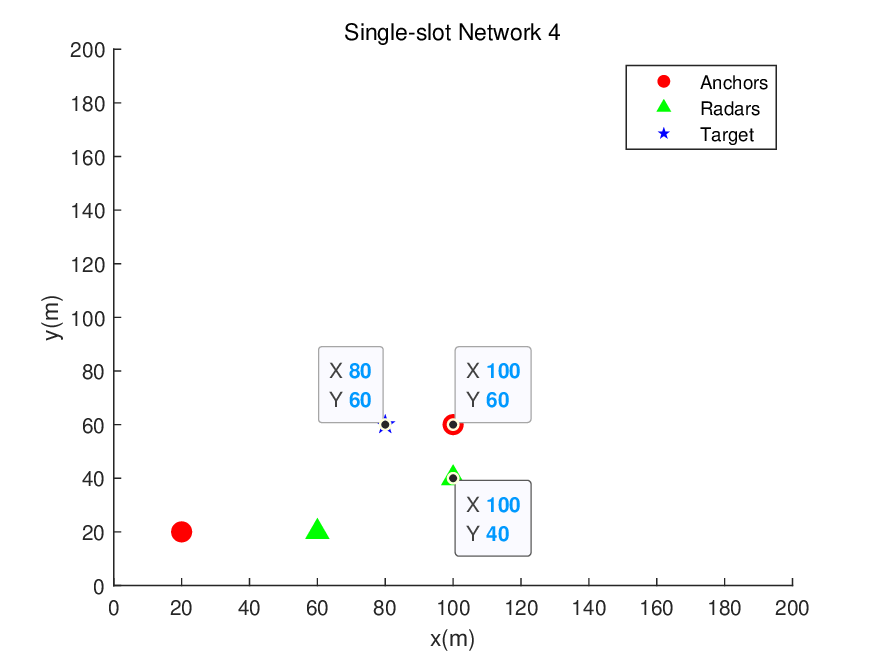
		}
	}
\end{minipage}
\vspace{0in}	
\end{adjustwidth}
\caption{Four single-slot networks.}
\vspace{0in}		
\label{singleslot_networks}
\end{figure}

For the single-slot ISAL networks, we present four networks with different topologies in the area of $200\text{m}\times200\text{m}$, as shown in the Figure \ref{singleslot_networks}. The red solid circles represent the anchors, the green solid triangles represent the radars, and the black solid pentagram represents the target. These four networks all contain 2 anchors, 2 radars, and 1 target. \par
\subsubsection{Dual-slot Network Settings}
\begin{figure}[htbp]
\begin{adjustwidth}{-\extralength}{0cm}
\centering
\vspace{0in}
\begin{minipage}{1\linewidth}	
    \captionsetup[subfloat]{justification=centering}
	\subfloat[]{
		\label{dualslot_network1}
		\includegraphics[width=0.49\linewidth]{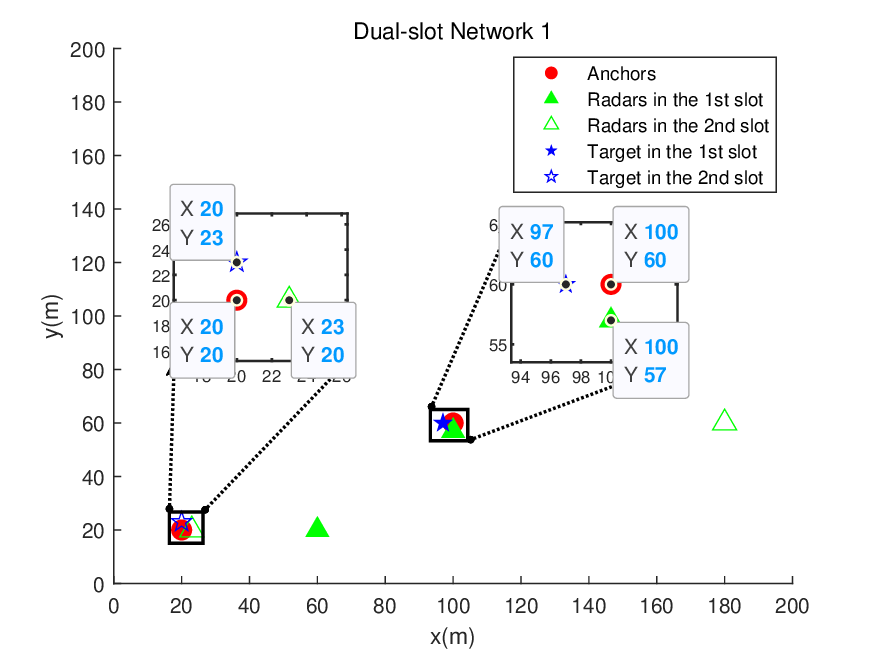}	
	}\noindent
	\subfloat[]{
		\label{dualslot_network2}
		\includegraphics[width=0.49\linewidth]{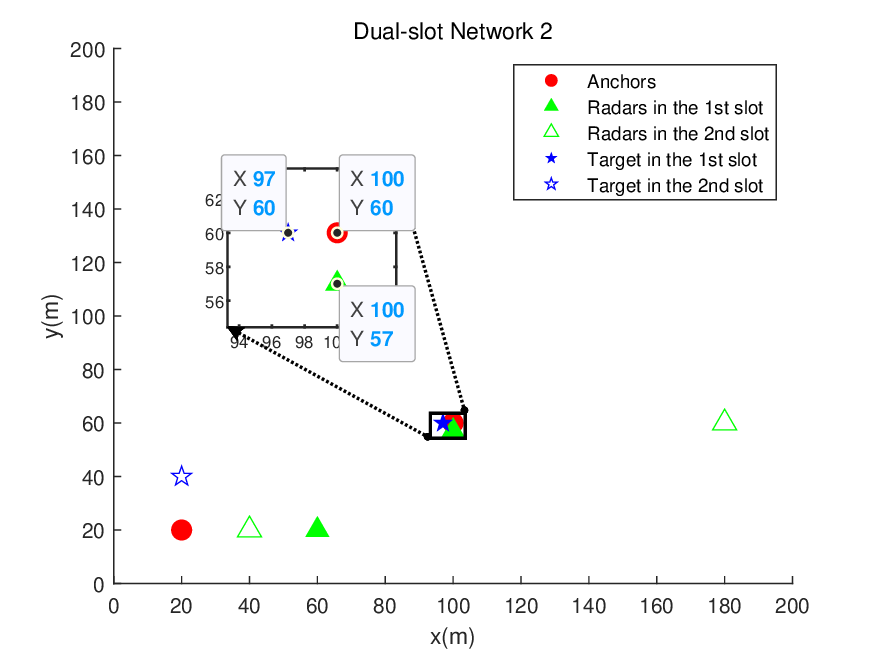}
	}
\end{minipage}
\vskip -0.3cm 
\begin{minipage}{1\linewidth }
    \captionsetup[subfloat]{justification=centering}
	\subfloat[]{
		\label{dualslot_network3}
		\includegraphics[width=0.49\linewidth]{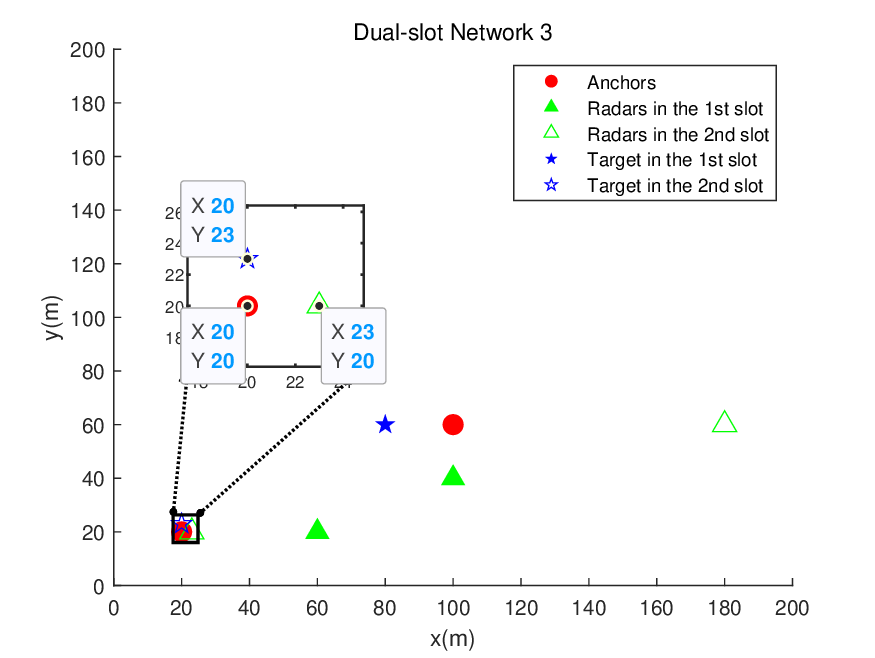
		}
	}\noindent
	\subfloat[]{
		\label{dualslot_network4}
		\includegraphics[width=0.49\linewidth]{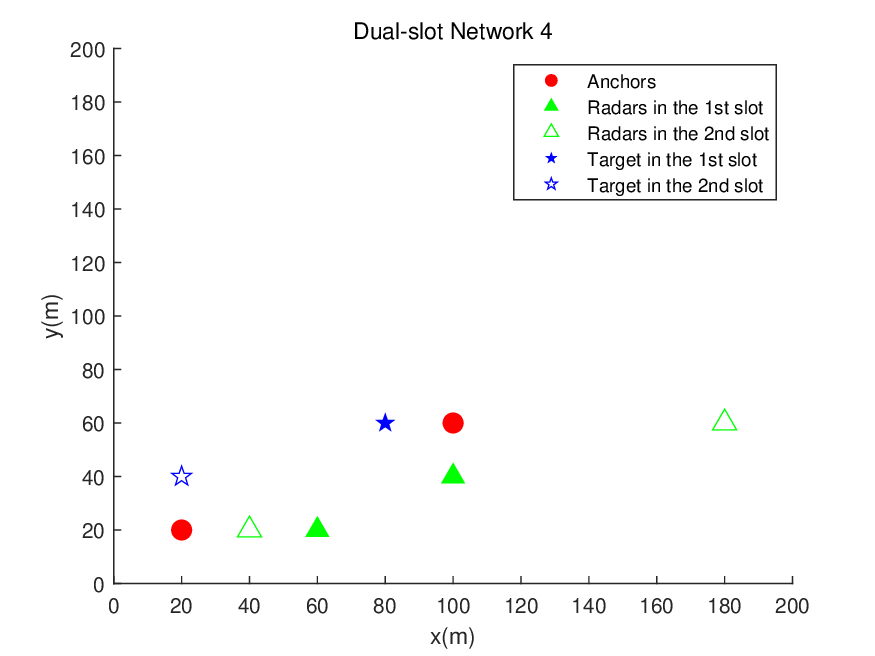
		}
	}
\end{minipage}
\vspace{0in}	
\end{adjustwidth}
\caption{Four dual-slot networks.}
\vspace{0in}		
\label{dualslot_networks}
\end{figure}

For the dual-slot ISAL networks, we present four networks with different topologies in the area of $200\text{m}\times200\text{m}$, as shown in the Figure \ref{dualslot_networks}. The red solid circles represent anchors whose positions remain unchanged within two time slots, the green solid triangles and the black solid pentagram respectively represent radars and target in slot 1, the green hollow triangles and the black hollow pentagram respectively represent radars and target in slot 2. Within each time slot, all these four networks contain 2 anchors, 2 radars, and 1 target. \par
\subsubsection{Simulation Parameters}
The relevant parameters in the simulation can be seen from Table \ref{Parameters_table}.
\begin{table}[h]
\captionsetup{justification=centering}
\caption{Simulation Parameters}
\begin{center}
\begin{tabular}{lll}
   \toprule
   Parameter & Symbol & Value \\
   \midrule
   Carrier frequency & $f_{\text{c}}$ & 77GHz \\
   Total energy & $E_{\text{total}}$ & 10J \\
   Total bandwidth & $B_{\text{total}}$ & 500MHz \\
   Antenna gain & $G$ & 10dB \\
   Noise power spectral density & $N_0$ & -174dBm/Hz \\
   Upper limit of anchor power & $P_{\text{a},\text{max}}$ & 1 \\
   Upper limit of radar power & $P_{\text{r},\text{max}}$ & 1 \\
   Radar cross section & $\sigma$ & 10$\text{m}^2$ \\
   System loss & $L$ & 3dB \\
   Variance of node velocity & $\eta^2$ & 0.01$\text{m}^2/\text{s}^2$ \\
   Variance of relative clock drift rate & $\rho^2$ & 1e-10 \\
   \bottomrule
\end{tabular}
\label{Parameters_table}
\end{center}
\end{table}
\subsection{Synchronous Networks}
\subsubsection{Single-slot Networks}
\begin{figure}[htbp]
\begin{adjustwidth}{-\extralength}{0cm}
\centering
\vspace{0in}
\begin{minipage}{1\linewidth}	
    \captionsetup[subfloat]{justification=centering}
	\subfloat[]{
		\label{syn_single_comparison1}
		\includegraphics[width=0.49\linewidth]{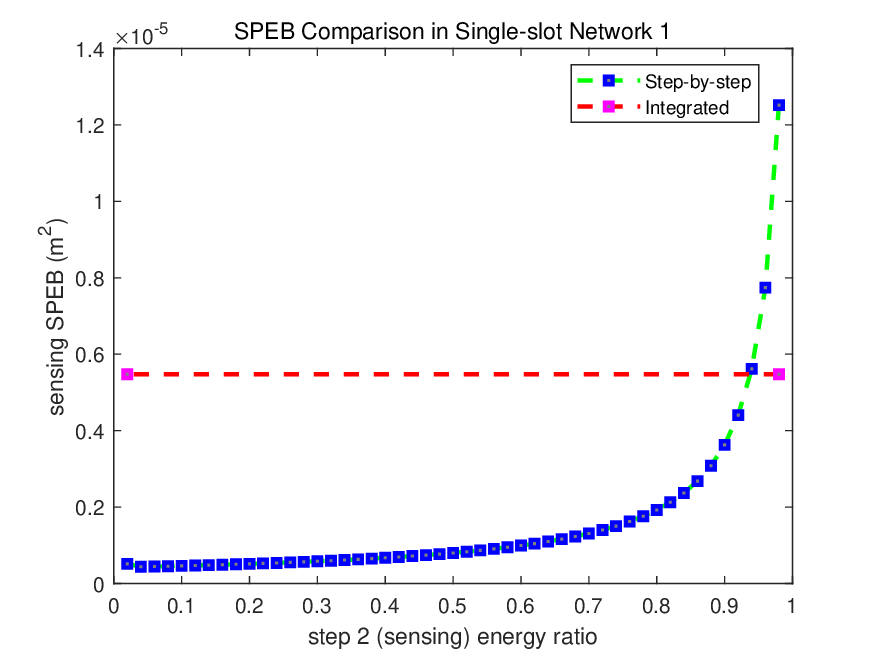}	
	}\noindent
	\subfloat[]{
		\label{syn_single_comparison2}
		\includegraphics[width=0.49\linewidth]{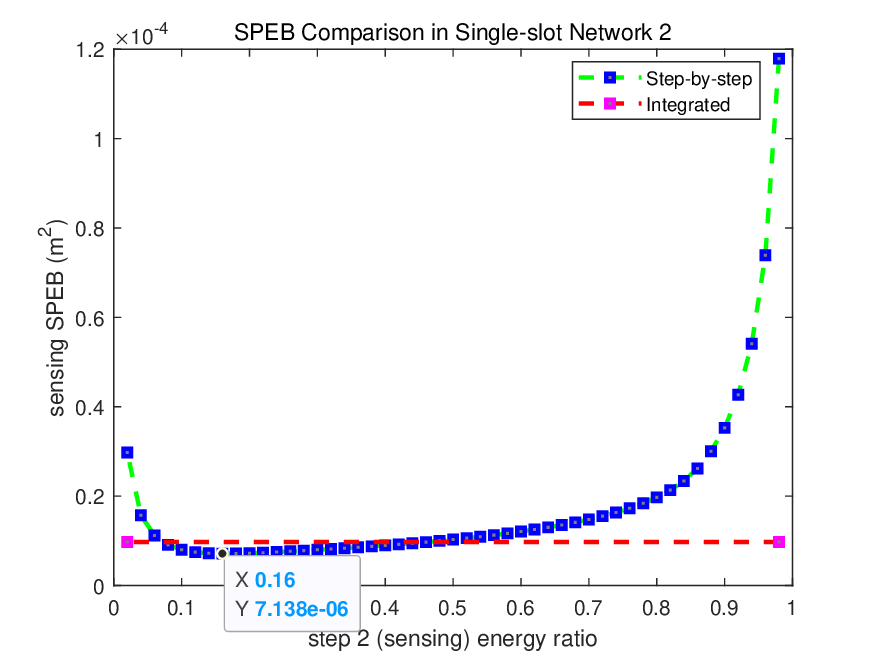}
	}
\end{minipage}
\vskip -0.3cm 
\begin{minipage}{1\linewidth }
    \captionsetup[subfloat]{justification=centering}
	\subfloat[]{
		\label{syn_single_comparison3}
		\includegraphics[width=0.49\linewidth]{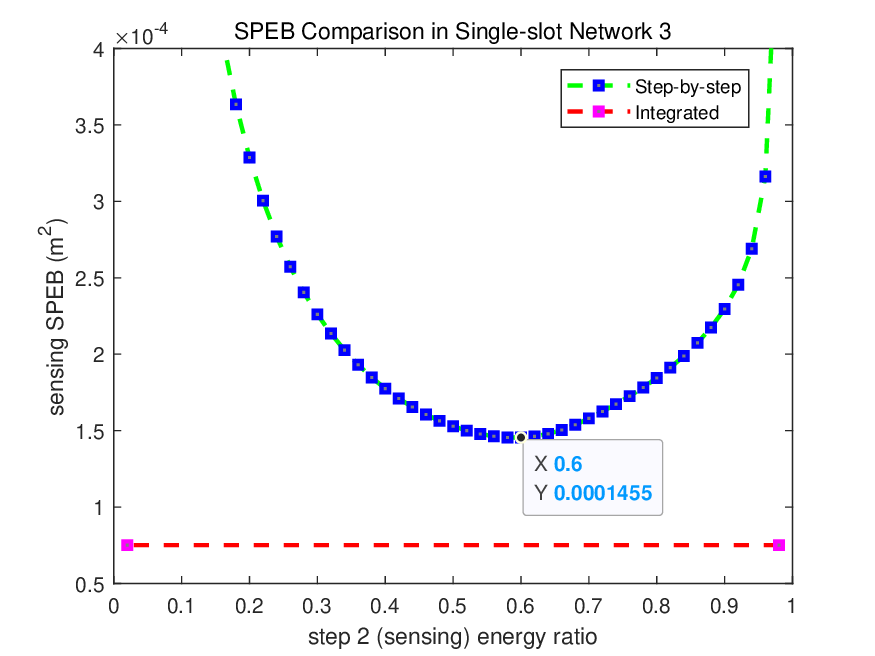
		}
	}\noindent
	\subfloat[]{
		\label{syn_single_comparison4}
		\includegraphics[width=0.49\linewidth]{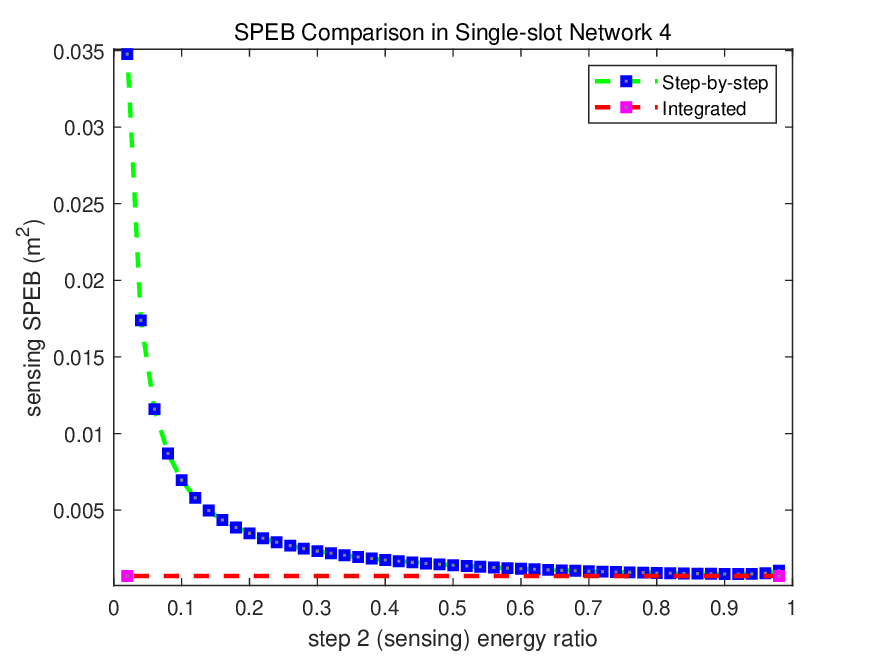
		}
	}
\end{minipage}
\vspace{0in}	
\end{adjustwidth}
\caption{SPEB comparison between step-by-step scheme and integrated scheme in four single-slot synchronous networks. (a) SPEB comparison in single-slot synchronous network (\ref{singleslot_network1}). (b) SPEB comparison in single-slot synchronous network (\ref{singleslot_network2}). (c) SPEB comparison in single-slot synchronous network (\ref{singleslot_network3}). (d) SPEB comparison in single-slot synchronous network (\ref{singleslot_network4}).}
\vspace{0in}		
\label{syn_single_comparison}
\end{figure}

For the synchronous single-slot ISAL networks, as shown in Figure \ref{syn_single_comparison}, we provide comparison of the sensing SPEB results between the step-by-step scheme and the integrated scheme in the four networks in Figure \ref{singleslot_networks}. \par
Due to the fact that all active nodes in this article can both transmit and receive signals, the target sensing can be achieved by only two active nodes through twice TOA and once TSOA. In Figure \ref{singleslot_networks}, we set up a variably-sized partial network with one target and two active nodes. The networks with relatively close distances among three nodes are called sensing-resource-abundant networks. On the contrary, the networks with relatively far distances among three nodes are called sensing-resource-deficient networks. \par
According to Figure \ref{syn_single_comparison}, we have the following observations. Firstly, when the distribution of nodes is relatively uniform in the ISAL network, in other words, when the length of the localization and sensing links in the network is basically the same level (such as 1e+1 $ \text {m} $ $\sim$ 1e+2 $ \text {m} $), the step-by-step optimization scheme will allocate the majority of the total energy to the second step for target sensing, as shown in the Figure \ref{syn_single_comparison4}. This indicates that in the networks with evenly distributed nodes, the difficulty of sensing passive targets is much higher than the difficulty of locating active nodes. Secondly, in the single-slot synchronous ISAL networks, when the optimization objective function is the sensing SPEB, the step-by-step optimization scheme is more suitable for the sensing-resource-abundant networks, as shown in Figure \ref{syn_single_comparison1} and \ref{syn_single_comparison2}. The integrated optimization scheme is more suitable for the sensing-resource-deficient networks, as shown in Figure \ref{syn_single_comparison3} and \ref{syn_single_comparison4}. In addition, from the SPEB results of the step-by-step scheme of the single-slot synchronous ISAL networks, we can observe the tradeoff relationship between sensing and localization, as shown in Figure \ref{syn_single_comparison2} and \ref{syn_single_comparison3}. \par

For the conclusions above, we give the following interpretations. 
Firstly, according to the RII expressions (\ref{RII_loc}-\ref{RII_sen}), it is not difficult to find that when the length of the localization and sensing links is basically the same level, the RII of the sensing link is much smaller than that of the localization link. Therefore, when the network nodes are evenly distributed, the difficulty of sensing passive targets is relatively higher. 
Secondly, in the sensing-resource-abundant networks, target sensing requires less energy than radar positioning. Due to the flexibility of energy allocation between two steps in the step-by-step scheme, the step-by-step scheme can reduce the energy allocated for step 2 (target sensing) to allow more energy to be used for radar positioning, which is beneficial for improving radar positioning accuracy. Additionally, the optimization results of step 1 (radar positioning) in the step-by-step scheme can serve as a prior information for step 2. Therefore, the improvement of radar positioning accuracy has a positive promoting effect on the improvement of target sensing accuracy. However, for the integrated optimization scheme, due to its simultaneous radar positioning and target sensing, and the optimization objective function being the sensing SPEB, it will lead to poor radar positioning accuracy obtained by the integrated scheme, which is detrimental to the accuracy of target sensing. Therefore, the step-by-step scheme is more suitable for the sensing-resource-abundant networks. 
For the sensing-resource-deficient networks, as radar positioning is easier than target sensing, the optimized power allocation results for target sensing in the integrated optimization scheme can meet the needs of radar positioning in the networks. In this situation, the accuracy of radar positioning obtained by the integrated scheme will have no adverse impact on its sensing accuracy. However, in the step-by-step scheme, in order to ensure that the accuracy of radar positioning does not have an adverse impact on sensing accuracy, a portion of energy needs to be allocated to step 1 for radar positioning, resulting in a reduction in the energy allocated to step 2 for target sensing. Due to the difficulty of target sensing being higher than radar positioning in the sensing-resource-deficient networks, the reduction in the energy allocated to step 2 is not conducive to the accuracy of target sensing in the step-by-step scheme. Therefore, the integrated scheme is more suitable for the sensing-resource-deficient networks. 
Finally, due to the spatial cooperation between radar positioning and target sensing, the accuracy of radar positioning can affect the accuracy of target sensing, resulting in the tradeoff mentioned above.
\subsubsection{Dual-slot Networks}
\begin{figure}[htbp]
\begin{adjustwidth}{-\extralength}{0cm}
\centering
\vspace{0in}
\begin{minipage}{1\linewidth}	
    \captionsetup[subfloat]{justification=centering}
	\subfloat[]{
		\label{syn_dual_comparison1}
		\includegraphics[width=0.49\linewidth]{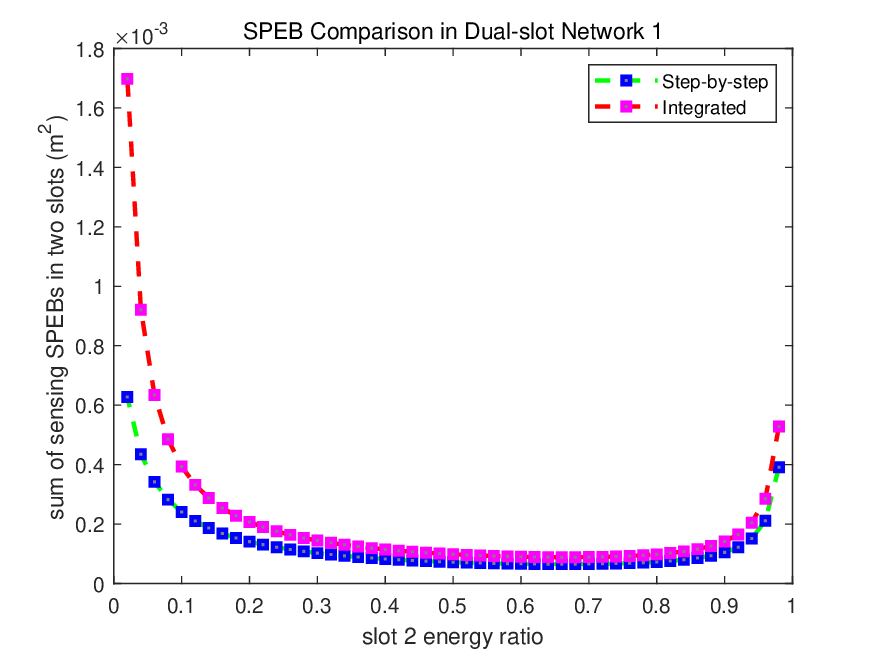}	
	}\noindent
	\subfloat[]{
		\label{syn_dual_comparison2}
		\includegraphics[width=0.49\linewidth]{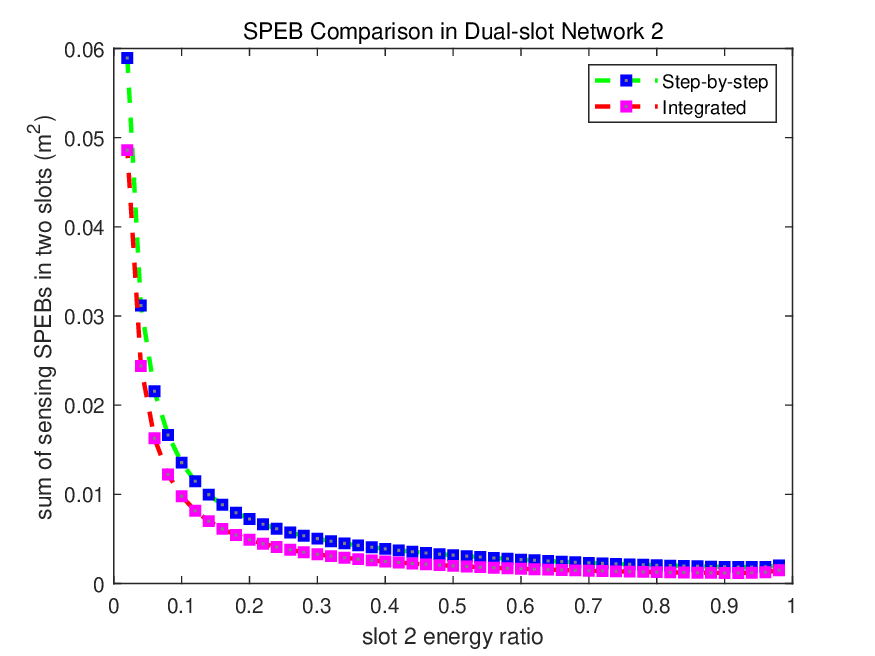}
	}
\end{minipage}
\vskip -0.3cm 
\begin{minipage}{1\linewidth }
    \captionsetup[subfloat]{justification=centering}
	\subfloat[]{
		\label{syn_dual_comparison3}
		\includegraphics[width=0.49\linewidth]{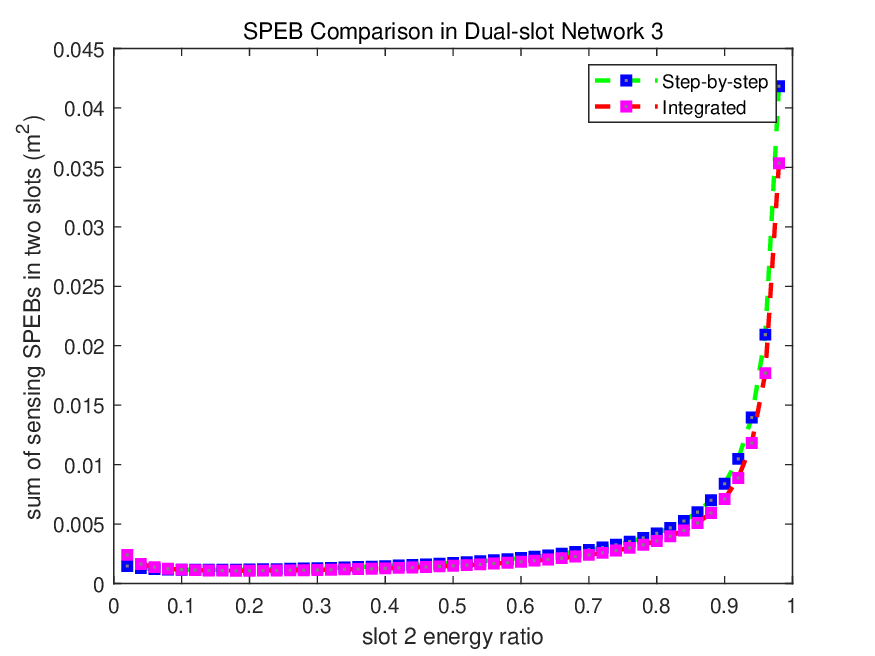
		}
	}\noindent
	\subfloat[]{
		\label{syn_dual_comparison4}
		\includegraphics[width=0.49\linewidth]{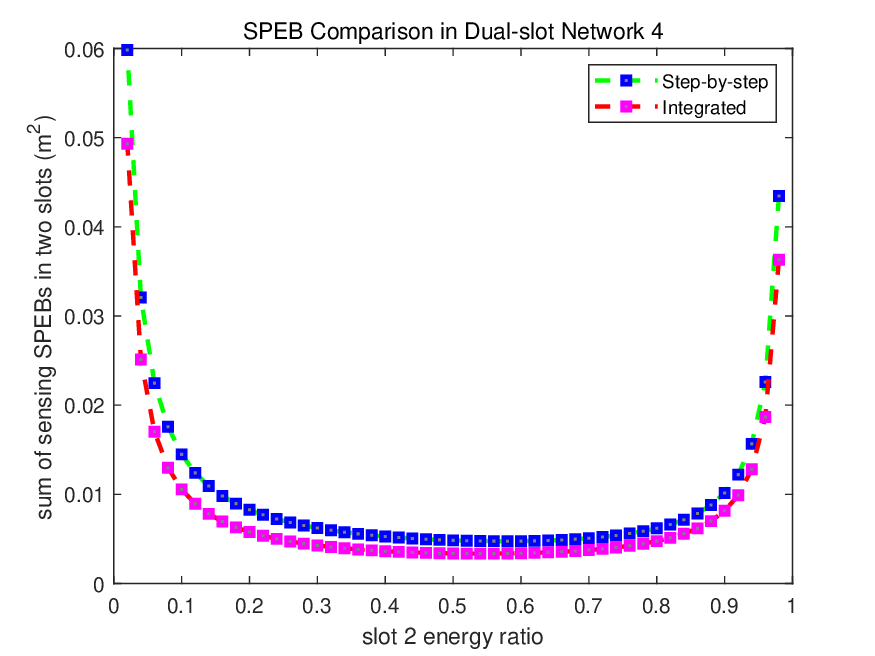
		}
	}
\end{minipage}
\vspace{0in}	
\end{adjustwidth}
\caption{SPEB comparison between step-by-step scheme and integrated scheme in four dual-slot synchronous networks. (a) SPEB comparison in dual-slot synchronous network (\ref{dualslot_network1}). (b) SPEB comparison in dual-slot synchronous network (\ref{dualslot_network2}). (c) SPEB comparison in dual-slot synchronous network (\ref{dualslot_network3}). (d) SPEB comparison in dual-slot synchronous network (\ref{dualslot_network4}).}
\vspace{0in}		
\label{syn_dual_comparison}
\end{figure}

For the dual-slot synchronous ISAL networks, as shown in Figure \ref{syn_dual_comparison}, we give the sensing SPEB of the step-by-step scheme and the integrated scheme in the four networks with different topologies in Figure \ref{dualslot_networks}.\par
According to Figure \ref{syn_dual_comparison}, we can draw the following conclusions: when the topologies of the partial networks used for target sensing in the two slots is similar, such as the topologies shown in Figure \ref{dualslot_network1} and \ref{dualslot_network4}, the energy allocation between the two time slots tends to be equal, as shown in Figure \ref{syn_dual_comparison1} and \ref{syn_dual_comparison4}; when there is a significant difference in the topologies of the partial networks used for target sensing in the two time slots, as shown in Figure \ref{dualslot_network2} and \ref{dualslot_network3}, more energy is allocated to the time slot corresponding to the sensing-resource-deficient network, as shown in Figure \ref{syn_dual_comparison2} and \ref{syn_dual_comparison3}.
\subsection{Asynchronous Networks}
\subsubsection{Single-slot Networks}
\begin{figure}[htbp]
\begin{adjustwidth}{-\extralength}{0cm}
\centering
\vspace{-0.4in}
\begin{minipage}{1\linewidth}	
    \captionsetup[subfloat]{justification=centering}
	\subfloat[]{
		\label{asyn_single_comparison1_locsen1}
		\includegraphics[width=0.24\linewidth]{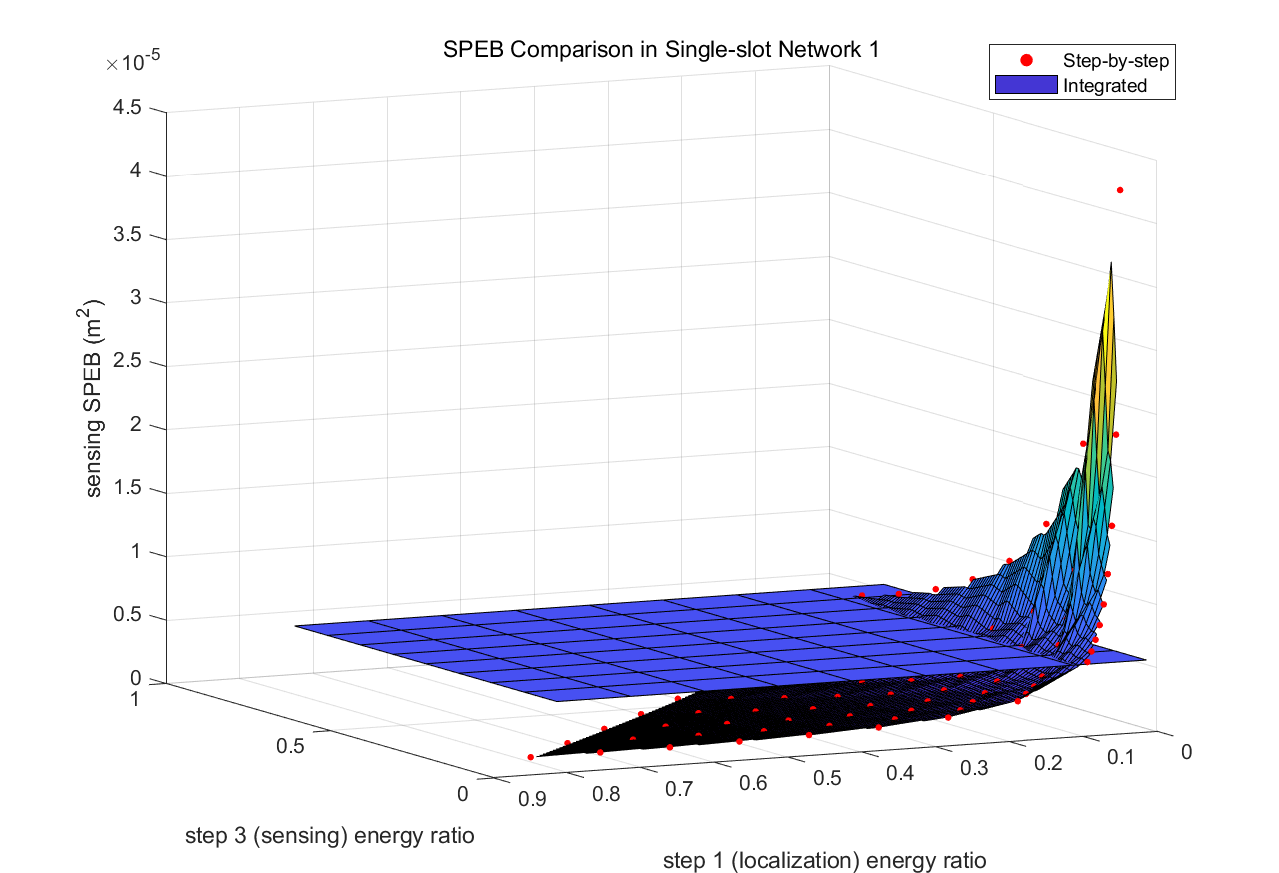}	
	}\noindent
	\subfloat[]{
		\label{asyn_single_comparison1_locsen2}
		\includegraphics[width=0.24\linewidth]{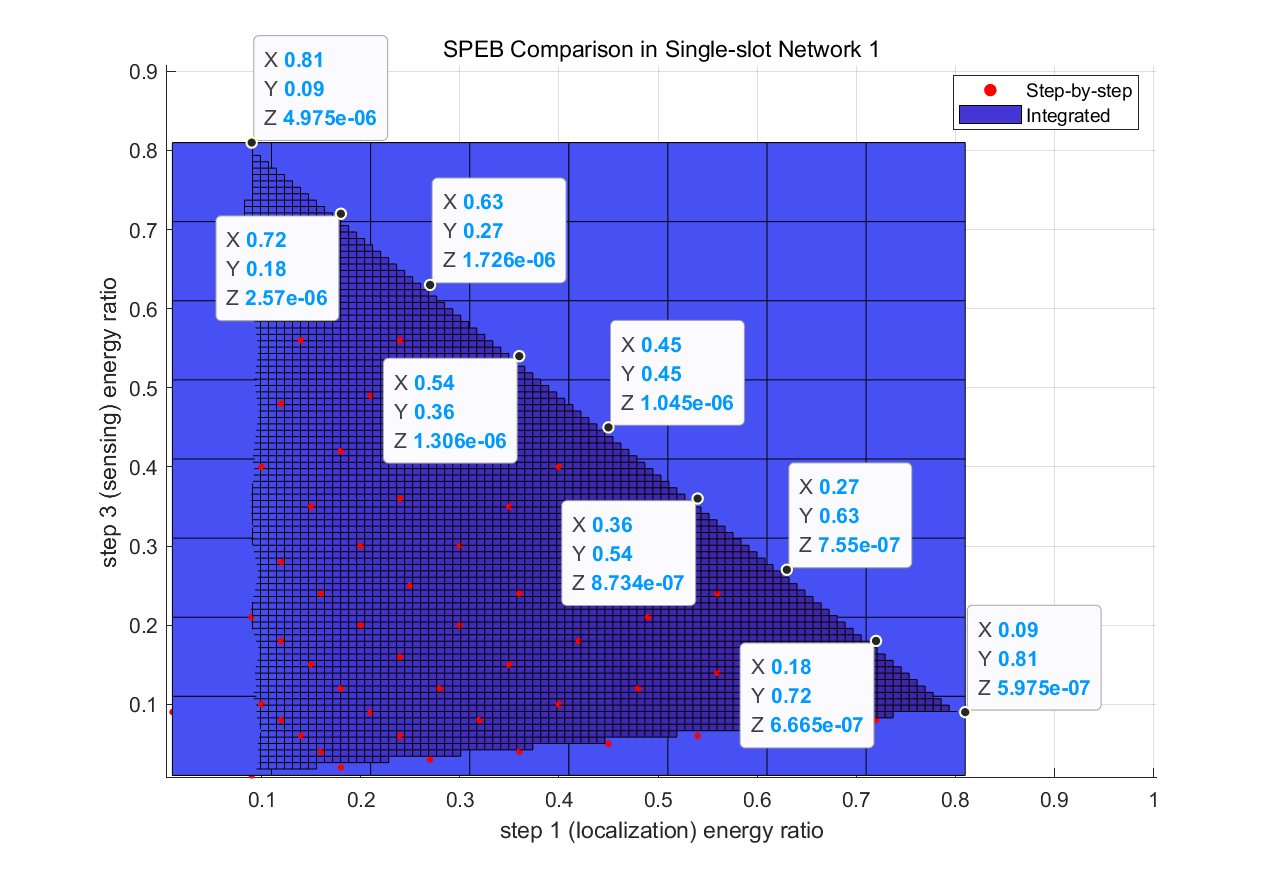}	
	}\noindent
	\subfloat[]{
		\label{asyn_single_comparison2_locsen1}
		\includegraphics[width=0.24\linewidth]{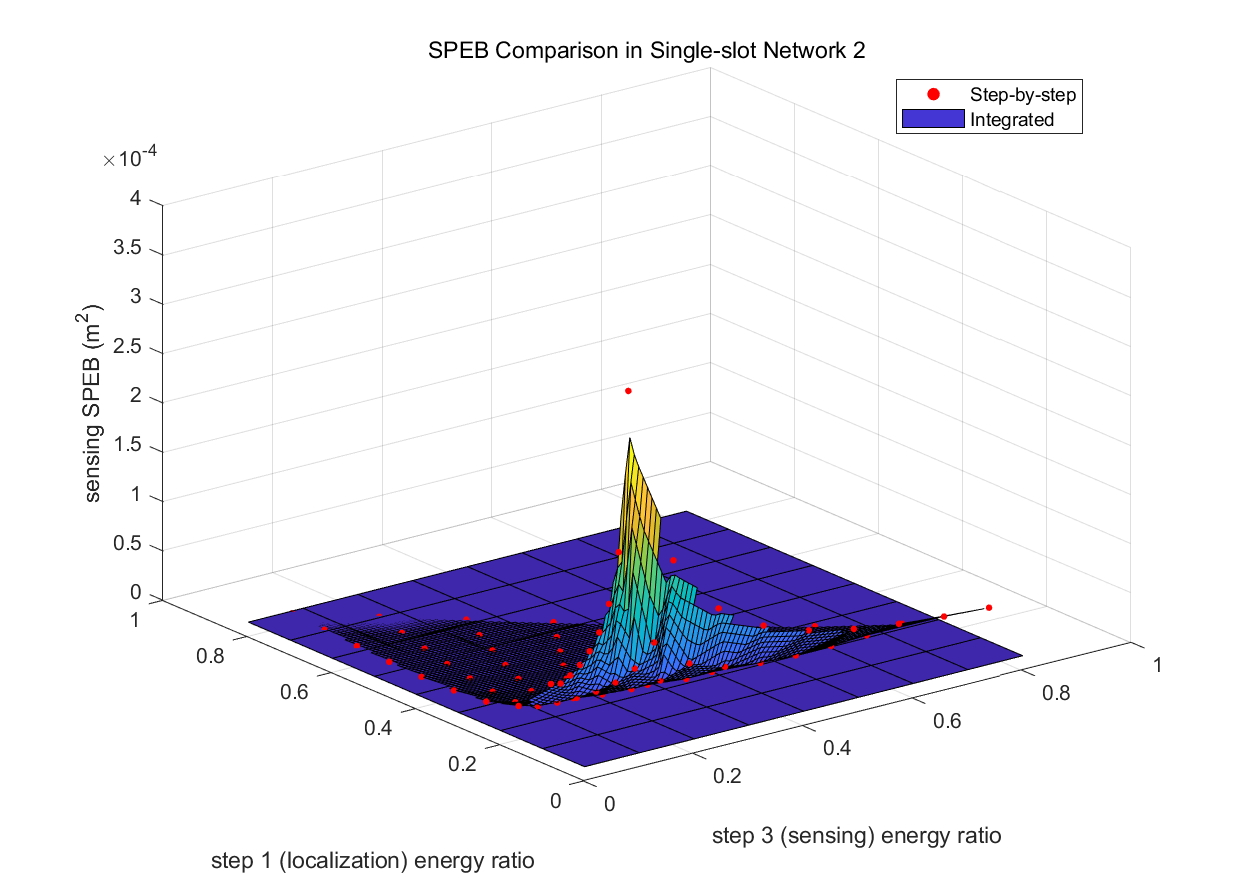}	
	}\noindent
	\subfloat[]{
		\label{asyn_single_comparison2_locsen2}
		\includegraphics[width=0.24\linewidth]{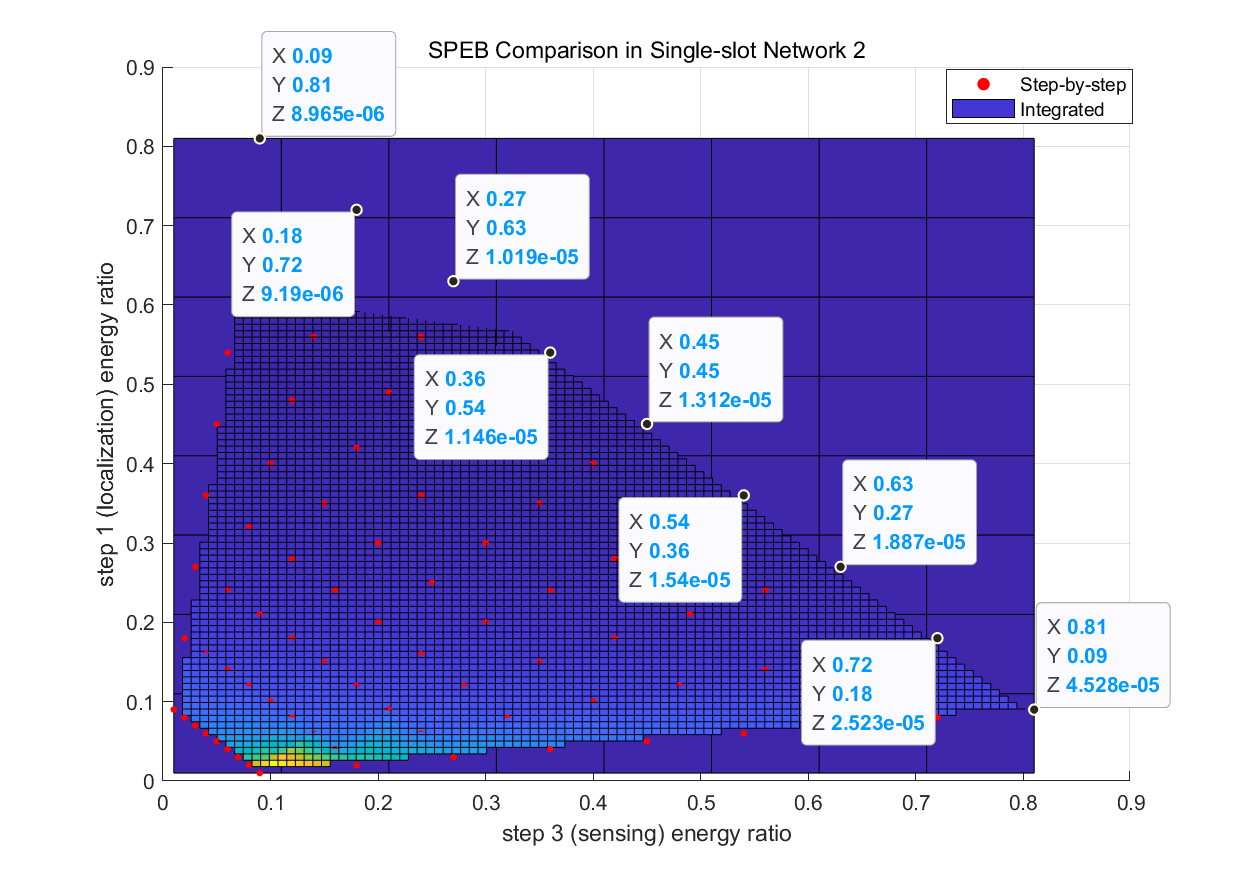}
	}
\end{minipage}
\vskip -0.4cm 
\begin{minipage}{1\linewidth}	
    \captionsetup[subfloat]{justification=centering}
	\subfloat[]{
		\label{asyn_single_comparison1_locdrift1}
		\includegraphics[width=0.24\linewidth]{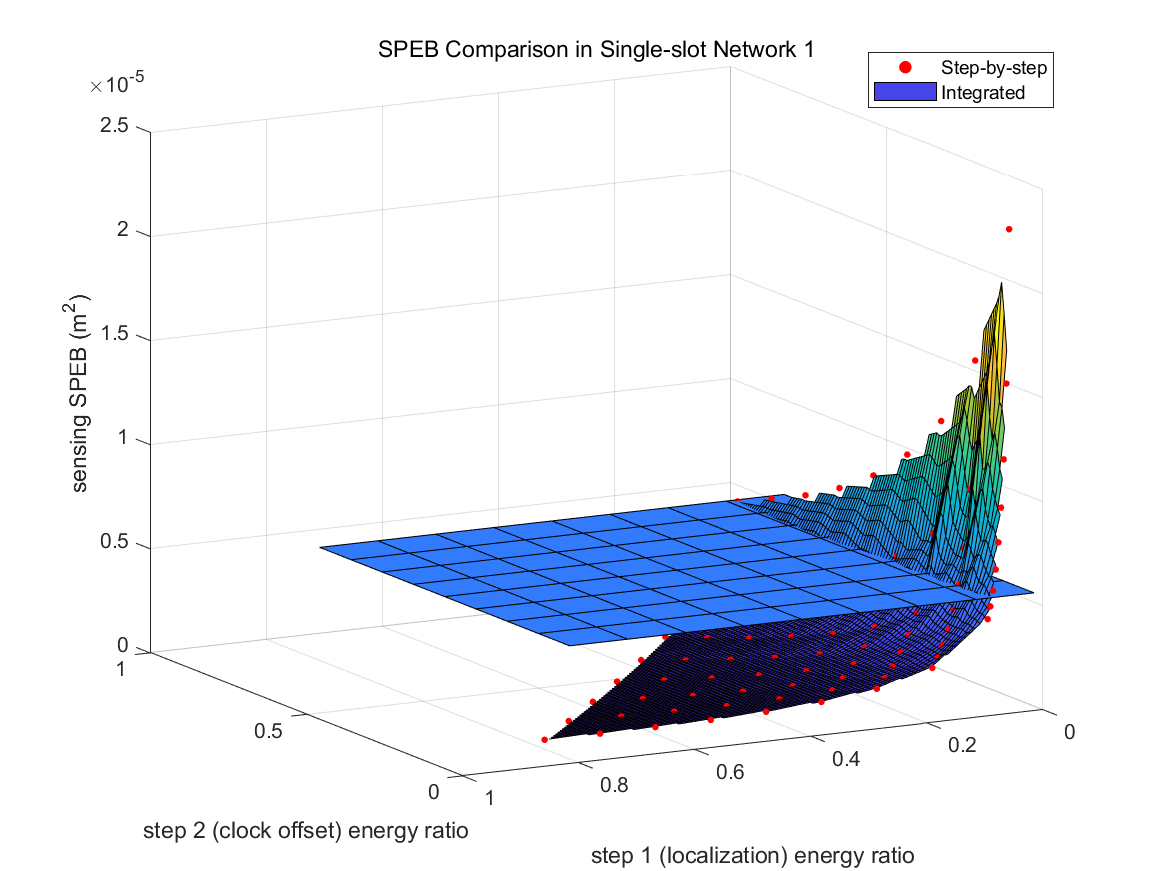}	
	}\noindent
	\subfloat[]{
		\label{asyn_single_comparison1_locdrift2}
		\includegraphics[width=0.24\linewidth]{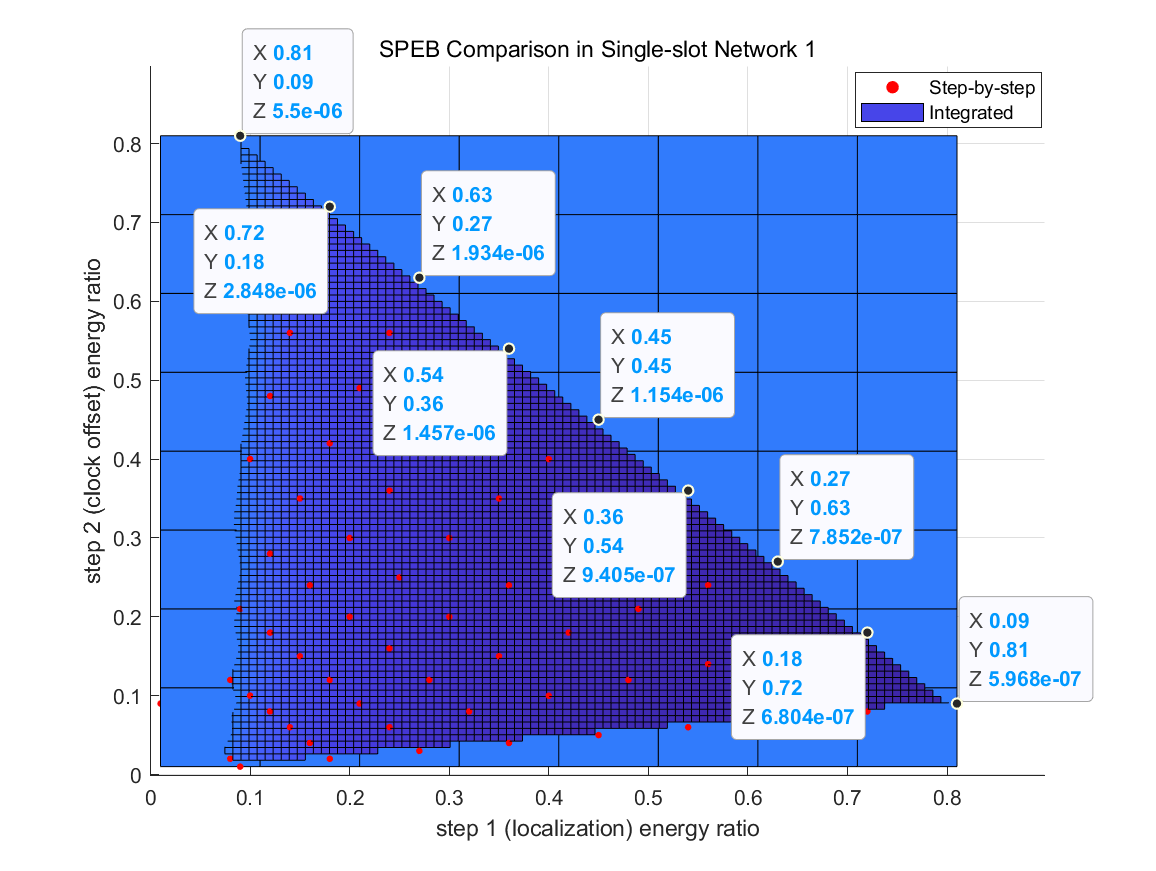}	
	}\noindent
	\subfloat[]{
		\label{asyn_single_comparison2_locdrift1}
		\includegraphics[width=0.24\linewidth]{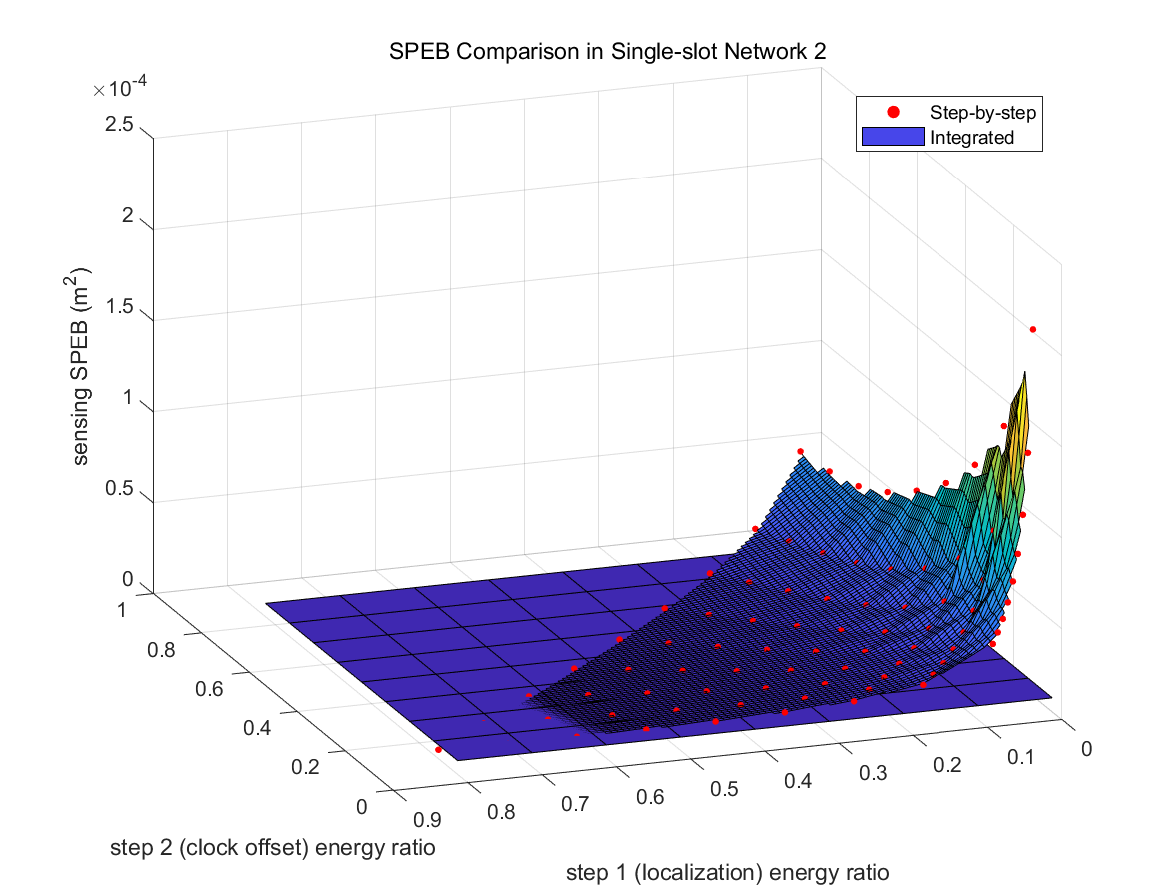}	
	}\noindent
	\subfloat[]{
		\label{asyn_single_comparison2_locdrift2}
		\includegraphics[width=0.24\linewidth]{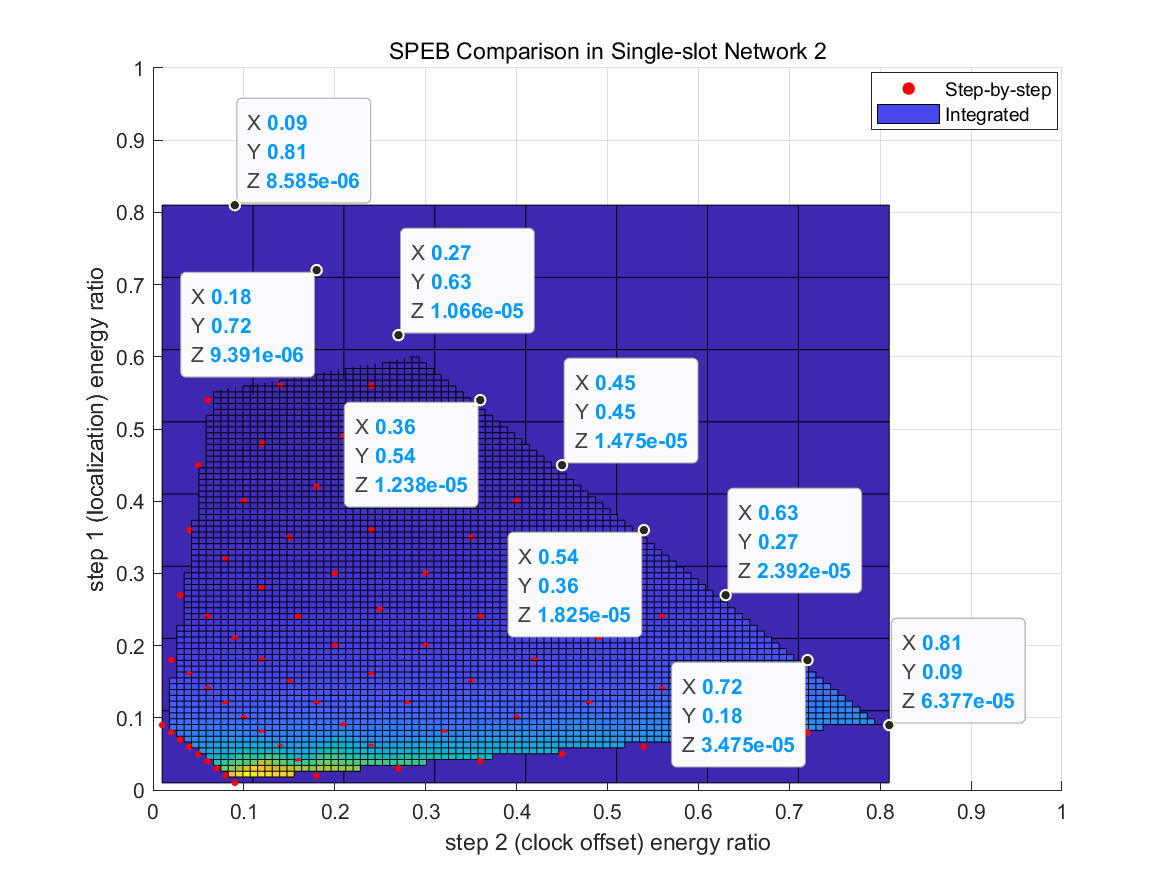}
	}
\end{minipage}
\vskip -0.4cm 
\begin{minipage}{1\linewidth}	
    \captionsetup[subfloat]{justification=centering}
	\subfloat[]{
		\label{asyn_single_comparison1_sendrift1}
		\includegraphics[width=0.24\linewidth]{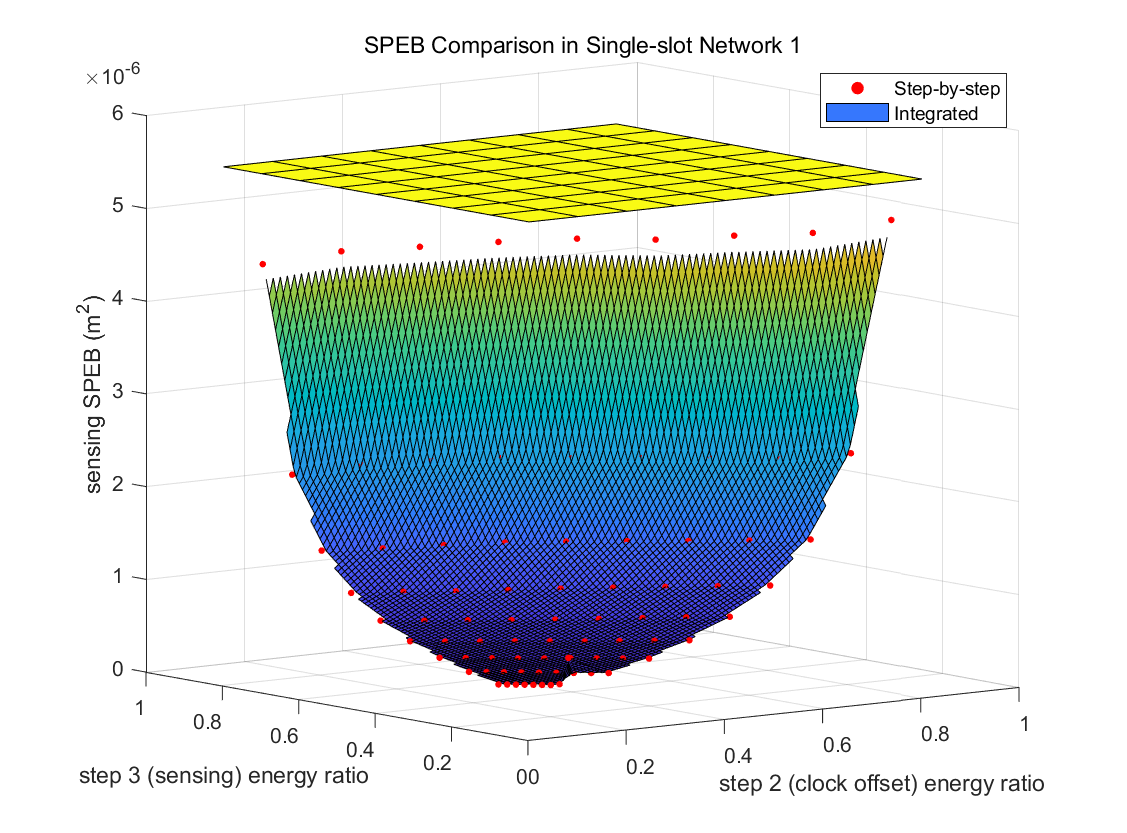}	
	}\noindent
	\subfloat[]{
		\label{asyn_single_comparison1_sendrift2}
		\includegraphics[width=0.24\linewidth]{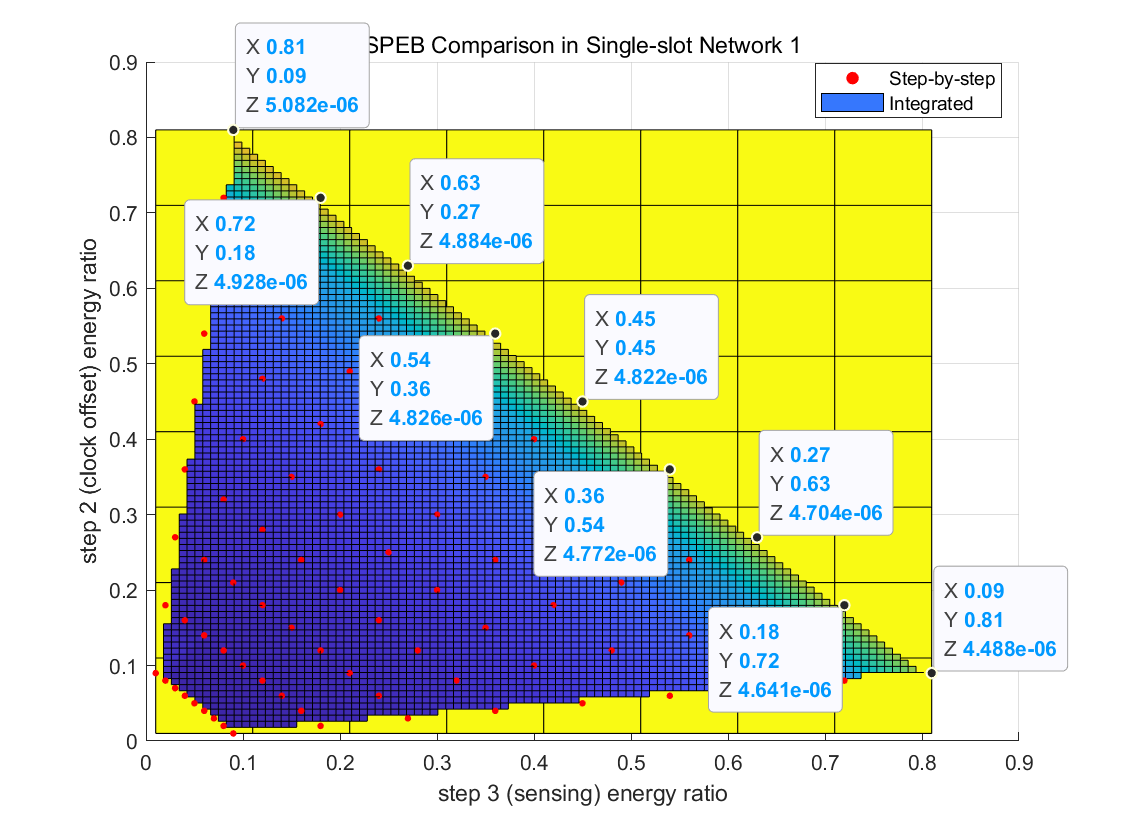}	
	}\noindent
	\subfloat[]{
		\label{asyn_single_comparison2_sendrift1}
		\includegraphics[width=0.24\linewidth]{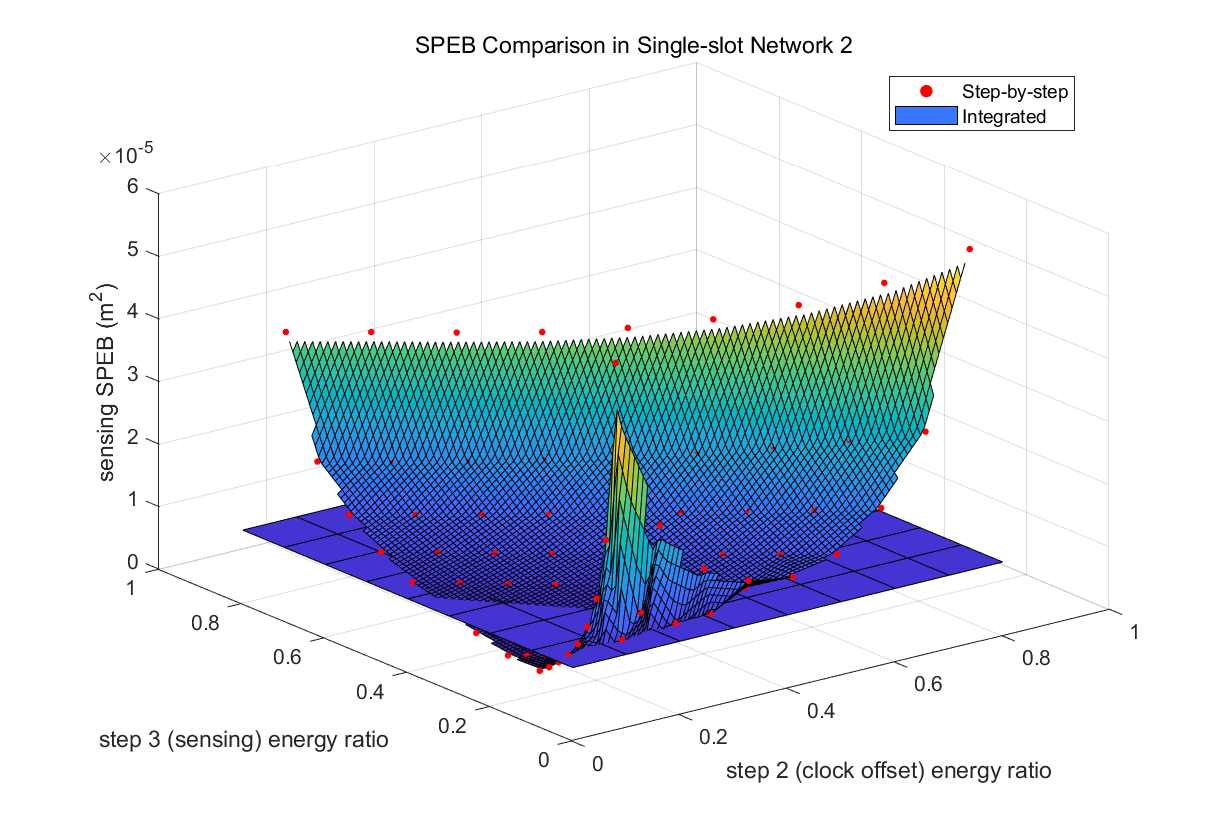}	
	}\noindent
	\subfloat[]{
		\label{asyn_single_comparison2_sendrift2}
		\includegraphics[width=0.24\linewidth]{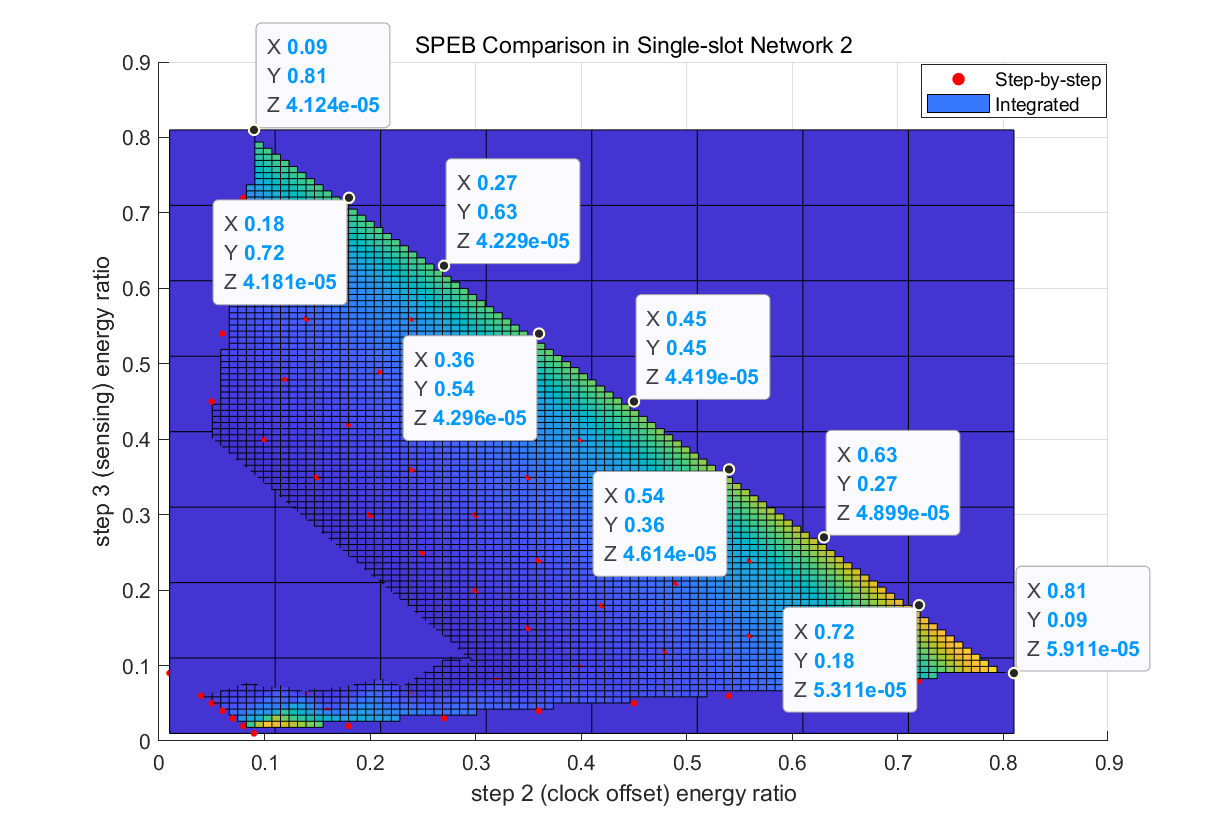}
	}
\end{minipage}
\vskip -0.2cm 
\begin{minipage}{1\linewidth}	
    \captionsetup[subfloat]{justification=centering}
	\subfloat[]{
		\label{asyn_single_comparison3_locsen1}
		\includegraphics[width=0.24\linewidth]{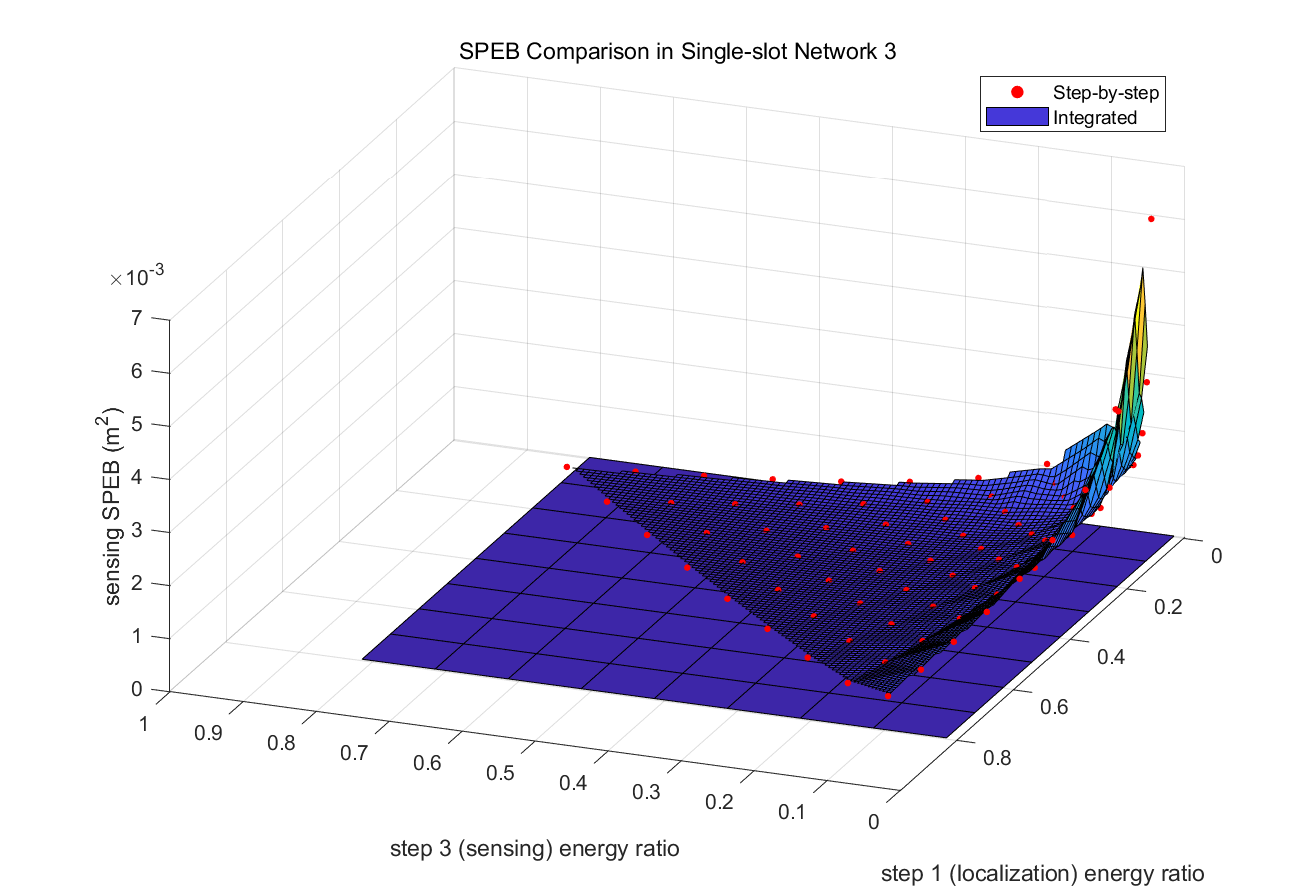}	
	}\noindent
	\subfloat[]{
		\label{asyn_single_comparison3_locsen2}
		\includegraphics[width=0.24\linewidth]{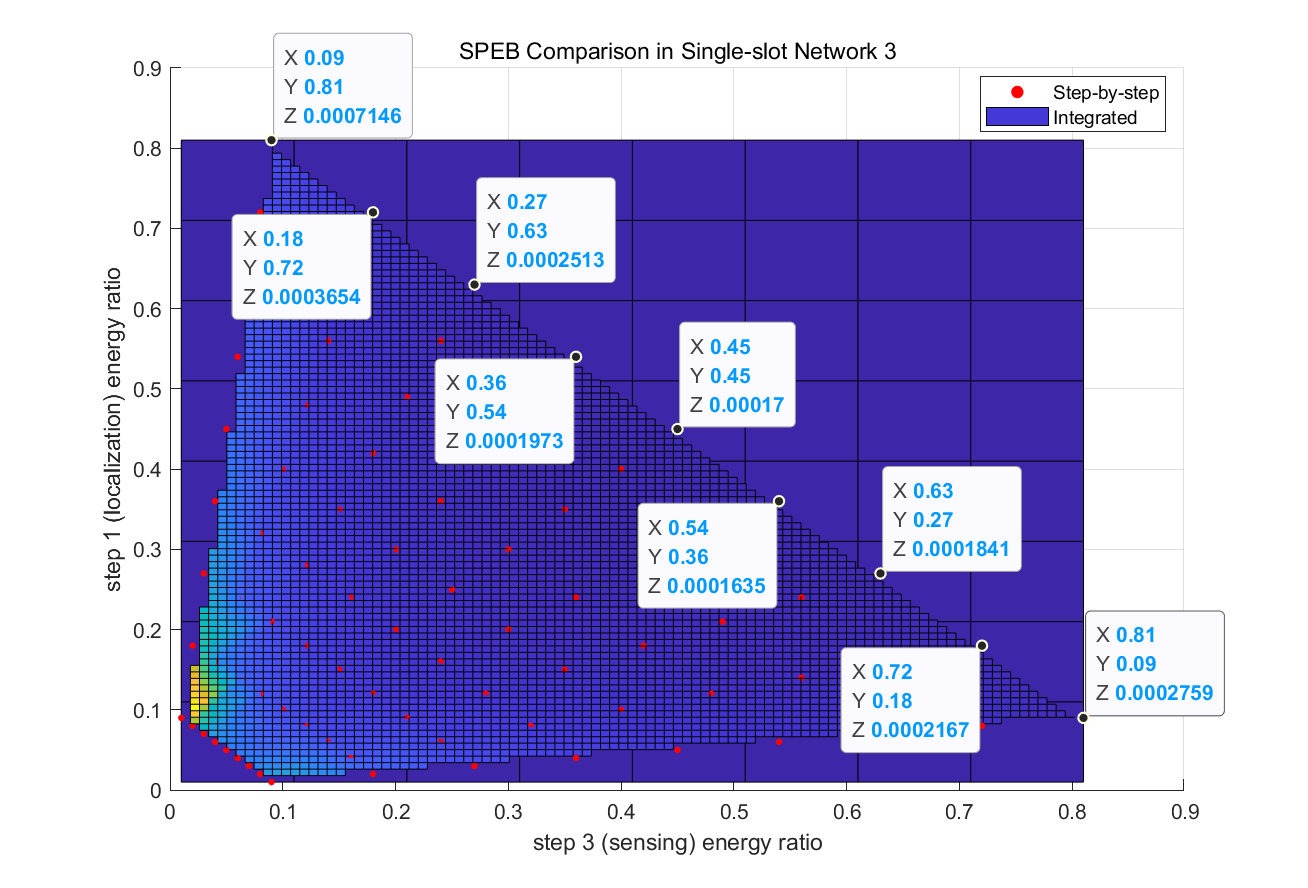}	
	}\noindent
	\subfloat[]{
		\label{asyn_single_comparison4_locsen1}
		\includegraphics[width=0.24\linewidth]{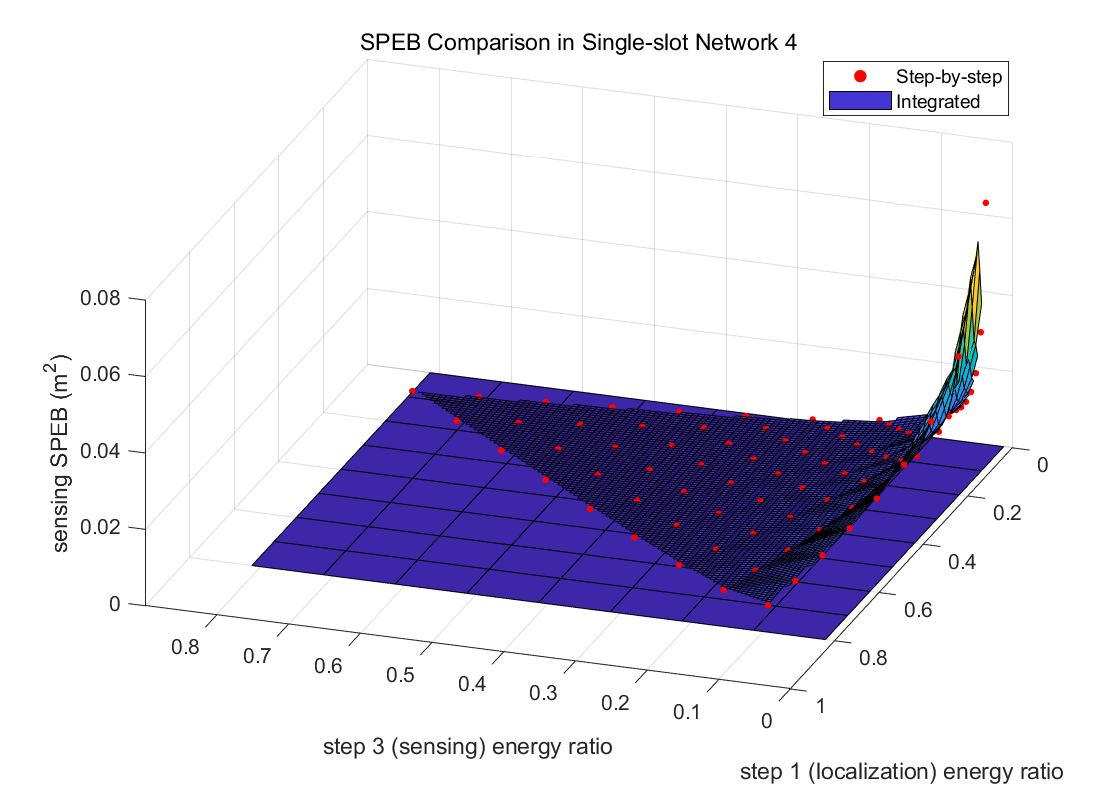}	
	}\noindent
	\subfloat[]{
		\label{asyn_single_comparison4_locsen2}
		\includegraphics[width=0.24\linewidth]{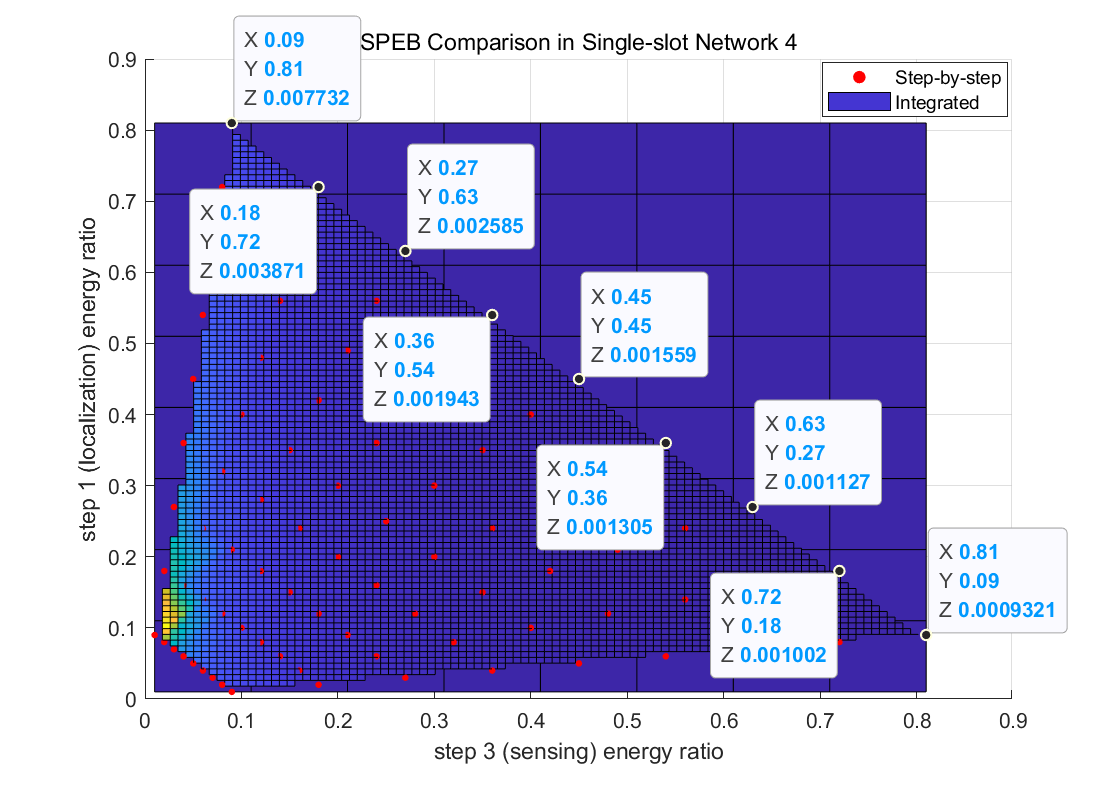}
	}
\end{minipage}
\vskip -0.4cm 
\begin{minipage}{1\linewidth}	
    \captionsetup[subfloat]{justification=centering}
	\subfloat[]{
		\label{asyn_single_comparison3_locdrift1}
		\includegraphics[width=0.24\linewidth]{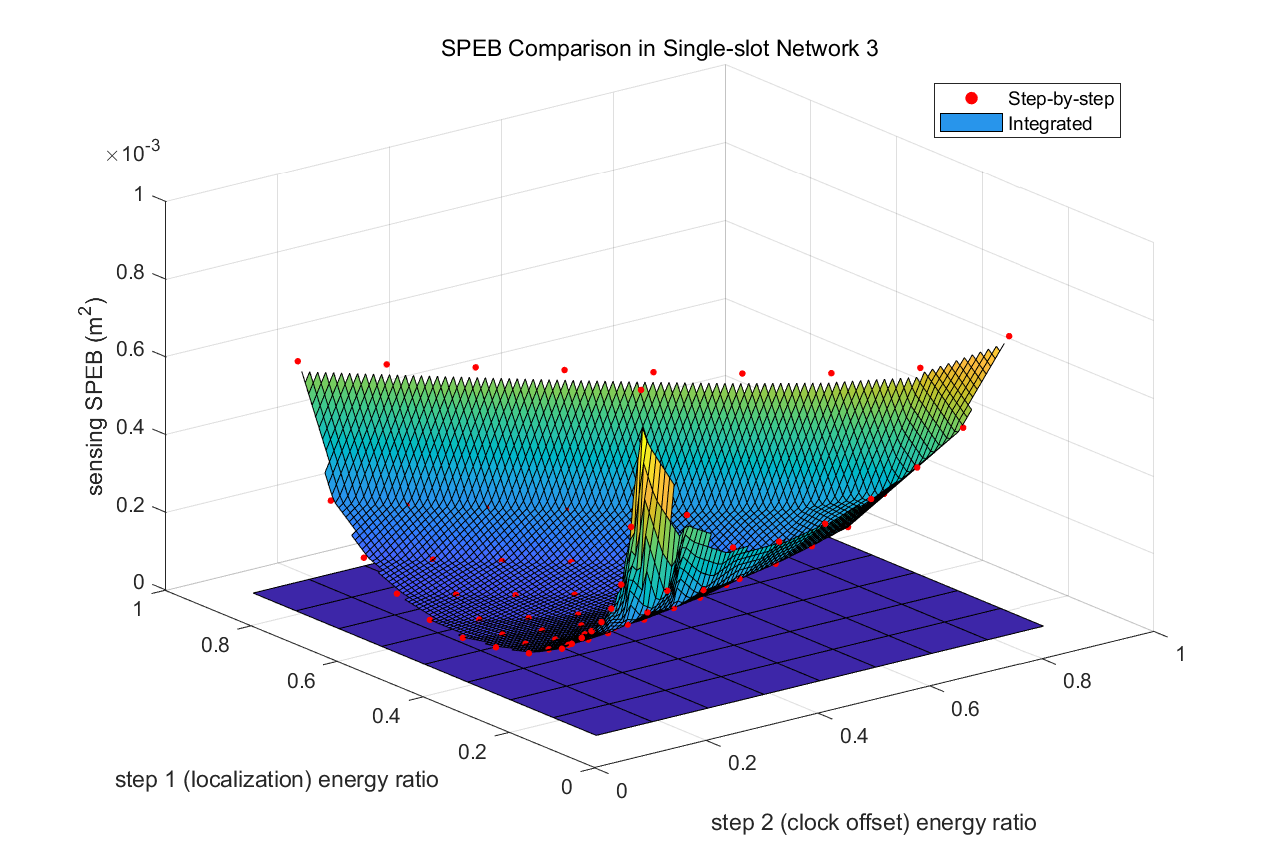}	
	}\noindent
	\subfloat[]{
		\label{asyn_single_comparison3_locdrift2}
		\includegraphics[width=0.24\linewidth]{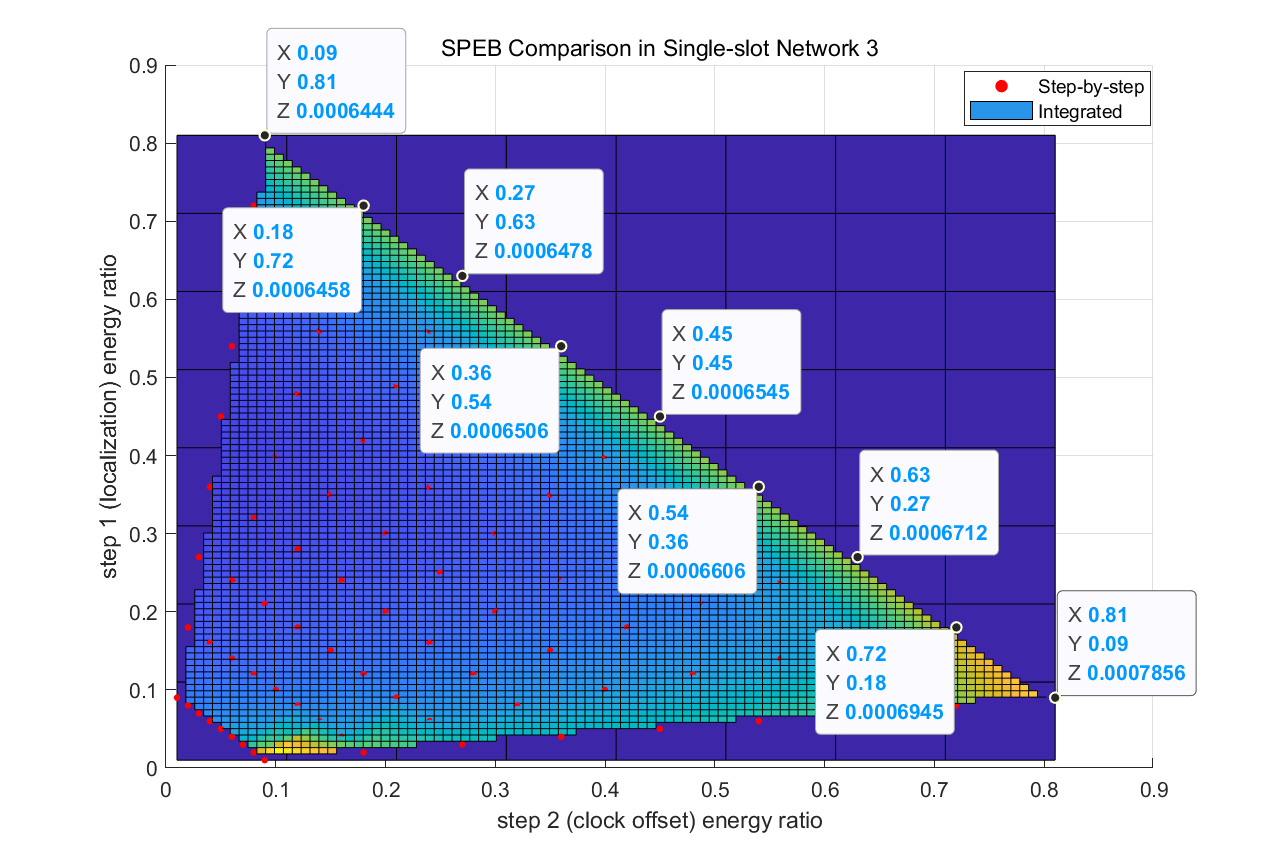}	
	}\noindent
	\subfloat[]{
		\label{asyn_single_comparison4_locdrift1}
		\includegraphics[width=0.24\linewidth]{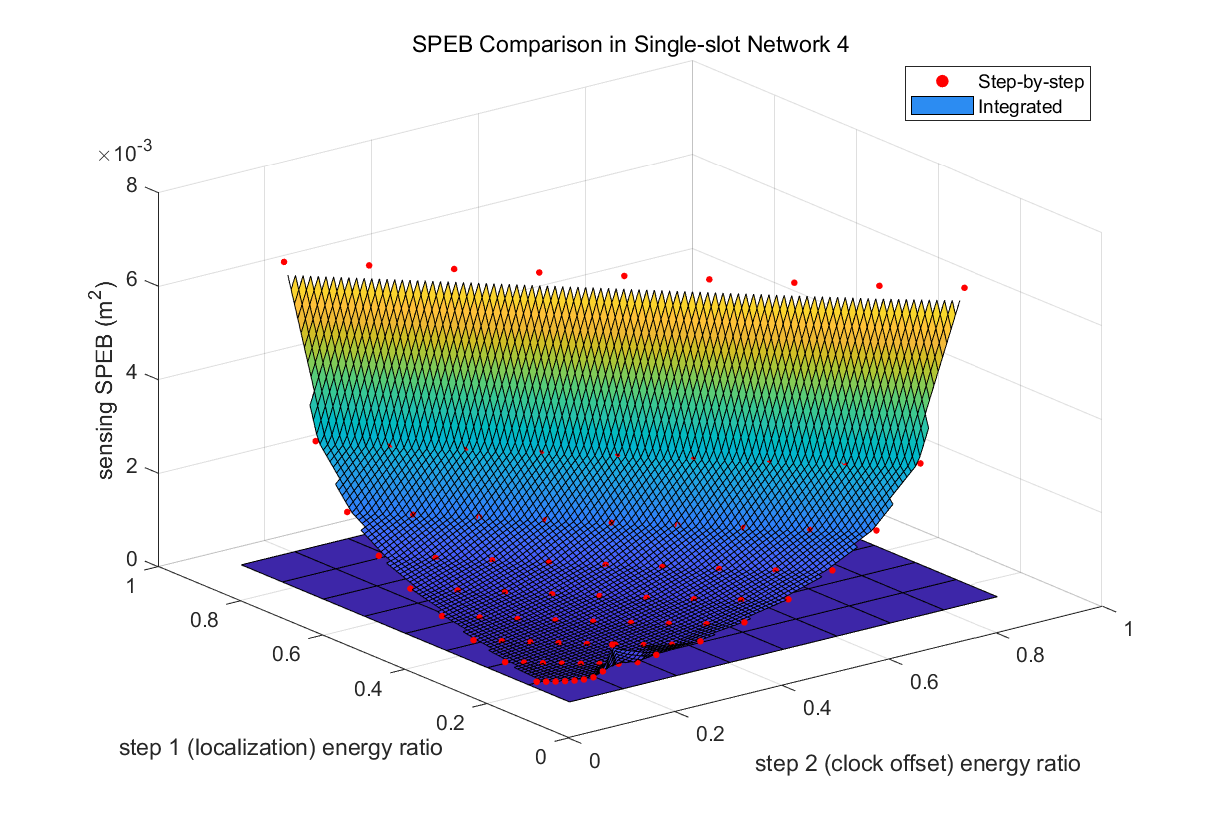}	
	}\noindent
	\subfloat[]{
		\label{asyn_single_comparison4_locdrift2}
		\includegraphics[width=0.24\linewidth]{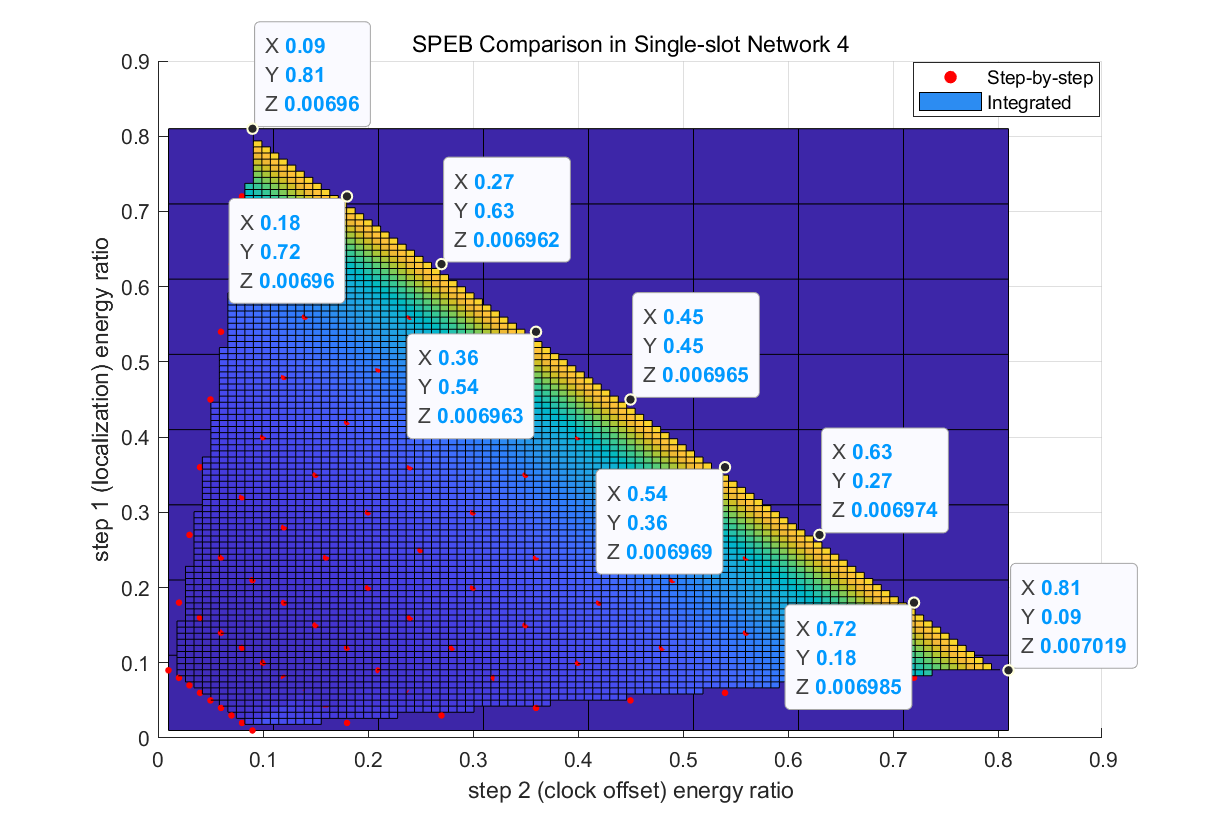}
	}
\end{minipage}
\vskip -0.4cm 
\begin{minipage}{1\linewidth}	
    \captionsetup[subfloat]{justification=centering}
	\subfloat[]{
		\label{asyn_single_comparison3_sendrift1}
		\includegraphics[width=0.24\linewidth]{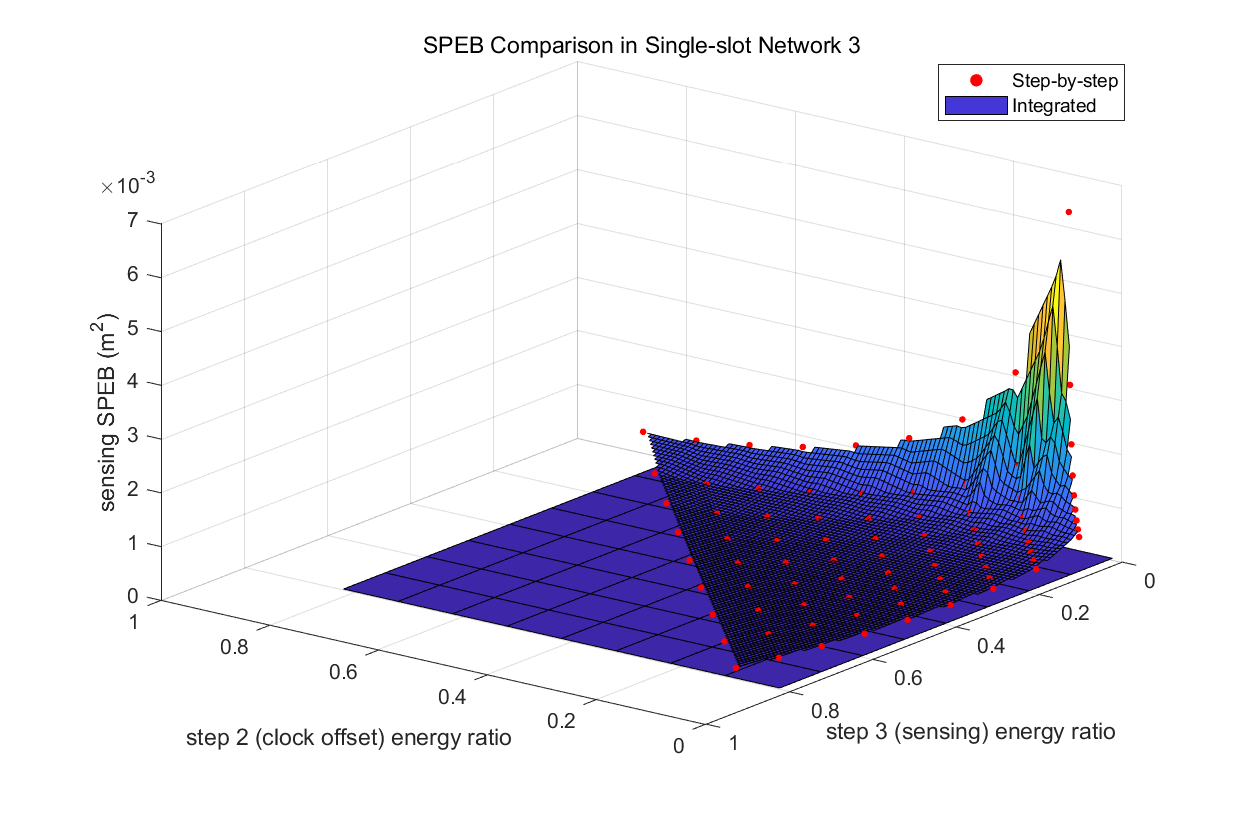}	
	}\noindent
	\subfloat[]{
		\label{asyn_single_comparison3_sendrift2}
		\includegraphics[width=0.24\linewidth]{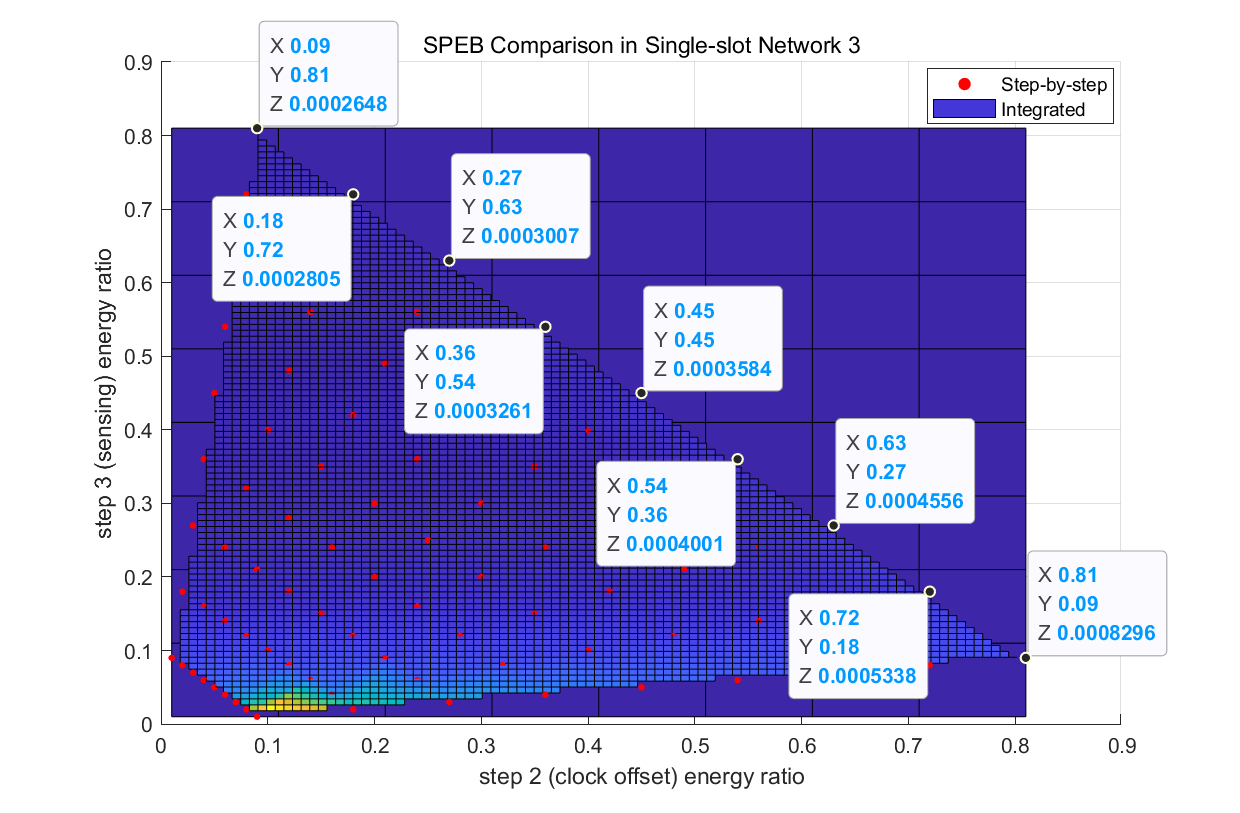}	
	}\noindent
	\subfloat[]{
		\label{asyn_single_comparison4_sendrift1}
		\includegraphics[width=0.24\linewidth]{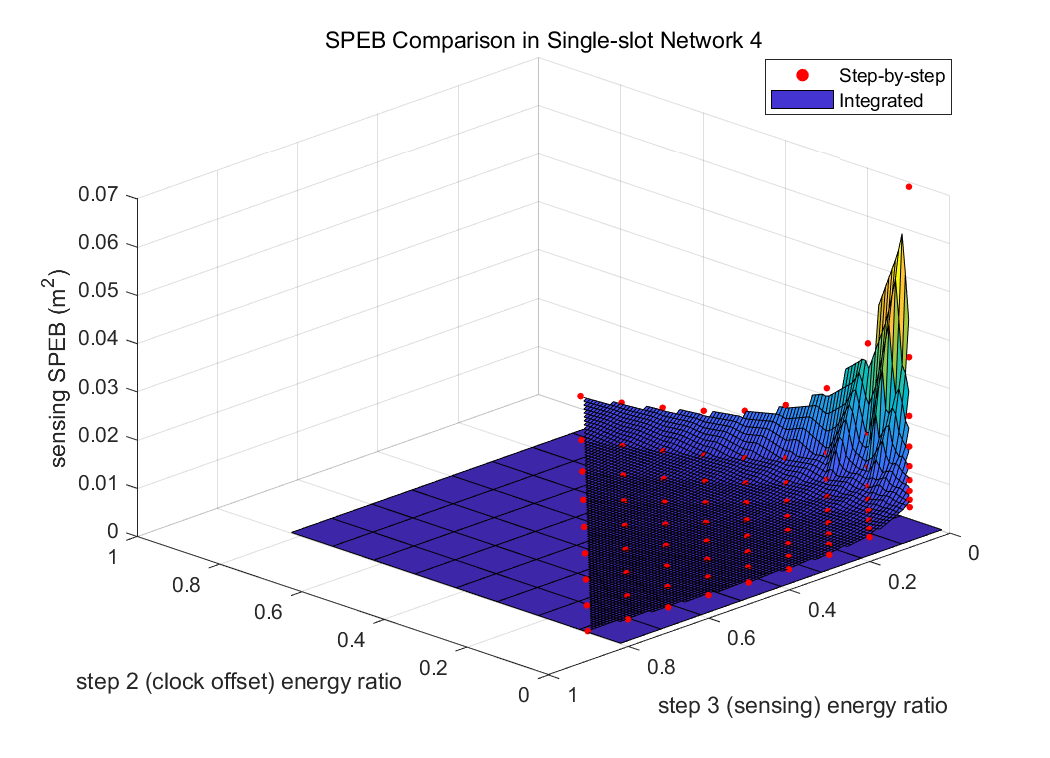}	
	}\noindent
	\subfloat[]{
		\label{asyn_single_comparison4_sendrift2}
		\includegraphics[width=0.24\linewidth]{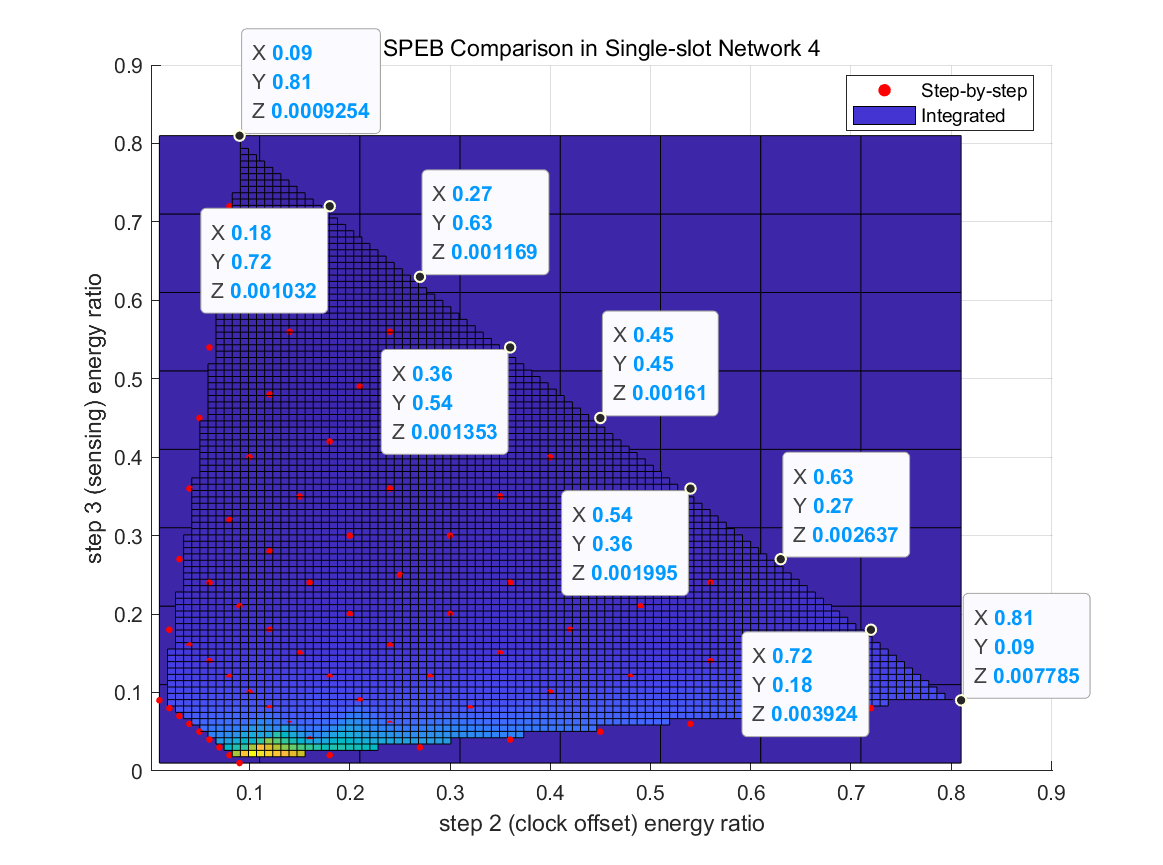}
	}
\end{minipage}
\vspace{-0.1in}	
\end{adjustwidth}
\caption{SPEB comparison between step-by-step scheme and integrated scheme in four single-slot asynchronous networks. The images in the second and fourth columns are the plans of the stereograms in the first and third columns, respectively. (a-b), (e-f) and (i-j) are respectively SPEB comparison with step 2 energy, step 3 energy and step 1 energy fixed in single-slot asynchronous network (\ref{singleslot_network1}). (c-d), (g-h) and (k-l) are respectively SPEB comparison with step 2 energy, step 3 energy and step 1 energy fixed in single-slot asynchronous network (\ref{singleslot_network2}). (m-n), (q-r) and (u-v) are respectively SPEB comparison with step 2 energy, step 3 energy and step 1 energy fixed in single-slot asynchronous network (\ref{singleslot_network3}). (o-p), (s-t) and (w-x) are respectively SPEB comparison with step 2 energy, step 3 energy and step 1 energy fixed in single-slot asynchronous network (\ref{singleslot_network4}).}
\vspace{0in}		
\label{asyn_single_comparison}
\end{figure}

In the asynchronous single-slot ISAL networks, due to the increase of clock difference vector $\bm{\tau}$ in the parameter vectors to be estimated, when analyzing the energy optimization allocation between each step of the step-by-step scheme, we adopt the method of fixing the energy allocated to one parameter vector and analyzing the energy allocation relationship between the other two parameter vectors. For the asynchronous single-slot ISAL networks, as shown in Figure \ref{asyn_single_comparison}, we provide comparison of the sensing SPEB results of the step-by-step scheme and the integrated scheme in the four networks with different topologies in Figure \ref{singleslot_networks}. \par
According to Figure \ref{asyn_single_comparison}, we can have the following observations. Firstly, as shown in Figure \ref{asyn_single_comparison1_locsen1}-\ref{asyn_single_comparison2_locsen2},\ref{asyn_single_comparison3_locsen1}-\ref{asyn_single_comparison4_locsen2}, when the energy allocated to step 2 for estimating clock difference is fixed, there is still a tradeoff between localization and sensing. Moreover, when the network nodes are relatively evenly distributed, the difficulty of target sensing is still much higher than that of radar positioning, which is consistent with the conclusion of synchronous single-slot networks. From Figure \ref{asyn_single_comparison1_locdrift1}-\ref{asyn_single_comparison2_locdrift2},\ref{asyn_single_comparison3_locdrift1}-\ref{asyn_single_comparison4_locdrift2}, it can be seen that when the energy allocated to step 3 for target sensing is fixed, the more energy allocated to step 1 for radar positioning, the better the sensing accuracy. That means, in the four single-slot asynchronous networks with different topologies, the difficulty of radar positioning is higher than that of estimating clock difference. Similarly, from Figure \ref{asyn_single_comparison1_sendrift1}-\ref{asyn_single_comparison2_sendrift2},\ref{asyn_single_comparison3_sendrift1}-\ref{asyn_single_comparison4_sendrift2}, it can be seen that when the energy allocated to step 1 for radar positioning is fixed, the more energy allocated to step 3 for target sensing, the better the sensing accuracy. That means, in the four single-slot asynchronous networks with different topologies, the difficulty of target sensing is higher than that of estimating clock difference as well. In addition, similar to single-slot synchronous networks, in single-slot asynchronous networks, when the optimization objective function is the sensing SPEB, the step-by-step optimization scheme is more suitable for sensing-resource-abundant networks, as shown in Figure \ref{asyn_single_comparison1_locsen1}-\ref{asyn_single_comparison2_sendrift2}. The integrated optimization scheme is more suitable for sensing-resource-deficient networks, as shown in Figure \ref{asyn_single_comparison3_locsen1}-\ref{asyn_single_comparison4_sendrift2}. \par
For the conclusion that the difficulty of estimating clock difference in the single-slot asynchronous networks is lower than that of radar positioning and target sensing, we give the following interpretation: as the example given in the Appendix A, compared to (\ref{partial_radar}-\ref{partial_target}), the elements of (\ref{partial_clock}) do not contain the coefficient $\frac{1}{c}$. Therefore, step 2 in which the clock difference is estimated in the step-by-step scheme requires less energy than the other two steps. \par
Above all, for the practical asynchronous ISAL networks with relatively uniform node distribution, if the step-by-step optimization scheme is adopted, in order to reduce the time complexity of the step-by-step scheme, we can set the energy allocated to each of the three steps in the step-by-step scheme according to the size relationship as: energy for clock offset $<$ energy for radar localization $<$ energy for target sensing, thereby reducing the times of traversal.
\subsubsection{Dual-slot Networks}
\begin{figure}[htbp]
\begin{adjustwidth}{-\extralength}{0cm}
\centering
\vspace{0in}
\begin{minipage}{1\linewidth}	
    \captionsetup[subfloat]{justification=centering}
	\subfloat[]{
		\label{asyn_dual_comparison1}
		\includegraphics[width=0.49\linewidth]{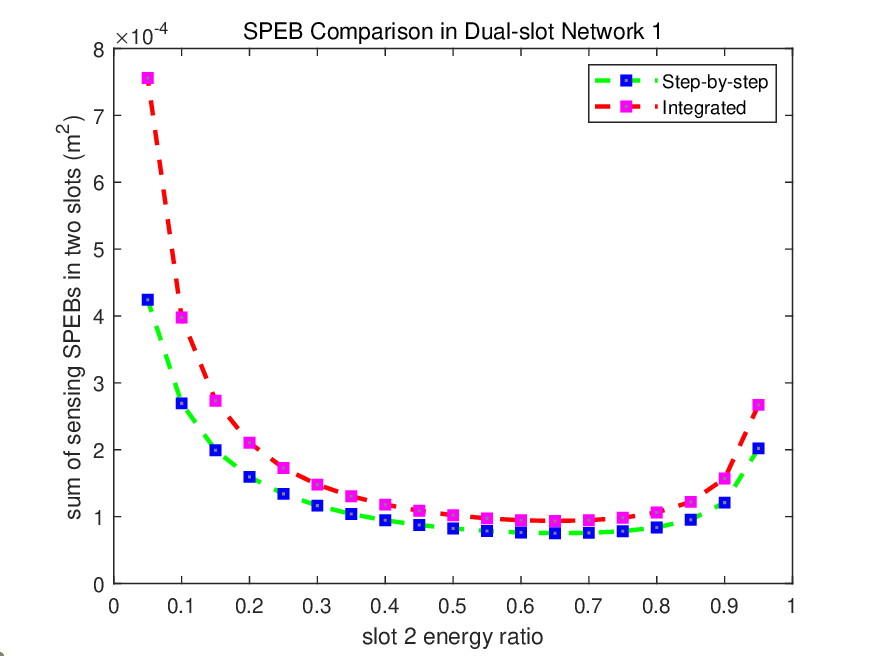}	
	}\noindent
	\subfloat[]{
		\label{asyn_dual_comparison2}
		\includegraphics[width=0.49\linewidth]{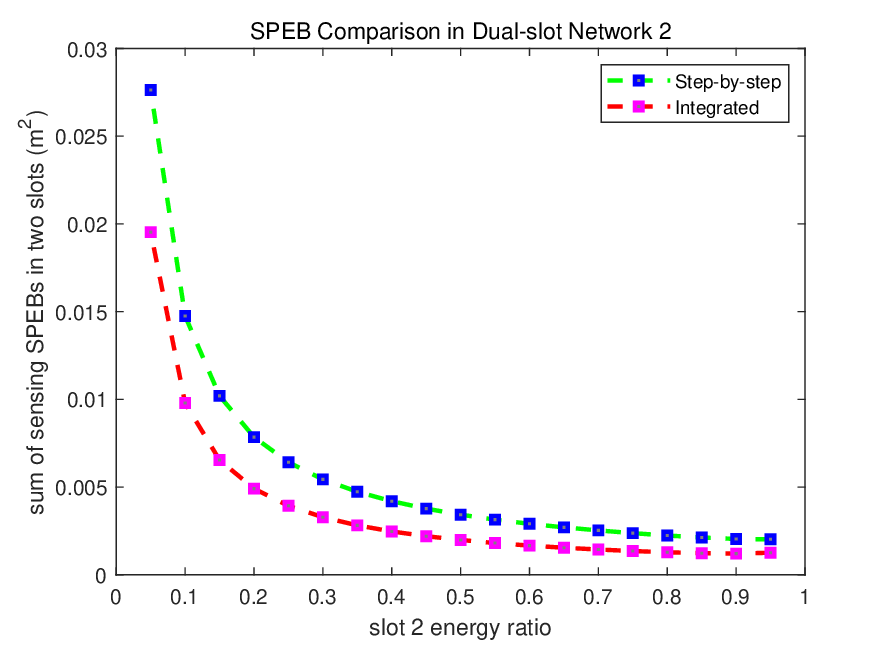}
	}
\end{minipage}
\vskip -0.3cm 
\begin{minipage}{1\linewidth }
    \captionsetup[subfloat]{justification=centering}
	\subfloat[]{
		\label{asyn_dual_comparison3}
		\includegraphics[width=0.49\linewidth]{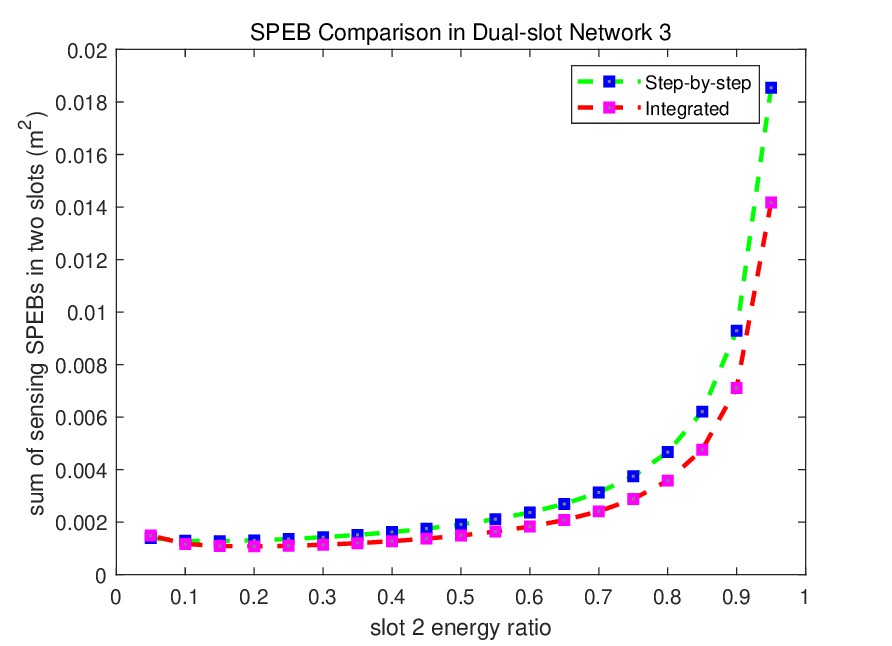
		}
	}\noindent
	\subfloat[]{
		\label{asyn_dual_comparison4}
		\includegraphics[width=0.49\linewidth]{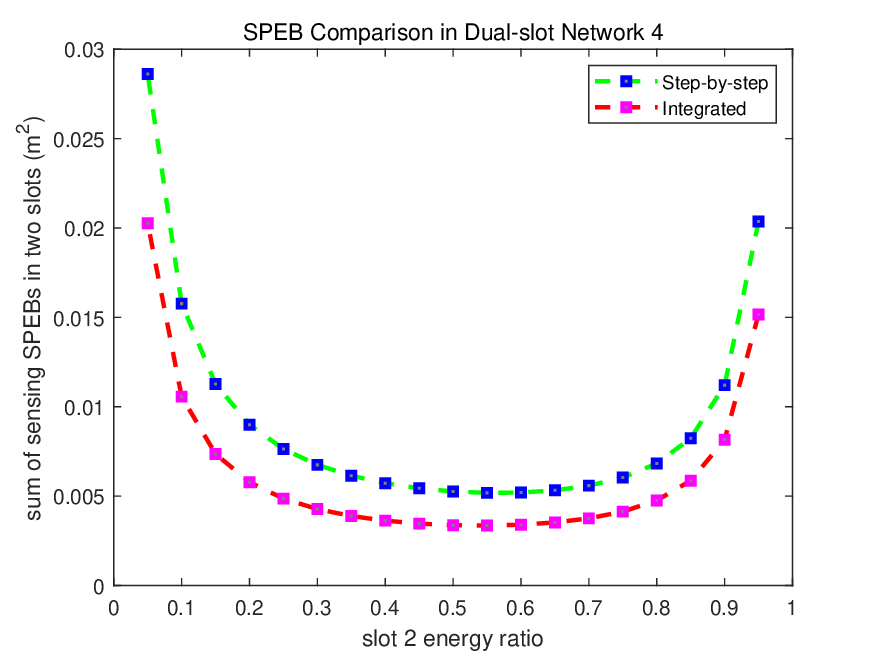
		}
	}
\end{minipage}
\vspace{0in}	
\end{adjustwidth}
\caption{SPEB comparison between step-by-step scheme and integrated scheme in four dual-slot asynchronous networks. (a) SPEB comparison in dual-slot asynchronous network (\ref{dualslot_network1}). (b) SPEB comparison in dual-slot asynchronous network (\ref{dualslot_network2}). (c) SPEB comparison in dual-slot asynchronous network (\ref{dualslot_network3}). (d) SPEB comparison in dual-slot asynchronous network (\ref{dualslot_network4}).}
\vspace{0in}		
\label{asyn_dual_comparison}
\end{figure}

For asynchronous dual-slot ISAL networks, as shown in Figure \ref{asyn_dual_comparison}, we give the sensing SPEB of the step-by-step scheme and the integrated scheme in the four networks with different topologies in Figure \ref{dualslot_networks}. \par
It is easy to find from Figure \ref{asyn_dual_comparison} that the energy optimization allocation strategy between the two slots in the dual-slot asynchronous networks is consistent with that in the dual-slot synchronous networks.
\section{Conclusions}
In this article, we propose single-slot and dual-slot ISAL network models, and provide the derivation of fundamental limits in synchronous and asynchronous networks respectively. For the optimization allocation of energy and power in the resource-constrained networks, this article proposes two optimization schemes: a step-by-step scheme and an integrated scheme, and provides solutions from the perspective of energy optimization allocation to solve the problem of the high time complexity of the step-by-step scheme. In addition, we also summarized the suitable scenarios for the step-by-step scheme and the integrated scheme by comparing the optimization results of the two optimization schemes in the networks with different topologies. From the simulation results, we can draw the following regular conclusions: (i) In the sensing-resource-deficient networks with uniformly distributed nodes, if a step-by-step scheme is used for resource optimization, in order to reduce the time complexity of the step-by-step scheme, we can allocate energy for each step in the step-by-step scheme according to the size relationship as: energy for clock offset $<$ energy for radar localization $<$ energy for target sensing, thereby reducing the times of traversal. (ii) In the multi-slot ISAL networks, when the optimization objective function is the sum of sensing SPEBs in the multiple time slots, more energy will be allocated to the time slots in which the networks are sensing-resource-deficient. (iii) The step-by-step optimization scheme is more suitable for sensing-resource-abundant networks, while the integrated optimization scheme is more suitable for sensing-resource-deficient networks. The regular conclusions summarized in this article provide valuable references for the resource optimization allocation in practical ISAL networks.


\vspace{6pt} 


\authorcontributions{Conceptualization, R.Z. and T.Z.; methodology, R.Z. and J.Y.; software, R.Z. and M.J.; validation, J.Y. and T.Z.; formal analysis, R.Z. and J.Y.; investigation, R.Z. and M.J.; resources, T.Z.; data curation, R.Z.; writing---original draft preparation, R.Z.; writing---review and editing, R.Z., T.Z. and J.Y.; visualization, R.Z.; supervision, T.Z.; project administration, T.Z.; funding acquisition, T.Z. All authors have read and agreed to the published version of the manuscript.}

\funding{This research received no external funding.}

\institutionalreview{Not applicable.}

\informedconsent{Not applicable.}

\dataavailability{The data presented in this research are available upon request from the corresponding author.} 



\conflictsofinterest{The authors declare no conflict of interest.} 

\appendixtitles{yes} 
\appendixstart
\appendix
\section{Partial derivative matrices derivation}
\subsection{Synchronous networks}
Taking a synchronous ISAL network containing 2 radars (numbered 1 and 2), 2 anchors (numbered 3 and 4) and 1 target (denoted as $tar$) as an example, we give the expressions of $\frac{\partial {\bm{\gamma}}}{\partial {\mathbf{p}_{\text{r}}}}$ and $\frac{\partial {\bm{\gamma}}}{\partial {\mathbf{p}_{\text{t}}}}$ as (\ref{partial_radar}-\ref{partial_target}).
\begin{equation}
\setlength{\arraycolsep}{0pt}
\frac{\partial {\bm{\gamma}}}{\partial {\mathbf{p}_{\text{r}}}} = \left[
\begin{array}{cccc}
\frac{\partial {\tau_{1,1}^{\text{sen}}}}{\partial x_1} & \frac{\partial {\tau_{1,1}^{\text{sen}}}}{\partial y_1} & \frac{\partial {\tau_{1,1}^{\text{sen}}}}{\partial x_2} & \frac{\partial {\tau_{1,1}^{\text{sen}}}}{\partial y_2}\\
\vdots & \vdots & \vdots & \vdots\\
\frac{\partial {\tau_{4,4}^{\text{sen}}}}{\partial x_1} & \frac{\partial {\tau_{4,4}^{\text{sen}}}}{\partial y_1} & \frac{\partial {\tau_{4,4}^{\text{sen}}}}{\partial x_2} & \frac{\partial {\tau_{4,4}^{\text{sen}}}}{\partial y_2}\\
\frac{\partial {\tau_{1,2}^{\text{ran}}}}{\partial x_1} & \frac{\partial {\tau_{1,2}^{\text{ran}}}}{\partial y_1} & \frac{\partial {\tau_{1,2}^{\text{ran}}}}{\partial x_2} & \frac{\partial {\tau_{1,2}^{\text{ran}}}}{\partial y_2}\\
\vdots & \vdots & \vdots & \vdots\\
\frac{\partial {\tau_{4,2}^{\text{ran}}}}{\partial x_1} & \frac{\partial {\tau_{4,2}^{\text{ran}}}}{\partial y_1} & \frac{\partial {\tau_{4,2}^{\text{ran}}}}{\partial x_2} & \frac{\partial {\tau_{4,2}^{\text{ran}}}}{\partial y_2}\\
\end{array}\right]_{26\times4} 
= \left[
\begin{array}{cccc}
 \frac{2 \cos \left(\emptyset _{\text{tar},1}\right)}{c} & \frac{2 \sin \left(\emptyset _{\text{tar},1}\right)}{c} & 0 & 0 \\
 \frac{\cos \left(\emptyset _{\text{tar},1}\right)}{c} & \frac{\sin \left(\emptyset _{\text{tar},1}\right)}{c} & \frac{\cos \left(\emptyset _{\text{tar},2}\right)}{c} & \frac{\sin \left(\emptyset _{\text{tar},2}\right)}{c} \\
 \frac{\cos \left(\emptyset _{\text{tar},1}\right)}{c} & \frac{\sin \left(\emptyset _{\text{tar},1}\right)}{c} & 0 & 0 \\
 \frac{\cos \left(\emptyset _{\text{tar},1}\right)}{c} & \frac{\sin \left(\emptyset _{\text{tar},1}\right)}{c} & 0 & 0 \\
 \frac{\cos \left(\emptyset _{\text{tar},1}\right)}{c} & \frac{\sin \left(\emptyset _{\text{tar},1}\right)}{c} & \frac{\cos \left(\emptyset _{\text{tar},2}\right)}{c} & \frac{\sin \left(\emptyset _{\text{tar},2}\right)}{c} \\
 0 & 0 & \frac{2 \cos \left(\emptyset _{\text{tar},2}\right)}{c} & \frac{2 \sin \left(\emptyset _{\text{tar},2}\right)}{c} \\
 0 & 0 & \frac{\cos \left(\emptyset _{\text{tar},2}\right)}{c} & \frac{\sin \left(\emptyset _{\text{tar},2}\right)}{c} \\
 0 & 0 & \frac{\cos \left(\emptyset _{\text{tar},2}\right)}{c} & \frac{\sin \left(\emptyset _{\text{tar},2}\right)}{c} \\
 \frac{\cos \left(\emptyset _{\text{tar},1}\right)}{c} & \frac{\sin \left(\emptyset _{\text{tar},1}\right)}{c} & 0 & 0 \\
 0 & 0 & \frac{\cos \left(\emptyset _{\text{tar},2}\right)}{c} & \frac{\sin \left(\emptyset _{\text{tar},2}\right)}{c} \\
 0 & 0 & 0 & 0 \\
 0 & 0 & 0 & 0 \\
 \frac{\cos \left(\emptyset _{\text{tar},1}\right)}{c} & \frac{\sin \left(\emptyset _{\text{tar},1}\right)}{c} & 0 & 0 \\
 0 & 0 & \frac{\cos \left(\emptyset _{\text{tar},2}\right)}{c} & \frac{\sin \left(\emptyset _{\text{tar},2}\right)}{c} \\
 0 & 0 & 0 & 0 \\
 0 & 0 & 0 & 0 \\
 \frac{\cos \left(\emptyset _{2,1}\right)}{c} & \frac{\sin \left(\emptyset _{2,1}\right)}{c} & \frac{\cos \left(\emptyset _{1,2}\right)}{c} & \frac{\sin \left(\emptyset _{1,2}\right)}{c} \\
 \frac{\cos \left(\emptyset _{3,1}\right)}{c} & \frac{\sin \left(\emptyset _{3,1}\right)}{c} & 0 & 0 \\
 \frac{\cos \left(\emptyset _{4,1}\right)}{c} & \frac{\sin \left(\emptyset _{4,1}\right)}{c} & 0 & 0 \\
 \frac{\cos \left(\emptyset _{2,1}\right)}{c} & \frac{\sin \left(\emptyset _{2,1}\right)}{c} & \frac{\cos \left(\emptyset _{1,2}\right)}{c} & \frac{\sin \left(\emptyset _{1,2}\right)}{c} \\
 0 & 0 & \frac{\cos \left(\emptyset _{3,2}\right)}{c} & \frac{\sin \left(\emptyset _{3,2}\right)}{c} \\
 0 & 0 & \frac{\cos \left(\emptyset _{4,2}\right)}{c} & \frac{\sin \left(\emptyset _{4,2}\right)}{c} \\
 \frac{\cos \left(\emptyset _{3,1}\right)}{c} & \frac{\sin \left(\emptyset _{3,1}\right)}{c} & 0 & 0 \\
 0 & 0 & \frac{\cos \left(\emptyset _{3,2}\right)}{c} & \frac{\sin \left(\emptyset _{3,2}\right)}{c} \\
 \frac{\cos \left(\emptyset _{4,1}\right)}{c} & \frac{\sin \left(\emptyset _{4,1}\right)}{c} & 0 & 0 \\
 0 & 0 & \frac{\cos \left(\emptyset _{4,2}\right)}{c} & \frac{\sin \left(\emptyset _{4,2}\right)}{c} \\
\end{array}
\right]
\label{partial_radar}
\end{equation}
\begin{equation}
\frac{\partial {\bm{\gamma}}}{\partial {\mathbf{p}_{\text{t}}}} = \left[
\begin{array}{cc}
\frac{\partial {\tau_{1,1}^{\text{sen}}}}{\partial x_{tar}} & \frac{\partial {\tau_{1,1}^{\text{sen}}}}{\partial y_{tar}}\\
\vdots & \vdots\\
\frac{\partial {\tau_{4,4}^{\text{sen}}}}{\partial x_{tar}} & \frac{\partial {\tau_{4,4}^{\text{sen}}}}{\partial y_{tar}}\\
\frac{\partial {\tau_{1,2}^{\text{ran}}}}{\partial x_{tar}} & \frac{\partial {\tau_{1,2}^{\text{ran}}}}{\partial y_{tar}}\\
\vdots & \vdots\\
\frac{\partial {\tau_{4,2}^{\text{ran}}}}{\partial x_{tar}} & \frac{\partial {\tau_{4,2}^{\text{ran}}}}{\partial y_{tar}}\\
\end{array}\right]_{26\times2}
=\left[
\begin{array}{cc}
 \frac{2 \cos \left(\emptyset _{1,\text{tar}}\right)}{c} & \frac{2 \sin \left(\emptyset _{1,\text{tar}}\right)}{c} \\
 \frac{\cos \left(\emptyset _{1,\text{tar}}\right)+\cos \left(\emptyset _{2,\text{tar}}\right)}{c} & \frac{\sin \left(\emptyset _{1,\text{tar}}\right)+\sin \left(\emptyset _{2,\text{tar}}\right)}{c} \\
 \frac{\cos \left(\emptyset _{1,\text{tar}}\right)+\cos \left(\emptyset _{3,\text{tar}}\right)}{c} & \frac{\sin \left(\emptyset _{1,\text{tar}}\right)+\sin \left(\emptyset _{3,\text{tar}}\right)}{c} \\
 \frac{\cos \left(\emptyset _{1,\text{tar}}\right)+\cos \left(\emptyset _{4,\text{tar}}\right)}{c} & \frac{\sin \left(\emptyset _{1,\text{tar}}\right)+\sin \left(\emptyset _{4,\text{tar}}\right)}{c} \\
 \frac{\cos \left(\emptyset _{1,\text{tar}}\right)+\cos \left(\emptyset _{2,\text{tar}}\right)}{c} & \frac{\sin \left(\emptyset _{1,\text{tar}}\right)+\sin \left(\emptyset _{2,\text{tar}}\right)}{c} \\
 \frac{2 \cos \left(\emptyset _{2,\text{tar}}\right)}{c} & \frac{2 \sin \left(\emptyset _{2,\text{tar}}\right)}{c} \\
 \frac{\cos \left(\emptyset _{2,\text{tar}}\right)+\cos \left(\emptyset _{3,\text{tar}}\right)}{c} & \frac{\sin \left(\emptyset _{2,\text{tar}}\right)+\sin \left(\emptyset _{3,\text{tar}}\right)}{c} \\
 \frac{\cos \left(\emptyset _{2,\text{tar}}\right)+\cos \left(\emptyset _{4,\text{tar}}\right)}{c} & \frac{\sin \left(\emptyset _{2,\text{tar}}\right)+\sin \left(\emptyset _{4,\text{tar}}\right)}{c} \\
 \frac{\cos \left(\emptyset _{1,\text{tar}}\right)+\cos \left(\emptyset _{3,\text{tar}}\right)}{c} & \frac{\sin \left(\emptyset _{1,\text{tar}}\right)+\sin \left(\emptyset _{3,\text{tar}}\right)}{c} \\
 \frac{\cos \left(\emptyset _{2,\text{tar}}\right)+\cos \left(\emptyset _{3,\text{tar}}\right)}{c} & \frac{\sin \left(\emptyset _{2,\text{tar}}\right)+\sin \left(\emptyset _{3,\text{tar}}\right)}{c} \\
 \frac{2 \cos \left(\emptyset _{3,\text{tar}}\right)}{c} & \frac{2 \sin \left(\emptyset _{3,\text{tar}}\right)}{c} \\
 \frac{\cos \left(\emptyset _{3,\text{tar}}\right)+\cos \left(\emptyset _{4,\text{tar}}\right)}{c} & \frac{\sin \left(\emptyset _{3,\text{tar}}\right)+\sin \left(\emptyset _{4,\text{tar}}\right)}{c} \\
 \frac{\cos \left(\emptyset _{1,\text{tar}}\right)+\cos \left(\emptyset _{4,\text{tar}}\right)}{c} & \frac{\sin \left(\emptyset _{1,\text{tar}}\right)+\sin \left(\emptyset _{4,\text{tar}}\right)}{c} \\
 \frac{\cos \left(\emptyset _{2,\text{tar}}\right)+\cos \left(\emptyset _{4,\text{tar}}\right)}{c} & \frac{\sin \left(\emptyset _{2,\text{tar}}\right)+\sin \left(\emptyset _{4,\text{tar}}\right)}{c} \\
 \frac{\cos \left(\emptyset _{3,\text{tar}}\right)+\cos \left(\emptyset _{4,\text{tar}}\right)}{c} & \frac{\sin \left(\emptyset _{3,\text{tar}}\right)+\sin \left(\emptyset _{4,\text{tar}}\right)}{c} \\
 \frac{2 \cos \left(\emptyset _{4,\text{tar}}\right)}{c} & \frac{2 \sin \left(\emptyset _{4,\text{tar}}\right)}{c} \\
 0 & 0 \\
 0 & 0 \\
 0 & 0 \\
 0 & 0 \\
 0 & 0 \\
 0 & 0 \\
 0 & 0 \\
 0 & 0 \\
 0 & 0 \\
 0 & 0 \\
\end{array}
\right]
\label{partial_target}
\end{equation}\par
\subsection{Asynchronous Networks}
Taking an asynchronous ISAL network containing 2 radars (numbered 1 and 2), 2 anchors (numbered 3 and 4) and 1 target (denoted as $tar$) as an example, $\bm {\tau} = [\tau_{1,2},\tau_{1,3}]^{\text{T}}$, $\frac{\partial {\bm{\zeta}}}{\partial {\mathbf{p}_{\text{r}}}}$ and $\frac{\partial {\bm{\zeta}}}{\partial {\mathbf{p}_{\text{t}}}}$ are expressed as (\ref{partial_radar}-\ref{partial_target}). We give the expression of $\frac{\partial {\bm{\zeta}}}{\partial {\bm{\tau}}}$ as (\ref{partial_clock}).
\begin{equation}
\frac{\partial {\bm{\zeta}}}{\partial {\bm{\tau}}} = \left[
\begin{array}{cc}
\frac{\partial \zeta_{1,1}^{\text{sen}}}{\partial {\tau_{1,2}}} & \frac{\partial \zeta_{1,1}^{\text{sen}}}{\partial {\tau_{1,3}}}\\
\vdots & \vdots\\
\frac{\partial \zeta_{4,4}^{\text{sen}}}{\partial {\tau_{1,2}}} & \frac{\partial \zeta_{4,4}^{\text{sen}}}{\partial {\tau_{1,3}}}\\
\frac{\partial \zeta_{1,2}^{\text{ran}}}{\partial {\tau_{1,2}}} & \frac{\partial \zeta_{1,2}^{\text{ran}}}{\partial {\tau_{1,3}}}\\
\vdots & \vdots\\
\frac{\partial \zeta_{4,2}^{\text{ran}}}{\partial {\tau_{1,2}}} & \frac{\partial \zeta_{4,2}^{\text{ran}}}{\partial {\tau_{1,3}}}\\
\end{array}\right]_{26\times2}
= \begin{small}
\left[
\begin{array}{cc}
 0 & 0 \\
 1 & 0 \\
 0 & 1 \\
 0 & 1 \\
 -1 & 0 \\
 0 & 0 \\
 -1 & 1 \\
 -1 & 1 \\
 0 & -1 \\
 1 & -1 \\
 0 & 0 \\
 0 & 0 \\
 0 & -1 \\
 1 & -1 \\
 0 & 0 \\
 0 & 0 \\
 1 & 0 \\
 0 & 1 \\
 0 & 1 \\
 -1 & 0 \\
 -1 & 1 \\
 -1 & 1 \\
 0 & -1 \\
 1 & -1 \\
 0 & -1 \\
 1 & -1 \\
\end{array}
\right]
\end{small}
\label{partial_clock}
\end{equation}\par

\begin{adjustwidth}{-\extralength}{0cm}

\reftitle{References}


\bibliography{energyallocation}

%


\PublishersNote{}
\end{adjustwidth}
\end{document}